\documentclass[11pt,tightenlines,onecolumn,preprintnumbers,superscriptaddress,nofootinbib]{revtex4}

\usepackage[a4paper,left=2.5cm,right=2.5cm, top=2.0cm,bottom=1.5cm]{geometry}
\usepackage[english]{babel}
\usepackage[applemac]{inputenc}

\usepackage{hyperref}
\usepackage{amsfonts, amsmath, amssymb}
\usepackage{graphicx}
\usepackage{booktabs, multirow}
\usepackage{color}
\usepackage{afterpage}
\usepackage[detect-all]{siunitx}

\makeatletter
\g@addto@macro\bfseries{\boldmath}
\makeatother

\usepackage{relsize}
\newcommand{\babar}{\mbox{\slshape B\kern-0.1em{\smaller A}\kern-0.1em B\kern-0.1em{\smaller A\kern-0.2em R}}}

\newcommand{\be}{\begin{equation}}
\newcommand{\ee}{\end{equation}}
\newcommand{\ba}{\begin{eqnarray}}
\newcommand{\ea}{\end{eqnarray}}
\newcommand{\bi}{\begin{itemize}}
\newcommand{\ei}{\end{itemize}}

\newcommand{\Tr}{{\rm Tr\,}}

\newcommand{\<}{\langle}
\renewcommand{\>}{\rangle}

\newcommand{\la}{\label}

\newcommand{\Q}{{\cal Q}}

\newcommand{\dof}{\mathrm{d.o.f.}}

\newcommand{\MTT}{\mathcal{M}_{TT}}
\newcommand{\MTTa}{\mathcal{M}_{TT}^{a}}
\newcommand{\MTTt}{\mathcal{M}_{TT}^{\tau}}
\newcommand{\MTL}{\mathcal{M}_{TL}}
\newcommand{\MLT}{\mathcal{M}_{LT}}
\newcommand{\MTLt}{\mathcal{M}_{TL}^{\tau}}
\newcommand{\MTLa}{\mathcal{M}_{TL}^{a}}
\newcommand{\MLL}{\mathcal{M}_{LL}}

\newcommand{\MsTT}{\overline{\mathcal{M}}_{TT}}
\newcommand{\MsTTa}{\overline{\mathcal{M}}_{TT}^{a}}
\newcommand{\MsTTt}{\overline{\mathcal{M}}_{TT}^{\tau}}
\newcommand{\MsTL}{\overline{\mathcal{M}}_{TL}}
\newcommand{\MsLT}{\overline{\mathcal{M}}_{LT}}
\newcommand{\MsTLt}{\overline{\mathcal{M}}_{TL}^{\tau}}
\newcommand{\MsTLa}{\overline{\mathcal{M}}_{TL}^{a}}
\newcommand{\MsLL}{\overline{\mathcal{M}}_{LL}}

\newcommand{\fm}{\mathrm{fm}}
\newcommand{\MeV}{\mathrm{MeV}}
\newcommand{\GeV}{\mathrm{GeV}}

\newcommand{\phys}{\mathrm{phys}}       
\newcommand{\lat}{\mathrm{lat}}

\newcommand{\GammaGG}{\Gamma_{\gamma\gamma}}
\newcommand{\GammaGGt}{\tilde{\Gamma}_{\gamma\gamma}}

\preprint{MITP/17-088, DESY 17-202}

\begin{document}
%----------------------

\title{Hadronic light-by-light scattering amplitudes from lattice QCD \\ versus dispersive sum rules}

\author{Antoine G\'erardin}
\affiliation{PRISMA Cluster of Excellence \& Institut f\"ur Kernphysik,  Johannes Gutenberg-Universit\"at Mainz, 55099 Mainz, Germany}
\author{Jeremy Green}
\affiliation{John von Neumann Institute for Computing (NIC), DESY, Platanenallee 6, 15738 Zeuthen, Germany}
\author{Oleksii Gryniuk}
\affiliation{PRISMA Cluster of Excellence \& Institut f\"ur Kernphysik,  Johannes Gutenberg-Universit\"at Mainz, 55099 Mainz, Germany}
\author{Georg von Hippel} 
\affiliation{PRISMA Cluster of Excellence \& Institut f\"ur Kernphysik,  Johannes Gutenberg-Universit\"at Mainz, 55099 Mainz, Germany}
\author{Harvey B.\ Meyer}
\affiliation{PRISMA Cluster of Excellence \& Institut f\"ur Kernphysik,  Johannes Gutenberg-Universit\"at Mainz, 55099 Mainz, Germany}
\affiliation{Helmholtz Institut Mainz,  55099 Mainz, Germany}
\author{Vladimir Pascalutsa}
\affiliation{PRISMA Cluster of Excellence \& Institut f\"ur Kernphysik,  Johannes Gutenberg-Universit\"at Mainz, 55099 Mainz, Germany}
\author{Hartmut Wittig} 
\affiliation{PRISMA Cluster of Excellence \& Institut f\"ur Kernphysik,  Johannes Gutenberg-Universit\"at Mainz, 55099 Mainz, Germany}
\affiliation{Helmholtz Institut Mainz,  55099 Mainz, Germany}

\date{\today}

\begin{abstract}
The hadronic contribution to the eight forward amplitudes of light-by-light scattering
 ($\gamma^*\gamma^*\to \gamma^*\gamma^*$)  is computed in lattice QCD.
Via dispersive sum rules, the amplitudes are compared to a model
of the $\gamma^*\gamma^*\to {\rm hadrons}$ cross sections in which the fusion process
is described by hadronic resonances.
Our results thus provide an important test for the model estimates of hadronic light-by-light scattering 
in the anomalous magnetic moment of the muon, $a_\mu^{\rm HLbL}$.
Using simple parametrizations of the resonance $M\to \gamma^*\gamma^*$ transition form factors,
we determine the corresponding monopole and dipole masses by performing a global fit to all eight amplitudes.
Together with a previous dedicated calculation of the $\pi^0\to \gamma^*\gamma^*$ transition form factor,
our calculation provides valuable information for phenomenological estimates of $a_\mu^{\rm HLbL}$.
The presented calculations are performed in two-flavor QCD with pion masses extending down to 190\,MeV
at two different lattice spacings. In addition to the fully connected Wick contractions, 
on two lattice ensembles we also compute the (2+2) disconnected class of diagrams,
and find that their overall size is compatible with a parameter-free, large-$N$ inspired prediction,
where $N$ is the number of colors.
Motivated by this observation, we estimate in the same way the disconnected contribution 
to $a_\mu^{\rm HLbL}$.
\end{abstract}

\maketitle
\newpage

\tableofcontents

%%%%%%%%%%%%%%%%%%%%%%%%%%%%%%%%%%%%%%%
\section{Introduction}
%%%%%%%%%%%%%%%%%%%%%%%%%%%%%%%%%%%%%%%

The non-vanishing probability of two photons scattering off each other
is a striking prediction of quantum electrodynamics (QED)~\cite{Heisenberg1936,Karplus:1951uo}.  The
smallness of the cross section has so far prohibited a direct
experimental observation, although evidence for the phenomenon has
recently been found by the ATLAS experiment in relativistic heavy-ion
collisions~\cite{Aaboud:2017bwk}.  Equally interesting is the
scattering of spacelike virtual photons, $\gamma^*\gamma^*\to\gamma^*\gamma^*$.  
While the contributions of virtual leptons
are calculable in QED perturbatively, the hadronic contributions require
a nonperturbative approach. 
When the photons are real or spacelike,
dispersive sum rules can be used to express the forward $\gamma^*\gamma^*$ scattering amplitudes
in terms of experimentally more accessible $\gamma^*\gamma^*$-fusion 
cross sections~\cite{Pascalutsa:2010sj,Pascalutsa:2012pr,Danilkin:2016hnh,Dai:2017cvz}.
The hadronic contributions to the  $\gamma^*\gamma^*\to\gamma^*\gamma^*$ amplitudes
can also be computed \emph{ab initio} using lattice QCD~\cite{Green:2015sra}.

One very timely application of hadronic light-by-light (HLbL) scattering is the anomalous magnetic
moment of the muon, $a_\mu =\frac12 (g-2)_\mu$.  The current
discrepancy between the direct measurement of $a_\mu$ and the Standard
Model prediction amounts to about 3.6 standard
deviations~\cite{Blum:2013xva}.  While the current theory and
experimental errors are comparable in size, two new $(g-2)_\mu$
experiments~\cite{Venanzoni:2014ixa,Otani:2015lra}  in preparation at Fermilab and J-PARC 
aim at reducing the experimental error by a factor of four.  
The largest sources of theory error are contributions from the hadronic vacuum polarization (HVP) and from HLbL scattering. 
The latter is expected to dominate in the 
future in view of the dedicated measurements at $e^+e^-$ colliders
ever better constraining the former. Although an active area of
research, the experimental data needed for the recently proposed data-driven
dispersive approaches to HLBL~\cite{Colangelo:2014dfa,Colangelo:2014pva,Colangelo:2015ama,Pauk:2014rfa,Hagelstein:2017obr}  
are harder to obtain, and lattice QCD calculations are in particularly high demand.

Lattice QCD calculations of hadron structure have been steadily
advancing in recent years.  Several collaborations calculate hadronic
observables directly at physical values of the quark masses.  At least
two collaborations are addressing the HLbL contribution to $(g-2)_\mu$
on the lattice~\cite{Blum:2014oka,Blum:2015gfa,Blum:2016lnc,Green:2015sra,Green:2015mva,Asmussen:2016lse,Asmussen:2017bup}.
Although the calculation poses serious challenges due to the
complexity of the four-point function and the long-range nature of the
dominant contribution, as a spacelike quantity it is well suited for a
first-principles treatment directly in the Euclidean theory.  A second
 role of lattice QCD is to provide the necessary hadronic
input for the model and dispersive approaches to $a_\mu^{\rm HLbL}$.
Model calculations (see e.g.\ \cite{Nyffeler:2017ohp} for a recent
overview) consistently suggest that the contribution of the
pseudoscalar mesons ($\pi^0,\eta,\eta'$) is dominant, and therefore
determining their respective transition form factors is of primary
importance.  A first calculation of the $\pi^0\to \gamma^*\gamma^*$
transition form factor in the range of photon virtualities relevant to
$(g-2)_\mu$ has been carried out on the
lattice~\cite{Gerardin:2016cqj}.  An extension of this calculation to
the $\eta,\eta'$ mesons is possible, though more demanding, due to the
appearance of disconnected Wick-contraction diagrams.  Computing the
spectrum and two-photon coupling of the scalar, axial-vector and
tensor mesons is qualitatively more complicated in lattice QCD, since
these states are resonances and require a dedicated treatment.

In many ways, the light-by-light scattering
amplitudes are the more accessible observable in lattice calculations,
because they involve spacelike photons that can be treated directly in
the Euclidean theory. The lattice calculation of the cross section
$\gamma^*\gamma^*\to\pi\pi$ is for instance more complex than
calculating the cross section $\gamma^*\gamma^*\to\gamma^*\gamma^*$
for spacelike photons. In experiments, the weakness of the
electromagnetic coupling would make such a measurement impractical,
but in lattice QCD the factor $e^4$ merely multiplies a four-point
correlation function at the end of the calculation.

In this article, we compute the HLbL scattering amplitude for spacelike photons in lattice QCD.
Being parametrized by functions of six Lorentz invariants, it is a complicated object.
We focus on the forward amplitudes because they are simpler functions of three invariants and,
using the optical theorem, they are related to $\gamma^*\gamma^*\to \,{\rm hadrons}$ cross sections.
Our objectives are:
\begin{itemize}
\item[(i)] Provide a stringent test that the light-by-light amplitude
 for spacelike photons is correctly described by the type of hadronic model used so far to estimate  $a_\mu^{\rm HLbL}$.
The model includes the exchange of pseudoscalar, scalar, axial-vector and tensor mesons.
\item[(ii)] Provide information on their two-photon transition form factors via global fits.
\item[(iii)] Compare the transition form factors to phenomenological determinations based on
light-by-light sum rules.
\end{itemize}
In~\cite{Green:2015sra}, we laid out the method and computed the
forward amplitude sensitive to the total transverse $\gamma^*\gamma^*$
cross section, ${\cal M}_{\rm TT}$, via a dispersive sum rule. Here we
extend the comparison between lattice data and phenomenological
parametrizations of the $\gamma^*\gamma^*\to \,{\rm hadrons}$
cross sections to encompass all eight forward amplitudes. This more
extensive analysis allows us to place much stronger constraints on the
size of the contributions of different resonances, because they
contribute to different amplitudes with different weight factors,
often even with opposite signs.

Our lattice calculation is performed in QCD with two flavors of light
quarks; it involves pion masses down to 190\,MeV and two lattice
spacings. While a fully realistic lattice calculation would have to
include at least a dynamical strange quark, the present calculation does provide a
suitable test of hadronic models via dispersive sum rules, since at
the required level of precision it is fairly straightforward to adapt
these models to QCD without the strange quark, as discussed in section \ref{sec:model}.

Here we have computed the fully connected and the dominant
disconnected Wick-contraction diagrams.  We discuss what these classes
of diagrams correspond to in terms of the quantum numbers of the
exchanged resonances. The large-$N$ inspired approximation that a
quark loop containing a single, vector-current insertion gives a
negligible contribution, corresponds, in two-flavor QCD, to including
only isovector resonances, enhanced by a factor of 34/9.  This
interpretation of the fully connected class of diagrams was first
pointed out in~\cite{Bijnens:2015jqa,Bijnens:2016hgx}, mainly
concerning the pseudoscalar sector; see also the arguments presented
in~\cite{Jin:2016rmu}.  We rederive the result in detail in the SU(2)
and in the SU(3) flavor symmetric theory under slightly weaker
assumptions; see section~\ref{sec:34ov9}.

The paper is organized as follows.  We begin by introducing the
theoretical background for light-by-light scattering in section
\ref{sec:theobkd}.  We then describe the lattice method for computing
the hadronic light-by-light amplitude, including the analytic
continuation and the numerical method to obtain the fully connected
and the (2+2) disconnected four-point function (section~\ref{sec:sec3latmethod}).  Section \ref{sec:res} presents our
numerical results in two-flavor QCD, while some additional material is provided in appendix \ref{sec:addmat}.  After introducing the details of
the hadronic model for the $\gamma^*\gamma^*\to \,{\rm hadrons}$
cross sections in section \ref{sec:model} and appendix \ref{app:CS},
we perform fits to the lattice data in section \ref{sec:fit}. We
compare the results for the transition form factors to existing
phenomenological estimates. In section \ref{sec:disc}, 
we discuss our results on the leading disconnected diagrams to the 
HLbL amplitude and present what results would have to be obtained in
lattice QCD for the connected and the leading disconnected diagram
contributions to $a_\mu^{\rm HLbL}$
in order to confirm the state-of-the-art model estimate.
We conclude in section~\ref{sec:concl}.

%%%%%%%%%%%%%%%%%%%%%%%%%%%%%%%%%%%%%%%
\section{Forward light-by-light scattering and sum rules\label{sec:theobkd}}
%%%%%%%%%%%%%%%%%%%%%%%%%%%%%%%%%%%%%%%

In order to establish our notation, we start by recalling the dispersive sum rules
for the scattering of spacelike photons \cite{Pascalutsa:2010sj,Pascalutsa:2012pr}. Just as for real photons \cite{Gerasimov:1973ja},
they are based on unitarity and analyticity of the forward scattering amplitude. More specifically,
the optical theorem allows one to relate the absorptive part of the $\gamma^\ast(\lambda_1,q_1) \gamma^\ast(\lambda_2,q_2) 
\to \gamma^\ast(\lambda_1^{\prime},q_1) \gamma^\ast(\lambda_2^{\prime},q_2)$ forward scattering amplitude to fusion cross sections 
for the process $\gamma^\ast \gamma^\ast \to \mathrm{X}$, where X stands for any $C$-parity even final state. 
The relevant kinematic variables are the photon virtualities, $q_i^2=-Q_i^2$, ($i=1,2$),
and the crossing-symmetric variable $\nu = q_1 \cdot q_2$, which is related to the squared center-of-mass energy by $s=2\nu-Q_1^2-Q_2^2$.
Denoting the absorptive part of the helicity amplitude $\mathcal{M}_{\lambda^\prime_1 \lambda^\prime_2, \lambda_1 \lambda_2}$ by
\begin{equation}
W_{\lambda^\prime_1 \lambda^\prime_2, \lambda_1 \lambda_2} =  \mathrm{Im} \left( \mathcal{M}_{\lambda^\prime_1 \lambda^\prime_2, \lambda_1 \lambda_2} \right) \,, 
\end{equation} 
the optical theorem yields (with a factor of one half because both photons are identical, 
and $\mathrm{d}\Gamma_{\mathrm X}$ is the phase space for a final state X)
\begin{equation}
W_{\lambda^\prime_1 \lambda^\prime_2, \lambda_1 \lambda_2} =  \frac{1}{2} \int 
% \frac{ \mathrm{d}^3 p_X}{ (2\pi)^3 2E_X} 
\mathrm{d}\Gamma_X
\ (2 \pi)^4 \delta^4(q_1 + q_2 - p_\mathrm{X}) \,  {\cal M}_{\lambda_1 \lambda_2} (q_1, q_2; p_\mathrm{X}) \,  {\cal M}^\ast_{\lambda^\prime_1 \lambda^\prime_2} (q_1, q_2; p_\mathrm{X}) \,, 
\label{eq:abspart}
\end{equation}
where ${\cal M}_{\lambda_1 \lambda_2} (q_1, q_2; p_\mathrm{X})$ denotes the invariant helicity amplitude for the fusion process
\begin{equation}
\gamma^\ast(\lambda_1, q_1) + \gamma^\ast(\lambda_2, q_2) \to \mathrm{X}(p_\mathrm{X}) \,.
\label{eq:fusionCS}
\end{equation}
The helicity amplitudes are related to the Feynman amplitudes by
\begin{equation}
\mathcal{M}_{\lambda_1^{\prime}\lambda_2^{\prime}\lambda_1\lambda_2}(q_1,q_2)  = \mathcal{M}_{\mu\nu\rho\sigma}(q_1,q_2) \ \epsilon^{*\mu}(\lambda_1^{\prime},q_1)  \ \epsilon^{*\nu}(\lambda_2^{\prime},q_2) \ \epsilon^{\rho}(\lambda_1,q_1) \ \epsilon^{\sigma}(\lambda_2,q_2)     \,.
\end{equation} 
Using parity and time-reversal invariance, we are left with only eight independent amplitudes $\mathcal{M}_{\lambda^\prime_1 \lambda^\prime_2, \lambda_1 \lambda_2}$ \cite{Budnev:1971sz}. Forming linear combinations, we can consider eight amplitudes which are either even (first six amplitudes) or odd (last two amplitudes) with respect to the variable $\nu$:
\begin{gather*}
\mathcal{M}_{TT} = \frac{1}{2} (\mathcal{M}_{++,++} + \mathcal{M}_{+-,+-}) \,,\ \mathcal{M}_{TT}^{\tau} = \mathcal{M}_{++,--} \,,\   \\ 
\mathcal{M}_{TL} =\mathcal{M}_{+0,+0} \,,\ \mathcal{M}_{LT} ={\cal M}_{0+,0+} \,,\ \mathcal{M}_{TL}^{\tau} = \frac{1}{2} (\mathcal{M}_{++,00} + \mathcal{M}_{0+,-0}) \,, \ \mathcal{M}_{LL} =\mathcal{M}_{00,00} \,,
 \\ \mathcal{M}_{TT}^{a} = \frac{1}{2} (\mathcal{M}_{++,++} - \mathcal{M}_{+-,+-})  \,, \ \mathcal{M}_{TL}^{a} = \frac{1}{2} (\mathcal{M}_{++,00} - \mathcal{M}_{0+,-0}) \,.  
\end{gather*}
In terms of the Feynman amplitudes, the eight independent helicity amplitudes are then given 
by~\cite{Budnev:1971sz}\footnote{Our definitions of $\mathcal{M}_{TL}^{a}$ and $\mathcal{M}_{TL}^{\tau}$ are swapped relative to~\cite{Budnev:1971sz},
so that our $\mathcal{M}_{TL}^{a}$ is odd in $\nu$ and our  $\mathcal{M}_{TL}^{\tau}$ is even.}
\begin{subequations}
\label{eq:decomp}
\begin{eqnarray}
\mathcal{M}_{TT}  &= & \frac{1}{4} R^{\mu\mu^{\prime}} R^{\nu\nu^{\prime}} \mathcal{M}_{\mu^{\prime} \nu^{\prime}\mu \nu} \,, \\ 
\mathcal{M}_{TT}^{\tau} &= & \frac{1}{4} \left[ R^{\mu\nu} R^{\mu^{\prime}\nu^{\prime}} + R^{\mu\nu^{\prime}} R^{\mu^{\prime}\nu} - R^{\mu\mu^{\prime}} R^{\nu\nu^{\prime}} \right]  \mathcal{M}_{\mu^{\prime} \nu^{\prime} \mu \nu} \,, \\ 
\mathcal{M}_{TT}^a &= & \frac{1}{4} \left[ R^{\mu\nu} R^{\mu^{\prime}\nu^{\prime}} - R^{\mu\nu^{\prime}} R^{\mu^{\prime}\nu} \right] \mathcal{M}_{\mu^{\prime} \nu^{\prime} \mu \nu} \,, \\  
\mathcal{M}_{TL}  &= & \frac{1}{2}  R^{\mu\mu^{\prime}} k_2^{\nu} k_2^{\nu^{\prime}}  \mathcal{M}_{\mu^{\prime} \nu^{\prime} \mu \nu} \,, \\ 
 \mathcal{M}_{LT} &= & \frac{1}{2}  k_1^{\mu} k_1^{\mu^{\prime}} R^{\nu\nu^{\prime}}  \mathcal{M}_{\mu^{\prime} \nu^{\prime} \mu \nu} \,, \\ 
\mathcal{M}_{LL} &= & k_1^{\mu} k_1^{\mu^{\prime}} k_2^{\nu} k_2^{\nu^{\prime}} \mathcal{M}_{\mu^{\prime} \nu^{\prime} \mu \nu} \,, \\ 
\mathcal{M}_{TL}^{a}& = & - \frac{1}{8} \left[ R^{\mu\nu} k_1^{\mu^{\prime}} k_2^{\nu^{\prime}} + R^{\mu\nu^{\prime}} k_1^{\mu^{\prime}} k_2^{\nu} + (\mu\nu \leftrightarrow \mu^{\prime} \nu^{\prime} )  \right]  \mathcal{M}_{\mu^{\prime} \nu^{\prime} \mu \nu} \,, \\ 
\mathcal{M}_{TL}^{\tau} &= & - \frac{1}{8} \left[ R^{\mu\nu} k_1^{\mu^{\prime}} k_2^{\nu^{\prime}} - R^{\mu\nu^{\prime}} k_1^{\mu^{\prime}} k_2^{\nu} + (\mu\nu \leftrightarrow \mu^{\prime} \nu^{\prime} )  \right] \mathcal{M}_{\mu^{\prime} \nu^{\prime} \mu \nu} \,,
\end{eqnarray}
\end{subequations}
where the projector $R^{\mu\nu}$ onto the subspace orthogonal to $q_1$ and $q_2$,
 and the vectors $k_1$ and $k_2$ are defined in Appendix~\ref{app:CS}. 
The eight helicity amplitudes are functions of $(\nu,Q_1^2,Q_2^2)$.
Then, for fixed photon virtualities $Q_1^2$ and $Q_2^2$, the sum rules can be generically written as \cite{Pascalutsa:2012pr}
\begin{subequations}
\label{eq:dr}
\begin{eqnarray}
\mathcal{M}_{\rm even}(\nu)  & = & \frac{2}{\pi} \int_{\nu_0}^\infty \! d \nu^\prime \frac{\nu^\prime}{\nu^{\prime \, 2} - \nu^2-i \epsilon } W_{\rm even}(\nu^\prime) \,,\\
\label{eq:dreven}
\mathcal{M}_{\rm odd}(\nu) & = & \frac{2 \nu}{\pi} \int_{\nu_0}^\infty \! d \nu^\prime \frac{ 1}{\nu^{\prime \, 2} - \nu^2 - i \epsilon} W_{\rm odd}(\nu^\prime) \,, 
\label{eq:drodd}
\end{eqnarray}
\end{subequations}
assuming the convergence of the integral. Here $\nu_0 \equiv \frac{1}{2}(Q_1^2 + Q_2^2 )$. 
If the integral does not converge, it is necessary to introduce a subtraction 
\begin{subequations}
\begin{eqnarray}
\mathcal{M}_{\rm even}(\nu)  & = & \mathcal{M}_{\rm even}(0) + \frac{2\nu^2}{\pi} \int_{\nu_0}^\infty \! d \nu^\prime \frac{1}{\nu^\prime (\nu^{\prime \, 2} - \nu^2-i\epsilon)} W_{\rm even}(\nu^\prime) \,, \\
\mathcal{M}_{\rm odd}(\nu)  & = &  \nu \mathcal{M}^{\prime}_{\rm odd}(0) + \frac{2\nu^3}{\pi} \int_{\nu_0}^\infty \! d \nu^\prime \frac{1}{\nu^{\prime 2} (\nu^{\prime \, 2} - \nu^2-i\epsilon)} W_{\rm odd}(\nu^\prime) \,.
\label{eq:drsubt}
\end{eqnarray}
\end{subequations}
Finally, the absorptive parts $W_{\lambda^\prime_1 \lambda^\prime_2, \lambda_1 \lambda_2}$ of those eight independent amplitudes, given by Eq.~(\ref{eq:abspart}), are expressed in terms of the $\gamma^\ast \gamma^\ast \to \mathrm{X}$ fusion cross sections~\cite{Budnev:1974de},
\begin{subequations}
\label{eq:vcross}
\begin{eqnarray}
W_{++,++} + W_{+-,+-}  &\equiv& 2 \sqrt{X} \, \left(\sigma_0 + \sigma_2 \right) = 2 \sqrt{X} \,  \left(\sigma_\parallel + \sigma_\perp \right) \equiv 4 \sqrt{X} \, \sigma_{TT} \,,   \\
W_{++,--} &\equiv& 2 \sqrt{X} \,  \left(\sigma_\parallel - \sigma_\perp \right) \equiv 2 \sqrt{X} \, \tau_{TT} \,,  \\
W_{++,++} - W_{+-,+-} &\equiv& 2 \sqrt{X} \, \left(\sigma_0 - \sigma_2 \right) \equiv 4 \sqrt{X} \, \tau^a_{TT} \,,  \\
W_{+0,+0} &\equiv& 2 \sqrt{X} \, \sigma_{TL} \,,  \\
W_{0+,0+} &\equiv& 2 \sqrt{X} \, \sigma_{LT} \,,  \\
W_{++,00} + W_{0+,-0} &\equiv& 4 \sqrt{X} \, \tau_{TL} \,,  \\
W_{++,00} - W_{0+,-0} &\equiv& 4 \sqrt{X} \,  \tau^a_{TL} \,, \\
W_{00,00} &\equiv& 2 \sqrt{X} \, \sigma_{LL} \,,  
\end{eqnarray}
\end{subequations}
where $X = \nu^2 - Q_1^2 Q_2^2$ is the virtual-photon flux factor. Here, $L$ and $T$ refer to longitudinal and transverse polarizations respectively. The cross sections $\sigma$ are positive, but the interference terms $\tau$ are not sign-definite. The relevant cross sections for resonance contributions in each channel are explicitly given in Appendix~\ref{app:CS} in terms of transition form factors.

Thus, using Eqs.~(\ref{eq:dr}) and (\ref{eq:vcross}), we obtain the following dispersive sum rules, valid for fixed photon virtualities $Q_1^2, Q_2^2> 0$ 
\cite{Pascalutsa:2012pr}: 
\begin{subequations}
\begin{eqnarray}
\overline{\mathcal{M}}_{TT} = \frac{1}{2} \left( \overline{\mathcal{M}}_{++,++} (\nu)+ \overline{\mathcal{M}}_{+-,+-}(\nu) \right)   &=& 
\frac{4\nu^2}{\pi} \int_{\nu_0}^\infty \!\! d \nu^\prime \, \frac{ \sqrt{X^\prime}  \sigma_{TT}(\nu^\prime) }{ \nu^\prime ( \nu^{\prime \, 2} - \nu^2-i\epsilon )}   \,, \\
\overline{\mathcal{M}}_{TT}^{\tau}  =  \overline{\mathcal{M}}_{++,--} (\nu) &=& 
\frac{4\nu^2}{\pi} \int_{\nu_0}^\infty \!\!d \nu^\prime  \, \frac{ \sqrt{X^\prime} \tau_{TT}(\nu') }{ \nu^\prime ( \nu^{\prime \, 2} - \nu^2-i\epsilon )}  \,, \\
\overline{\mathcal{M}}_{TT}^{a} = \frac{1}{2} \left( \overline{\mathcal{M}}_{++,++}(\nu) - \overline{\mathcal{M}}_{+-,+-} (\nu) \right) &=& 
\frac{4\nu^3}{\pi} \int_{\nu_0}^\infty \!\!d \nu^\prime  \,\frac{\sqrt{X^\prime}\,  \tau_{TT}^a(\nu')  }{  \nu^{\prime \, 2} (  \nu^{\prime \, 2} - \nu^2-i\epsilon ) }  \,, \\
\overline{\mathcal{M}}_{TL} =  \overline{\mathcal{M}}_{+0,+0} (\nu) &=& 
\frac{4\nu^2}{\pi} \int_{\nu_0}^\infty \!\! d \nu^\prime \,  \frac{ \sqrt{X^\prime} \sigma_{TL}(\nu^\prime) }{ \nu^\prime ( \nu^{\prime \, 2} - \nu^2-i\epsilon )} \,, \\
\overline{\mathcal{M}}_{LT} =  \overline{\mathcal{M}}_{0+,0+} (\nu) &=& 
\frac{4\nu^2}{\pi} \int_{\nu_0}^\infty \!\! d \nu^\prime \,  \frac{ \sqrt{X^\prime} \sigma_{LT}(\nu^\prime) }{ \nu^\prime ( \nu^{\prime \, 2} - \nu^2-i\epsilon) } \,, \\
\overline{\mathcal{M}}_{TL}^{\tau} =\frac{1}{2} \left(  \overline{\mathcal{M}}_{++,00} (\nu)+ \overline{\mathcal{M}}_{0+,-0} (\nu) \right) &=& 
\frac{4 \nu^2}{\pi} \int_{\nu_0}^\infty \!\!d \nu^\prime \, \frac{ \sqrt{X^\prime}\, \tau_{TL} (\nu^\prime) }{\nu^{\prime} ( \nu^{\prime \, 2} - \nu^2-i\epsilon)} \,, \\
\overline{\mathcal{M}}_{TL}^{a} = \frac{1}{2} \left( \overline{\mathcal{M}}_{++,00}(\nu) - \overline{\mathcal{M}}_{0+,-0}(\nu) \right) &=& 
\frac{4\nu^3}{\pi} \int_{\nu_0}^\infty \!\!d \nu^\prime  \,\frac{\sqrt{X^\prime} \, \tau^a_{TL} (\nu^\prime)}{  \nu^{\prime \, 2} (\nu^{\prime \, 2} - \nu^2-i\epsilon )}  \,, \\
\overline{\mathcal{M}}_{LL} = \overline{\mathcal{M}}_{00,00}(\nu)  &=& 
\frac{4 \nu^2}{\pi} \int_{\nu_0}^\infty \!\!d \nu^\prime  \,\frac{ \sqrt{X^\prime} \, \sigma_{LL} (\nu^\prime)}{ \nu^{\prime} (\nu^{\prime \, 2} - \nu^2-i\epsilon) }  \,,
\end{eqnarray}
\label{eq:sumrules} 
\end{subequations}
where we use the notation $\overline{\mathcal{M}}(\nu) \equiv \mathcal{M}(\nu) - \mathcal{M}(0)$ 
or $\overline{\mathcal{M}}(\nu) \equiv \mathcal{M}(\nu) - \nu\mathcal{M}'(0)$ respectively for the even and odd amplitudes.
We always consider the subtracted sum rules, even when the unsubtracted version is well defined,
since the subtraction has the effect of suppressing the high-energy contributions.
Evaluating the sum rules using phenomenological
inputs on the two-photon fusion processes, one can confront the
results with the light-by-light forward amplitudes computed on the
lattice.  In section \ref{sec:model}, we will present an empirical model
for the description of the two-photon
fusion processes  and  subsequently, by comparing it with our lattice results, we
will be able to extract information about the $\gamma^{*}\gamma^{*} \to
M$ transition form factors. Before coming to that, we describe
the lattice QCD approach to calculating HLBL scattering amplitudes in the following two sections.
%%%%%%%%%%%%%%%%%%%%%%%%%%%%%%%%%%%%%%%
\section{Lattice QCD and light-by-light scattering\label{sec:sec3latmethod}}
%%%%%%%%%%%%%%%%%%%%%%%%%%%%%%%%%%%%%%%

\subsection{The scattering amplitude in Euclidean field theory}

The Feynman amplitudes can be obtained via the calculation of the following Euclidean four-point 
correlation function\footnote{We use capital letters to denote `Euclidean' vectors, 
i.e.\ the metric in the scalar product of two such vectors is understood to be Euclidean.}
\begin{equation}
\Pi^E_{\mu\nu\rho\sigma}(Q_1,Q_2) 
= \sum_{X_1,X_2,X_3} \, \langle J_{\mu}(X_1) J_{\nu}(X_2) J_{\rho}(X_3) J_{\sigma}(0) \rangle_E \, e^{i Q_1 (X_1-X_3)} \ e^{ i Q_2 X_2} \,,
\end{equation}
where $J_\mu(X)$ is the Euclidean electromagnetic vector current
($J_0=j_0$, $J_k=ij_k$) and $Q_i$ are the Euclidean four-momenta
($q_i^0 = -i Q_i^0$, $\vec{q}_i = \vec{Q}_i$).  Indeed, using the
Lehmann-Symanzik-Zimmermann reduction formula in Minkowski spacetime,
the relation of this Euclidean correlator to the Feynman forward
amplitudes in Minkowski spacetime is \cite{Green:2015sra}
\begin{equation}
\mathcal{M}_{\mu\nu\rho\sigma}(q_1,q_2) = e^4 i^{n_0} \Pi^E_{\mu\nu\rho\sigma}(Q_1,Q_2)  \,,
\end{equation}
where $n_0$ is the number of temporal indices. Each of the eight helicity amplitudes can be written as
\begin{equation}
\begin{aligned}
\mathcal{M}(q_1^2,q_2^2,\nu) &= T^{\mu\nu\mu^{\prime}\nu^{\prime}}(q_1,q_2) \, \mathcal{M}_{\mu\nu\mu^{\prime}\nu^{\prime}}(q_1,q_2) \\
&= e^4 \, T^E_{\mu\nu\mu^{\prime}\nu^{\prime}}(Q_1,Q_2) \, \Pi^E_{\mu\nu\mu^{\prime}\nu^{\prime}}(Q_1,Q_2) \,,
\end{aligned}
\end{equation}
for some Minkowski tensor $T$, defined above through Eq.~(\ref{eq:decomp}), and some Euclidean tensor $T^E$ given by
\begin{equation}
T^E_{\mu\nu\mu^{\prime}\nu^{\prime}}(Q_1,Q_2) = i^{n_0} T^{\mu\nu\mu^{\prime}\nu^{\prime}}(q_1,q_2) \,.
\end{equation}
Thus, we define
\begin{equation}
R^E_{\mu \nu}(Q_1, Q_2) = i^{n_0} R^{\mu \nu}(q_1, q_2) = \delta_{\mu \nu} - \frac{1}{X} \, \bigl \{ (Q_1 \cdot Q_2) \left( Q_{1\mu} \, Q_{2\nu} + Q_{1\nu} \, Q_{2\mu}  \right) - Q_1^2 \, Q_{2\mu} \, Q_{2\nu}  - Q_2^2 \, Q_{1\mu} \, Q_{1\nu} \bigr \} \,. 
\end{equation}
The case of $k_i$ in Eq.~(\ref{eq:decomp}) requires a bit more care, since their definitions contain $\sqrt{X}$ and in Euclidean space $X = (Q_1 \cdot Q_2)^2 - Q_1^2Q_2^2 \leq 0$. In the Minkowski center of mass frame, if $q_1 = (q_1^0, \vec{q})$ and $q_2 = (q_2^0, -\vec{q})$, then $X=(q_1^0 + q_2^0)^2 \vec{q}^{\, 2}$ and we can evaluate the ordinary positive square root. Performing the Wick rotation, $q_i^0 \to -iQ_i^0$, we get $\sqrt{X} \to -i(Q_1^0 + Q_2^0) |\vec{q}| = -i \sqrt{-X}$. Therefore, in Eq.~(\ref{eq:decomp}), we perform the following replacements to obtain the amplitudes in Euclidean space:
\begin{subequations}
\begin{align}
\mathcal{M}_{\mu^{\prime}\nu^{\prime},\mu\nu} &\to e^4 \Pi^E_{\mu^{\prime}\nu^{\prime},\mu\nu}(Q_1,Q_2) \,, \\
R^{\mu\nu} &\to R^E_{\mu\nu} \,, \\
k_1 &\to K_1 \equiv i \sqrt{\frac{Q_1^2}{-X}} \left( Q_2 - \frac{Q_1\cdot Q_2}{Q_1^2} Q_1 \right) \,, \\
k_2 &\to K_2 \equiv i \sqrt{\frac{Q_2^2}{-X}} \left( Q_1 - \frac{Q_1\cdot Q_2}{Q_2^2} Q_2 \right) \,.
\end{align}
\label{eq:MtoE}
\end{subequations}
These satisfy $K_i^2 = -1$, $K_i \cdot Q_i = 0$, $R^E_{\mu\nu} Q_{i\nu} = 0$, $R^E_{\mu\alpha} R^E_{\alpha\nu} = R^E_{\mu\nu}$ and $R^E_{\mu\mu} = 2$.

The largest value of $|\nu|$ that can be reached with Euclidean kinematics is
limited by the virtualities of the photons\footnote{One might be able to extend
the reach to $|\nu| = \nu_\pi$ with methods in the spirit of \cite{Ji:2001wha}.},
$ |\nu|  \leq  (Q_1^2 Q_2^2)^{1/2}\leq \frac{1}{2}(Q_1^2 + Q_2^2 )\equiv \nu_0$, while 
the nearest singularity is the s-channel $\pi^0$ pole located at
$ \nu_\pi = \frac{1}{2}(m_\pi^2 + Q_1^2 + Q_2^2)$.

%-------------------------------------------------------------------------
\subsubsection{Special case of (anti)parallel momenta}
%-------------------------------------------------------------------------

The tensors $T^E$, resulting from Eq.~(\ref{eq:decomp}) translated to Euclidean space using Eq.~(\ref{eq:MtoE}), are not defined for collinear $Q_1$ and $Q_2$, since in that case $X = 0$.
However, if we start with non-collinear momenta and rotate $Q_2$ toward being (anti)parallel with $Q_1$, then each tensor has such a limit. 
This limit depends on the initial direction of $Q_2$; we will use the average over this direction to define $T^E$ in the collinear limit.

We define the projector
\begin{equation}
\overline{R}_{\mu\nu} = \delta_{\mu\nu} - \frac{Q_{1\mu} Q_{1\nu}}{Q_1^2} \,,
\end{equation}
and find that $R^E_{\mu\nu} \to \overline{R}_{\mu\nu} - V_{\mu} V_{\nu}$. Here $V$ is a unit vector orthogonal to $Q_1$, pointing in the direction from which $Q_2$ approached being collinear with $Q_1$.
Since $K_2 \cdot Q_1 = i \sqrt{ -X/Q_2^2 } \to 0$, in the collinear limit any contraction of $T^E_{\mu\nu\mu^{\prime}\nu^{\prime}}$ with $Q_1$ will vanish. Thus, after averaging over all $V$ orthogonal to $Q_1$, each $T^E_{\mu\nu\mu^{\prime}\nu^{\prime}}$
will be a linear combination of
\begin{equation}
\overline{R}_{\mu\mu^{\prime}} \overline{R}_{\nu\nu^{\prime}} \,, \quad \overline{R}_{\mu\nu} \overline{R}_{\mu^{\prime}\nu^{\prime}} \,, \quad \overline{R}_{\mu\nu^{\prime}} \overline{R}_{\mu^{\prime}\nu} \,.
\end{equation}
We obtain the prefactors by contracting the indices in three different ways. For this, we will make use of 
\begin{equation}
K_1 \cdot K_2 = \frac{Q_1 \cdot Q_2}{ \sqrt{Q_1^2 Q_2^2}} \to s \equiv  \begin{cases}  1 & Q_1,Q_2 \text{ parallel} \\ -1 & Q_1,Q_2 \text{ antiparallel} \end{cases} \,.
\end{equation}
Denoting by $\<\dots\>_V$ the average over $V$, we find
\begin{subequations}
\begin{align}
\langle R^E_{\mu\mu^{\prime}} R^E_{\nu\nu^{\prime}} \rangle_V &= \frac{2}{5} \overline{R}_{\mu\mu^{\prime}} \overline{R}_{\nu\nu^{\prime}} + \frac{1}{15} \left( \overline{R}_{\mu\nu} \overline{R}_{\mu^{\prime}\nu^{\prime}}  +  \overline{R}_{\mu\nu^{\prime}} \overline{R}_{\mu^{\prime}\nu} \right) \,, \\
\langle R^E_{\mu\mu^{\prime}} K_{2\nu} K_{2\nu^{\prime}}  \rangle_V &= \frac{-4}{15} \overline{R}_{\mu\mu^{\prime}} \overline{R}_{\nu\nu^{\prime}} + \frac{1}{15} \left( \overline{R}_{\mu\nu} \overline{R}_{\mu^{\prime}\nu^{\prime}}  +  \overline{R}_{\mu\nu^{\prime}} \overline{R}_{\mu^{\prime}\nu} \right)   \\
& = \langle K_{1\mu} K_{1\mu^{\prime}} R^E_{\nu\nu^{\prime}}  \rangle_V \,, \\
\langle K_{1\mu} K_{1\mu^{\prime}} K_{2\nu} K_{2\nu^{\prime}}  \rangle_V &= \frac{1}{15} \left( \overline{R}_{\mu\mu^{\prime}} \overline{R}_{\nu\nu^{\prime}}  +  \overline{R}_{\mu\nu} \overline{R}_{\mu^{\prime}\nu^{\prime}}  + \overline{R}_{\mu\nu^{\prime}} \overline{R}_{\mu^{\prime}\nu}  \right) \,,\\
\langle R^E_{\mu\mu^{\prime}} K_{1\nu} K_{2\nu^{\prime}}  \rangle_V &= -s \, \langle R^E_{\mu\mu^{\prime}} K_{2\nu} K_{2\nu^{\prime}}  \rangle_V \,.
\end{align}
\end{subequations}

%%%%%%%%%%%%%%%%%%%%%%%%%%%%%%%%%%%%%%%%%%%%%%%%%%%%%%%%%%%%%%%%%%%%%%
\subsection{Flavor structure of the four-point function\label{sec:34ov9}}
%%%%%%%%%%%%%%%%%%%%%%%%%%%%%%%%%%%%%%%%%%%%%%%%%%%%%%%%%%%%%%%%%%%%%%

\begin{figure}
  \centering
  \includegraphics[width=0.7\textwidth]{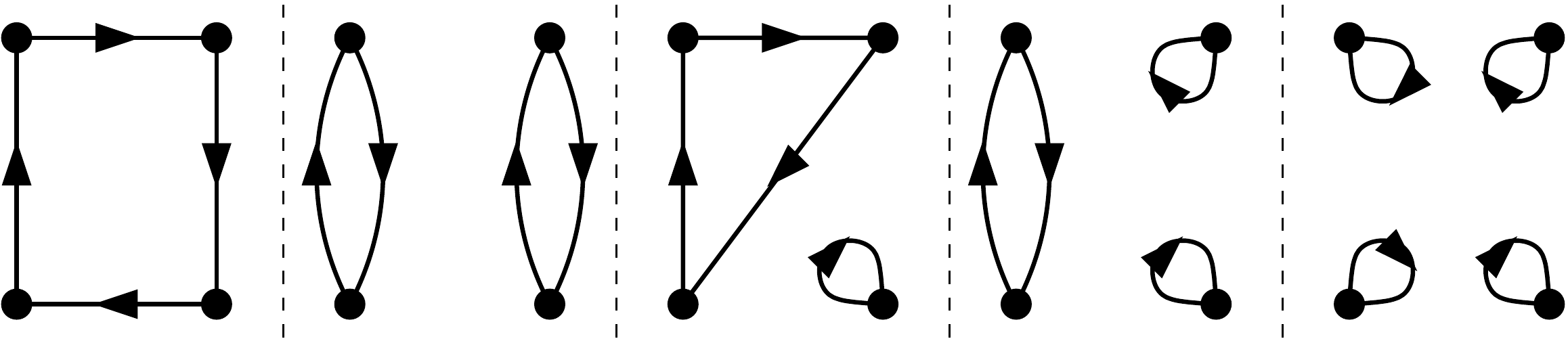}
  \caption{The five classes of quark contractions for four-point
    functions. In this work, we compute the leftmost, fully-connected
    set of contractions, as well as the (2+2) class of diagrams (second from the left).}
  \label{fig:contractions}
\end{figure}

In numerical lattice QCD calculations of $n$-point functions, the
quark path integral is evaluated analytically to yield a sum of
contractions of quark propagators. For the four-point function of
vector currents, these fall into five distinct topologies, illustrated
in Fig.~\ref{fig:contractions}.

The calculation of all Wick-contraction topologies is demanding.  In
many instances, disconnected diagram contributions have been found to
make numerically small contributions to hadronic matrix elements,
though not always~\cite{Gulpers:2013uca}.  Quark loops generated by a
single vector current have been empirically found to be particularly
suppressed (see for
instance~\cite{Green:2015wqa,Blum:2015you,Gerardin:2016cqj,DellaMorte:2017dyu}).
At short distances, perturbation theory provides an explanation for
the suppression of this type of contribution, since it requires the
exchange of at least three gluons~\cite{Chetyrkin:1994js}.  On the
other hand, it is well known that the disconnected diagram is
responsible for the difference between the pion and the $\eta'$ mass
in the pseudoscalar two-point function, and is therefore crucial at
long distances.

The importance of the disconnected diagrams in the HLbL amplitude has been pointed out
in~\cite{Bijnens:2015jqa,Bijnens:2016hgx}, showing that the pion and
$\eta'$ pole contributions would have the wrong weight factors if only
the connected diagrams were included.  Here,
\begin{enumerate}
\item we use (a) flavor symmetry, either SU(2) or SU(3), and (b) the assumption that 
Wick-contraction diagrams where a vector current appears as the only insertion in a quark loop, 
thus producing a factor ${\rm Tr}\{ \gamma_\mu S(X,X)\}$, are negligible;
\item we then derive the weight factors of non-singlet and singlet mesons in the fully connected
and the (2+2) disconnected contribution to the HLbL amplitude;
\item we show that whenever the HLbL amplitude is dominated by the pole-exchange
of an isovector resonance, isospin symmetry induces relations between
different Wick-diagram topologies. 
\end{enumerate}

The main result is that under the assumptions stated under (1.), 
in the fully connected diagrams the non-singlet meson poles over-contribute by a factor $34/9$
(respectively a factor 3) in  QCD with two (respectively three) degenerate flavors of quarks, 
while the singlet mesons do not contribute. The (2+2) disconnected diagrams contain the singlet-meson contribution 
and correct the fully connected diagram by compensating with  $(-25/9)$ (respectively $-2$) times the non-singlet meson contribution.
For QCD with a realistic quark spectrum, we expect the relations between the classes of diagrams  to lie between 
the quoted predictions.

The starting point is the observation, based on Fig.\ \ref{fig:contractions},
that the Wick contractions contributing to the HLbL amplitude can be written as
 (we  drop the space-time arguments and indices of the four-point amplitudes)
\ba\la{eq:Pi4Wick}
\Pi^{\rm HLbL}
 &=& \sum_f \Q_f^4 \Pi^{4}_f + \sum_{f_1,f_2} \Q_{f_1}^2 \Q_{f_2}^2 \Pi_{f_1,f_2}^{2+2}
+ \sum_{f_1,f_2} \Q_{f_1}^3 \Q_{f_2}\Pi_{f_1,f_2}^{3+1} 
\\ && + \sum_{f_1,f_2,f_3} \Q_{f_1}^2 \Q_{f_2} \Q_{f_3} \Pi^{2+1+1}_{f_1,f_2,f_3} 
+ \sum_{f_1,f_2,f_3,f_4} \Q_{f_1} \Q_{f_2} \Q_{f_3} \Q_{f_4}\Pi^{1+1+1+1}_{f_1,f_2,f_3,f_4} .
\nonumber
\ea 
In the case that the quark masses are all equal, one can drop the flavor indices, e.g.\ $\Pi_{f_1,f_2}^{2+2}\to \Pi^{2+2}$.
In particular, let $\Pi$ be the four-point function of the `up' current $\bar u\gamma_\mu u$.
Since in that case $\sum_f \Q_f^n = 1$ $\forall n$, and because all quark lines carry the same quark mass,
the $\Pi^{X}_f$ are precisely the Wick contractions appearing in $\Pi$ ($X=4,~2+2,~3+1,~2+1+1,~1+1+1+1$), including the normalization.
On the other hand, the HLbL amplitude is expressed as Wick contractions $\Pi^{X}_f$ weighted by polynomials in the quark charges.

An important observation is that with two flavors of quarks, 
there are only three linearly independent symmetric polynomials in $\Q_u$ and $\Q_d$, for instance
\be
(\Q_u^4+\Q_d^4),\qquad 2\Q_u^2 \Q_d^2, \qquad \Q_u\Q_d(\Q_u^2+\Q_d^2).
\ee
With three flavors, there are four such independent polynomials in $(\Q_u,\Q_d,\Q_s)$, for instance
\be
\begin{array}{ll}
 P_1\equiv (\Q_u^4+\Q_d^4+\Q_s^4), \qquad & P_2\equiv (\Q_u^2+\Q_d^2+\Q_s^2)^2,    
\phantom{\bigg|}\\
  P_3 \equiv (\Q_u^3+\Q_d^3+\Q_s^3)(\Q_u+\Q_d+\Q_s), \qquad & P_4 \equiv (\Q_u^2+\Q_d^2+\Q_s^2)(\Q_u+\Q_d+\Q_s)^2,
\end{array}
\ee
while $P_5 \equiv (\Q_u+\Q_d+\Q_s)^4 = 6P_1 -3P_2 - 8P_3 +6P_4$.

\subsubsection{The case of $N_f=2$ QCD}

We assume exact isospin SU(2) symmetry. The photon couples to the electromagnetic current, 
whose  isospin decomposition reads
\be
J_\mu^{\rm e.m.} = J^{1}_\mu + J^0_\mu, \qquad J^1_\mu = \frac{\Q_u-\Q_d}{2}(\bar u\gamma_\mu u - \bar d\gamma_\mu d),
\qquad  J^0_\mu = \frac{\Q_u+\Q_d}{2}(\bar u\gamma_\mu u + \bar d\gamma_\mu d).
\ee
The upper index on the current indicates the isospin quantum number $I=0,1$.

There are both isoscalar and isovector resonances that couple to two photons.
As is well known, the coupling of an \emph{isovector} resonance occurs only when one of the photons couples
via the isoscalar part, and one couples via the isovector part 
of the e.m.\ current\footnote{That an isovector resonance cannot decay into two isovector photons 
is shown by the Wigner-Eckart theorem:
\[
\<M^{I''=1, m''} | J^{1,m} | V^{1,m'}\> = C_{11}(1, m''; m, m') \<M^{1}|| J^1 || V^1\>,
\]
where $C_{I,I'}(I'', m''; m, m')$  is the Clebsch-Gordan coefficient for
composing isospin $I$ with isospin $I'$ and obtaining isospin $I''$ and $\<\pi|| J^1 || V^1\>$ is the
reduced matrix element. It so happens that $C_{11}(1, 0; 0, 0) = 0$.}.
Since the amplitude for a neutral pion to couple to two photons
vanishes if either both are isoscalar ($\Q_u=\Q_d$) or both are isovector ($\Q_u=-\Q_d$), and the
coupling must be quadratic in the charges, it must be proportional to
$(\Q_u^2-\Q_d^2)$.  
Then the contributions to $\Pi$ and $\Pi^{\rm HLbL}$ of an isovector resonance $M_1$ are related by
\be\la{eq:LbL,M1}
\Pi^{{\rm HLbL},M_1} = (\Q_u^2-\Q_d^2)^2 \;\Pi^{M_1}. 
\ee
Correspondingly, the dependence of the transition form factor of an \emph{isoscalar} resonance
on the quark charges is such that it contains two independent terms,
\be\la{eq:F_M0}
{F}_{M^0\gamma^*\gamma^*} = (\Q_u^2+\Q_d^2) {\cal F}_C + (\Q_u+\Q_d)^2 {\cal F}_D.
\ee
The notation indicates that ${\cal F}_D$ contains all the 
diagrams where at least one vector current appears isolated in a quark loop.
In view of the form (\ref{eq:F_M0}) of the $M^0\gamma\gamma$ vertex, the pole contribution of an isoscalar meson has the dependence 
\be\la{eq:LbL,M0}
\Pi^{\rm HLbL,M^0} = (\Q_u^2+\Q_d^2)^2 \Pi_A + (\Q_u+\Q_d)^2 (\Q_u^2+\Q_d^2) \Pi_B + (\Q_u+\Q_d)^4 \Pi_C.
\ee
on the quark charges, where the contributions $\Pi_B$ and $\Pi_C$ are only non-vanishing if 
the disconnected diagrams involving at least one isolated vector current inserted in a quark loop are non-vanishing.
As discussed below, $\Pi_B$ is O($1/N$) and $\Pi_C$ is O($1/N^2$) in the large-$N$ power counting.

By identifying the polynomials in $\Q_u$ and $\Q_d$ in Eqs.\ (\ref{eq:LbL,M1}) and (\ref{eq:Pi4Wick}), we obtain three conditions that relate
the contributions of an isovector resonance to the various Wick contractions:
\ba
\Pi^{M^1} &=& \Pi^{4,M^1} + \Pi^{2+2,M^1} + \Pi^{3+1,M^1} + \Pi^{2+1+1,M^1} + \Pi^{1+1+1+1,M^1},
\\
-\Pi^{M^1} &=&  \Pi^{2+2,M^1} + \Pi^{2+1+1,M^1} + 3\,\Pi^{1+1+1+1,M^1},
\\
0 &=&   \Pi^{3+1,M^1} + 2\,\Pi^{2+1+1,M^1} + 4\,\Pi^{1+1+1+1,M^1}.
\ea
Thus if for a specific kinematic regime one isovector resonance exchange dominates the HLbL amplitude,
it is sufficient to compute three of the five Wick-contraction classes. For instance,
$\Pi^{1+1+1+1,M^1}$ and $\Pi^{2+1+1,M^1} $ can be expressed in terms of
the classes $\Pi^{4,M^1},~ \Pi^{2+2,M^1}$ and $\Pi^{3+1,M^1}$ of diagrams.
We also note the exact expression
\ba\la{eq:Pi4M1}
\Pi^{4,M^1} &=& 2 ( \Pi^{M^1} + \Pi^{2+1+1,M^1} + 3\Pi^{1+1+1+1,M^1}),
\ea
for the fully connected class of diagrams.

Similarly, equating the expressions (\ref{eq:LbL,M0}) and (\ref{eq:Pi4Wick})
yields the relations
\ba
\Pi^{M^0} = \Pi_A + \Pi_B + \Pi_C
&=& \Pi^{4,M^0} + \Pi^{2+2,M^0} + \Pi^{3+1,M^0} + \Pi^{2+1+1,M^0} + \Pi^{1+1+1+1,M^0},
\\
\Pi_A + \Pi_B + 3\Pi_C 
&=& \Pi^{2+2,M^0} + \Pi^{2+1+1,M^0} + 3\,\Pi^{1+1+1+1,M^0},
\\
2\Pi_B + 4 \Pi_C
&=& \Pi^{3+1,M^0} + 2\,\Pi^{2+1+1,M^0} + 4\,\Pi^{1+1+1+1,M^0}.
\ea
Eliminating $\Pi^{2+2,M^0}$ and $\Pi^{3+1,M^0}$, we get
\be \la{eq:Pi4M0}
 \Pi^{4,M^0} = 2 ( -\Pi_B - 3\Pi_C  + \Pi^{2+1+1,M^0} + 3\Pi^{1+1+1+1,M^0}).
\ee

\subsubsection{Large-$N$ expectations}

In terms of large-$N$ counting, where $N$ is the number of colors, every
additional disconnected quark loop costs a factor $1/N$.  However it
is worth looking more closely how this property emerges.  For
instance, just keeping the leading class of diagrams $\Pi^{4,M}$ would
not reproduce the correct dependence on $\Q_u$ and $\Q_d$ for the
exchange of a single meson, be it isoscalar or isovector. Put in a different way, 
the relations (\ref{eq:Pi4M1}) and (\ref{eq:Pi4M0}) 
derived from isospin symmetry relate contributions to diagrams that scale differently with $N$.

The resolution of this apparent contradiction is that, at large $N$,
neutral mesons are expected to come in degenerate pairs, one
isoscalar, one isovector, due to the vanishing of the quark
annihilation diagrams. Only when the sum of the contributions of a
pair is considered, the large-$N$ counting should apply.
Here we only apply the large-$N$ counting rule to the insertion of a vector current.
Considering the sum of the contributions of such a pair,
we obtain from (\ref{eq:Pi4M1}--\ref{eq:Pi4M0}) the relation
\ba\la{eq:Pi4M0M1}
\Pi^{4,M^0+M^1}  &=& 2 \Big( \Pi^{M^1} -\Pi_B - 3\Pi_C
+ \Pi^{2+1+1,M^0+M^1} + 3\Pi^{1+1+1+1,M^0+M^1}\Big).\qquad
\ea
Large-$N$ counting should apply at this point, and we expect to be able to neglect in
first approximation $\Pi_B,\Pi_C$ (down by $1/N$ and $1/N^2$ respectively) and the terms $\Pi^{2+1+1,M^0+M^1}$ and
$\Pi^{1+1+1+1,M^0+M^1}$, since they are expected to be down by $1/N^2$ and $1/N^3$ respectively relative 
to $\Pi^{4,M^0+M^1}$.  All neglected terms contain at least one isolated vector current in a quark loop.
In that approximation, the leading diagram classes are given by 
\ba\label{eq:Pi422M0M1}
\Pi^{4,M^0+M^1}  &\approx &  2  \Pi^{M^1}, 
\\
\Pi^{2+2,M^0+M^1} & \approx &  - \Pi^{M^1} + \Pi^{M^0}.
\ea
Thus, using Eqs~(\ref{eq:Pi422M0M1}) and (\ref{eq:LbL,M1}) we obtain for the fully connected contribution in Eq.~(\ref{eq:Pi4Wick})
\be\label{eq:HLbL4M0M1}
(\Q_u^4+\Q_d^4) \Pi^{4,M^0+M^1}  \approx (\Q_u^4+\Q_d^4) 2\Pi^{M^1} = \frac{\Q_u^4+\Q_d^4}{(\Q_u^2-\Q_d^2)^2} 2\Pi^{{\rm HLbL},M^1} 
= \frac{34}{9} \Pi^{{\rm HLbL},M^1},
\ee
where we have included the physical charges of the $u$ and $d$ quarks in the last step.
The (2+2) disconnected diagrams complement the connected diagrams to yield the full contribution,
\be\label{eq:HLbL2+2M0M1}
(\Q_u^2+\Q_d^2)^2 \Pi^{2+2,M^0+M^1}  \approx  -\frac{25}{9} \Pi^{{\rm HLbL},M^1} + \Pi^{{\rm HLbL},M^0}.
\ee
The charge factors in Eqs.\ (\ref{eq:HLbL4M0M1}) and (\ref{eq:HLbL2+2M0M1}) agree with Refs.\ \cite{Bijnens:2015jqa,Bijnens:2016hgx}.

The stronger large-$N$ prediction, which however turns out not to be a
good approximation in QCD, is that isoscalar and isovector states in
each symmetry channel (pseudoscalars, scalars, tensors, etc.)
compensate each other in the (2+2) disconnected diagrams, making the
latter $1/N$ suppressed as compared to the fully connected diagrams.
The channel where the degeneracy expected at large $N$ is
most badly broken is the pseudoscalar channel, since $m_{\eta'}\gg m_{\pi^0}$. 
Therefore, we expect the $\Pi^{2+2}$ class of diagrams to be dominated by the $\pi^0$ and $\eta'$ contributions, 
since their contributions cancel each other to a far lesser extent than for other meson pairs such as  $a_2$ and $f_2$.
For instance, the empirical ratio of the two-photon widths of $a_2$ and $f_2$ are roughly as expected if one neglects disconnected diagrams
(see Table \ref{tab:particles} for the source of the phenomenological values),
\be
\frac{\Gamma_{f_2\gamma\gamma}}{\Gamma_{a_2\gamma\gamma}} \approx \frac{(2.93\pm0.40){\rm keV}}{(1.00\pm0.06){\rm keV}} = 2.93\pm0.44
\approx \frac{(\Q_u^2+\Q_d^2)^2}{(\Q_u^2-\Q_d^2)^2} = \frac{25}{9} \approx 2.7778.
\ee
On the other hand, using the phenomenological values~\cite{Olive:2016xmw,Larin:2010kq}
\be
\Gamma_{\eta'\gamma\gamma} = 4.35 {\rm\, keV},\qquad \Gamma_{\pi^0\gamma\gamma} = 7.82{\rm \, eV},
\ee
and the fact that $\Gamma\propto m_P^3 |F_{P\gamma^*\gamma^*}(0,0)|^2$, we obtain
\be
\frac{|{F}_{\eta'\gamma^*\gamma^*}(0,0)|^2}{|{ F}_{\pi^0\gamma^*\gamma^*}(0,0)|^2}
\approx 1.56 \neq \frac{(\Q_u^2+\Q_d^2)^2}{(\Q_u^2-\Q_d^2)^2} \approx 2.7778,
\ee
which does not agree well with the large-$N$ expectation and the approximation of $m_{\rm strange}= \infty$
implicitly made here by using relations derived in $N_f=2$ QCD.
A further channel in which  the large-$N$ expectations are not well fullfilled is the scalar sector;
in particular, there does not seem to be an isovector analogue of the $f_0(600)$ meson.
However, it turns out that the scalars make an overall small contribution to the light-by-light amplitudes.

\subsubsection{The case of $N_f=3$ QCD}

Here we assume exact SU(3) flavor symmetry, $m_u=m_d=m_s$. Since in
the previous paragraphs we have treated the case of $N_f=2$ QCD,
corresponding to $m_s=\infty$, one may hope that the real world lies
somewhere in between these two idealized cases.

In nature, $\Q_u=2/3$ and $\Q_d=\Q_s=-1/3$.
Somewhat more generally, for $\Q_u+\Q_d+\Q_s = 0$, the $SU(3)_{\rm f}$ decomposition reads
\be
J_\mu^{\rm e.m.} = J^{3}_\mu + J^8_\mu, \qquad J^3_\mu = \frac{\Q_u-\Q_d}{2}(\bar u\gamma_\mu u - \bar d\gamma_\mu d),
\qquad  J^8_\mu = \frac{\Q_u+\Q_d}{2}(\bar u\gamma_\mu u + \bar d\gamma_\mu d - 2\bar s\gamma_\mu s).
\ee
It follows directly from $\Q_u+\Q_d+\Q_s = 0$ that no diagrams with a single vector current inside a quark loop occurs in the HLbL amplitude:
$\Pi^{3,1}$, $\Pi^{2+1+1}$ and $\Pi^{1+1+1+1}$ do not contribute. However it is useful to keep the charges generic 
to derive relations from flavor symmetry and large-$N$ arguments.

In the SU(3) symmetric theory, mesons come in octets and 
singlets\footnote{Higher representations are allowed by symmetry, but do not seem to occur in QCD at low energies.}.
Taking the pseudoscalar sector as an example,
the transition form factor of the neutral pion is given by  $(\Q_u^2-\Q_d^2){\cal F}_8$;
neglecting again the single vector current inside a quark loop,
the $\eta$ meson has the transition form factor  $\sqrt{1/3}(\Q_u^2+\Q_d^2-2\Q_s^2) {\cal F}_8$.
Under the same assumption, the form factor of the $\eta'$ 
has a charge dependence given by $\sqrt{{2}/{3}}(\Q_u^2+\Q_d^2+\Q_s^2){\cal F}_0$.
Only in the strict large-$N$ limit would we have  ${\cal F}_0= {\cal F}_8$.

Thus the pole contribution of a meson octet to the four-point function of the electromagnetic current has
the following dependence on the quark charges,
\be\la{eq:oct}
\Pi^{{\rm HLbL,oct}} = \frac{3}{4}\Big[(\Q_u^2-\Q_d^2)^2 + \frac{1}{3}(\Q_u^2+\Q_d^2-2\Q_s^2)^2\Big]\Pi^{\rm oct},
\ee
where $\Pi^{\rm oct}$ is the octet contribution to $\Pi$.
The corresponding expression for the singlet meson reads
\be\la{eq:sgl}
\Pi^{{\rm HLbL,sgl}} = (\Q_u^2+\Q_d^2+\Q_s^2)^2\; \Pi^{\rm sgl}.
\ee
Matching the  expressions (\ref{eq:oct}) and (\ref{eq:Pi4Wick}), we obtain the relations
\ba
\frac{3}{2}\Pi^{\rm oct} &=& \Pi^{4,{\rm oct}},
\\
-\frac{1}{2} \Pi^{\rm oct} &=& \Pi^{2+2,{\rm oct}}.
\ea
where we have consistently neglected the octet contribution to $\Pi^{3+1}$, $\Pi^{2+1+1}$ and $\Pi^{1+1+1+1}$.
Under the corresponding assumption for the singlet contribution, we have 
\ba
0 &=& \Pi^{4,{\rm sgl}},
\\
\Pi^{\rm sgl} &=& \Pi^{2+2,{\rm sgl}}.
\ea
Combining these equations, we finally have the following expressions for the fully connected and the (2+2) disconnected class of diagrams,
\ba
  (\Q_u^4+\Q_d^4+\Q_s^4)\Pi^{4,{\rm oct+sgl}}&\approx& P_1\;\Pi^{4,{\rm oct}} = P_1\cdot \frac{3}{2} \Pi^{{\rm oct}} 
 = \frac{P_1}{P_1-\frac{1}{3}P_2} \Pi^{\rm HLbL,oct}
\nonumber\\ &=&  3 \Pi^{\rm HLbL,oct},
\\
 (\Q_u^2+\Q_d^2+\Q_s^2)^2 \Pi^{2+2,{\rm oct+sgl}} &\approx& - \frac{P_2}{3(P_1-\frac{1}{3}P_2)} \Pi^{\rm HLbL,oct} + \Pi^{\rm HLbL,sgl}
\nonumber\\ &=&  -2 \Pi^{\rm HLbL,oct} + \Pi^{\rm HLbL,sgl}.
\ea
In the last of these equations we have set $\Q_u=2/3$, $\Q_d=\Q_s=-1/3$.

%%%%%%%%%%%%%%%%%%%%%%%%%%%%%%%%%%%%%%%%%%%%%%%%%
\subsection{Lattice calculation of the fully-connected vector four-point function}
%%%%%%%%%%%%%%%%%%%%%%%%%%%%%%%%%%%%%%%%%%%%%%%%%

\newcommand{\cT}{\mathcal{T}}
\newcommand{\cV}{\mathcal{J}}

We now describe our method to calculate the six contractions that have 
fully connected quark lines (leftmost topology in
Fig.~\ref{fig:contractions}), whereas the dominant class of
disconnected diagrams (second diagram topology from the left in
Fig.~\ref{fig:contractions}) is discussed in the next section.

We discretize the Euclidean four-point function using the local and
conserved currents, as well as a contact operator,
\begin{subequations}
\begin{align}
  J_\mu^l(X) &= Z_V \bar\psi(X) \gamma_\mu \mathcal{Q} \psi(X),\\
  J_\mu^c(X) &= \frac{1}{2}\left[
    \bar\psi(X+a\hat\mu) (\gamma_\mu+1)U_\mu^\dagger(X) \mathcal{Q} \psi(X)
  + \bar\psi(X) (\gamma_\mu-1)U_\mu(X) \mathcal{Q} \psi(X+a\hat\mu) \right],\\
  T_\mu(X) &= \frac{1}{2}\left[
    \bar\psi(X+a\hat\mu) (\gamma_\mu+1)U_\mu^\dagger(X) \mathcal{Q} \psi(X)
  - \bar\psi(X) (\gamma_\mu-1)U_\mu(X) \mathcal{Q} \psi(X+a\hat\mu) \right],
\end{align}
\end{subequations}
where $\psi= (u,d)^T$ is the doublet of light quarks,
$\mathcal{Q}=\operatorname{diag}(\tfrac{2}{3},-\tfrac{1}{3})$ is the
quark charge matrix, $U_\mu(X)$ are the gauge links, and $Z_V$ is the
renormalization factor for the local current. We use one local and
three conserved currents. In position space, the lattice four-point
function is given by
\begin{equation}
\begin{aligned}
    \Pi^\text{pos}_{\mu_1\mu_2\mu_3\mu_4}(X_1,X_2,X_3,0) &= 
    \Bigl\langle J_{\mu_4}^l(0) \Bigl[ 
      J_{\mu_1}^c(X_1) J_{\mu_2}^c(X_2) J_{\mu_3}^c(X_3)
    + \delta_{\mu_1\mu_2} \delta_{X_1X_2} T_{\mu_1}(X_1) J_{\mu_3}^c(X_3) \\
    &\qquad 
    + \delta_{\mu_1\mu_3} \delta_{X_1X_3} T_{\mu_3}(X_3) J_{\mu_2}^c(X_2) 
    + \delta_{\mu_2\mu_3} \delta_{X_2X_3} T_{\mu_3}(X_3) J_{\mu_1}^c(X_1) \\
    &\qquad
    + \delta_{\mu_1\mu_3} \delta_{\mu_2\mu_3} \delta_{X_1X_3} \delta_{X_2X_3} J_{\mu_3}^c(X_3)
    \Bigr] \Bigr\rangle \,.
\end{aligned}
\end{equation}
The contact terms are present when two or three conserved currents
coincide and serve to ensure that the conserved-current Ward
identities hold. Using the backward lattice derivative $\Delta$, these
take the form
\begin{equation}
  \Delta_{\mu_1}^{X_1} \Pi^\text{pos}_{\mu_1\mu_2\mu_3\mu_4}
= \Delta_{\mu_2}^{X_2} \Pi^\text{pos}_{\mu_1\mu_2\mu_3\mu_4}
= \Delta_{\mu_3}^{X_3} \Pi^\text{pos}_{\mu_1\mu_2\mu_3\mu_4}
= 0 \,.
\end{equation}
In momentum space, we evaluate the Euclidean four-point function as
\begin{equation}\label{eq:4pt_lccc_mom}
  \Pi^E_{\mu\nu\rho\sigma}(Q_1,Q_2) = \sum_{X_1,X_2,X_3}
  e^{iQ_1(X_1+\tfrac{a}{2}\hat\mu)}e^{iQ_2(X_2+\tfrac{a}{2}\hat\nu)}
  e^{-iQ_1(X_3+\tfrac{a}{2}\hat\rho)}
  \Pi^\text{pos}_{\mu\nu\rho\sigma}(X_1,X_2,X_3,0) \,.
\end{equation}

The fully-connected contribution to Eq.~\eqref{eq:4pt_lccc_mom}, which
is the part proportional to $\Tr(\mathcal{Q}^4)=\frac{17}{81}$, is
evaluated using the method of sequential propagators. First, a
point-source propagator
\begin{equation}
  S_0(X) \equiv S(X,0),
\end{equation}
where $S$ is the single-flavor all-to-all quark propagator (degenerate
for $u$ and $d$), is computed from the origin. To concisely describe
the sequential propagators, we introduce the ``insertions''
$\cV_\mu(X)$ and $\cT_\mu(X)$ for the conserved vector current and
contact operator:
\begin{subequations}
\begin{align}
  J_\mu^c(X) &= \bar \psi \cV_\mu(X) \mathcal{Q} \psi,\\
  T_\mu(X)  &= \bar \psi \cT_\mu(X) \mathcal{Q} \psi.
\end{align}
\end{subequations}
Formally, these objects have the same size as an all-to-all quark
propagator, but they are exactly zero for all sites except those that are removed from $X$ by one lattice spacing. The
point-source propagator is then combined with a plane wave and the
conserved vector current insertion to form the source for new
(sequential) propagators,
\begin{equation}
  S_{Q,\mu}\equiv S \sum_X e^{-iQ(X+\tfrac{a}{2}\hat\mu)} \cV_\mu(X) S_0 .
\end{equation}
These, in turn, are used to form sources for double-sequential
propagators
\begin{eqnarray}
  S_{Q_1\mu_1;Q_2\mu_2} &\equiv&  S \sum_X\left[
     e^{-iQ_1(X+\tfrac{a}{2}\hat\mu_1)} \cV_{\mu_1}(X) S_{Q_2\mu_2}
  +  e^{-iQ_2(X+\tfrac{a}{2}\hat\mu_2)} \cV_{\mu_2}(X) S_{Q_1\mu_1} \right.
  \nonumber\\
  && \left. + \, \delta_{\mu_1\mu_2}  e^{-i(Q_1+Q_2)(X+\tfrac{a}{2}\hat\mu_1)} \cT_{\mu_1} S_0 \right].
\end{eqnarray}
Finally, noting that $\gamma_5\cV_\mu(X)$ is anti-Hermitian and
$\gamma_5\cT_\mu(X)$ is Hermitian, the fully-connected four-point
function is obtained\footnote{A more generic case was given in
  Ref.~\cite{Green:2015mva}.} as
\begin{equation}
\begin{aligned}
  \Pi^{E,\text{conn}}_{\mu\nu\rho\sigma}(Q_1,Q_2) &=
 -\Tr[\mathcal{Q}^4]Z_V\sum_{X_2}e^{iQ_2(X_2+\tfrac{a}{2}\hat\nu)}
\Bigl\langle\Tr\Bigl(\gamma_{\sigma}\gamma_5\\[-0.5em]
&\qquad\qquad\times\Bigl[
  S_{Q_1\mu;-Q_1\rho}^\dagger\gamma_5\cV_{\nu}(X_2)S_0
+ S_0^\dagger\gamma_5\cV_{\nu}(X_2)S_{-Q_1\mu;Q_1\rho}\\
&\qquad\qquad\qquad
  - S^\dagger_{-Q_1\rho}\gamma_5\cV_{\nu}(X_2)S_{-Q_1\mu}
  - S^\dagger_{Q_1\mu}\gamma_5\cV_{\nu}(X_2)S_{Q_1\rho}  \\
&\qquad\qquad\qquad
  + \delta_{\mu\nu}e^{iQ_1(X_2+\tfrac{a}{2}\hat\nu)}\bigl(
    S_0^\dagger\gamma_5\cT_{\nu}(X_2)S_{Q_1\rho}
    - S_{-Q_1\rho}^\dagger\gamma_5\cT_{\nu}(X_2)S_0\bigr) \\
&\qquad\qquad\qquad
  + \delta_{\rho\nu}e^{-iQ_1(X_2+\tfrac{a}{2}\hat\nu)}\bigl(
    S_0^\dagger\gamma_5\cT_{\nu}(X_2)S_{-Q_1\mu}
    - S_{Q_1\mu}^\dagger\gamma_5\cT_{\nu}(X_2)S_0\bigr) \\
&\qquad\qquad\qquad
  + \delta_{\mu\nu}\delta_{\rho\nu}
  S_0^\dagger\gamma_5\cV_{\nu}(X_2)S_0
  \Bigr]\Bigr)\Bigr\rangle_U,
\end{aligned}
\end{equation}
where $\langle\dots\rangle_U$ denotes the expectation value over gauge
fields. The sequential propagators depend on $Q_1$, so a separate
calculation must be done for each $Q_1$. However, none of the sources
for the propagators depend on $Q_2$; therefore, we are able to
efficiently evaluate $\Pi^{E,\text{conn}}_{\mu\nu\rho\sigma}(Q_1,Q_2)$
for all $Q_2$ available on the lattice. In momentum space, the
conserved-current Ward identities take the form
\begin{equation}
  \hat Q_{1\mu}\Pi^E_{\mu\nu\rho\sigma}(Q_1,Q_2) =
  \hat Q_{2\nu}\Pi^E_{\mu\nu\rho\sigma}(Q_1,Q_2) =
  \hat Q_{1\rho}\Pi^E_{\mu\nu\rho\sigma}(Q_1,Q_2) = 0,
\end{equation}
where $\hat Q_\mu\equiv \frac{2}{a}\sin\frac{aQ_\mu}{2}$. We have
verified that in our implementation these hold on each gauge
configuration.

%%%%%%%%%%%%%%%%%%%%%%%%%%%%%%%%%%%%%%%%%%%%%%%%%
\subsection{Lattice calculation of the (2+2) disconnected four-point function}
%%%%%%%%%%%%%%%%%%%%%%%%%%%%%%%%%%%%%%%%%%%%%%%%%

We also calculate one class of disconnected diagrams to obtain an
indication of their relevance. Based on the charge factor and the arguments
given in section \ref{sec:34ov9}, the second
class in Fig.~\ref{fig:contractions}, which we call $(2+2)$ and is
proportional to $\Tr(\mathcal{Q}^2)^2=\tfrac{25}{81}$, is the most
important. We evaluate this class of diagrams using a different lattice
expression that has two local and two conserved currents:
\begin{equation}
  \Pi^{\text{pos,}(2+2)}_{\mu_1\mu_2\mu_3\mu_4}(X_1,X_2,X_3,0) = 
\left\langle J_{\mu_1}^i(X_1) J_{\mu_2}^c(X_2) J_{\mu_3}^j(X_3) J_{\mu_4}^l(0)
\right\rangle,
\end{equation}
where $(i,j)=(l,c)$ or $(c,l)$ depending on the contraction, chosen
such that each quark loop contains one local and one conserved
current. Specifically, denoting the $Y$-to-all propagator as
$S_Y(X)\equiv S(X,Y)$, we use
\begin{equation}
\begin{aligned}
\Pi^{E,(2+2)}_{\mu\nu\rho\sigma} &= \Tr[\mathcal{Q}^2]^2 Z_V^2
\sum_{X_1,X_2,X_3}e^{iQ_1(X_1-X_3)}e^{iQ_2(X_2+\tfrac{a}{2}\hat\nu)}\\
&\qquad\times\Bigl\langle
e^{iQ_1\tfrac{a}{2}\hat\mu}
\Tr\left[\gamma_\sigma \gamma_5 S_0^\dagger \gamma_5 \mathcal{J}_\mu(X_1)S_0\right]
\Tr\left[\gamma_\rho \gamma_5 S_{X_3}^\dagger \gamma_5 \mathcal{J}_\nu(X_2)S_{X_3}\right]\\
&\qquad\quad
+ e^{-iQ_1\tfrac{a}{2}\hat\rho}
\Tr\left[\gamma_\sigma\gamma_5 S_0^\dagger \gamma_5 \mathcal{J}_\rho(X_3)S_0\right]
\Tr\left[\gamma_\mu\gamma_5 S_{X_1}^\dagger \gamma_5 \mathcal{J}_\nu(X_2)S_{X_1}\right]\\
&\qquad\quad
+ e^{-iQ_1\tfrac{a}{2}\hat\rho}
\Tr\left[\gamma_\mu\gamma_5 S_{X_1}^\dagger \gamma_5 \mathcal{J}_\rho(X_3)S_{X_1}\right]
\Tr\left[\gamma_\sigma\gamma_5 S_0^\dagger \gamma_5 \mathcal{J}_\nu(X_2)S_0\right]
\Bigr\rangle_c,
\end{aligned}
\end{equation}
where
\begin{equation}
  \left\langle \Tr A \Tr B \right\rangle_c = \Bigl\langle
    \bigl(\Tr A - \langle \Tr A\rangle_U\bigr)
    \bigl(\Tr B - \langle \Tr B\rangle_U\bigr)
    \Bigr\rangle_U
\end{equation}
is the QCD-connected expectation value over gauge fields. Each trace
corresponds to a quark loop in Fig.~\ref{fig:contractions}.

Since $X_1$ and $X_3$ are summed over, we evaluate the traces
involving $S_{X_1}$ and $S_{X_3}$ stochastically. To do this, we
introduce a color triplet, scalar noise field $\phi_a(X)$ with
randomly chosen $U(1)$ components, so that it has expectation value
$E[\phi_a(X)\phi^\dagger_b(Y)] = \delta_{ab}\delta_{XY}$. We use this
as the source for two quark propagators,
\begin{equation}
  S_\phi(X) = \sum_Y S(X,Y)\phi(Y),\qquad
  S_{\phi Q_1}(X)= \sum_Y S(X,Y)e^{-iQY}\phi(Y),
\end{equation}
where each spin component is solved independently using the same noise
source $\phi$~\cite{Boucaud:2008xu}. With these, we use the one-end
trick~\cite{Foster:1998vw} and obtain
\begin{equation}
\begin{aligned}
 \sum_{X_1}e^{iQ_1X_1} \Tr\left[
 \gamma_\mu\gamma_5 S^\dagger_{X_1}\gamma_5\mathcal{J}_\nu(X_2) S_{X_1} \right]
&= E\left( \Tr\left[ 
\gamma_\mu\gamma_5 S_{\phi Q_1}^\dagger\gamma_5\mathcal{J}_\nu(X_2) S_\phi
 \right]\right),\\
 \sum_{X_3}e^{-iQ_1X_3} \Tr\left[
 \gamma_\mu\gamma_5 S^\dagger_{X_3}\gamma_5\mathcal{J}_\nu(X_2) S_{X_3} \right]
&= E\left( \Tr\left[ 
\gamma_\mu\gamma_5 S_\phi^\dagger\gamma_5\mathcal{J}_\nu(X_2) S_{\phi Q_1}
 \right]\right).
\end{aligned}
\end{equation}
We reduce this stochastic noise by averaging over four noise fields
per configuration, as well as using color
dilution~\cite{Wilcox:1999ab,Foley:2005ac} and hierarchical
probing~\cite{Stathopoulos:2013aci} with 32 Hadamard vectors. We find
that for these two-point loops hierarchical probing has no benefit
over using additional noise fields; however, the noise-source
propagators can be reused for one-point loops relevant for the
hadronic vacuum polarization and for the other disconnected four-point
diagrams, and for those loops it is beneficial. We also find that it
can (if possible) in some cases be beneficial to average over the
exchange of the local and conserved currents in a quark loop. We
further reduce gauge noise by translating the origin of the
point-source propagator $S_0$ and averaging over 128 point sources per
gauge configuration.

%%%%%%%%%%%%%%%%%%%%%%%%%%%%%%%%%%%%%%%
\section{Lattice results}
\label{sec:res}
%%%%%%%%%%%%%%%%%%%%%%%%%%%%%%%%%%%%%%%

%-------------------------------------------
\subsection{Lattice setup}
%-------------------------------------------

\begin{table}[t]
\caption{Parameters of the simulations: the bare coupling $\beta = 6/g_0^2$, the lattice resolution, the hopping parameter $\kappa$, the lattice spacing $a$ in physical units extracted from \cite{Fritzsch:2012wq}, the pion mass $m_{\pi}$, the rho mass $m_{\rho}$ and the number of gauge configurations.}
\vskip 0.1in
\begin{tabular}{lcl@{\hskip 02em}l@{\hskip 01em}c@{\hskip 01em}c@{\hskip 01em}c@{\hskip 01em}c@{\hskip 01em}c}
	\hline
	CLS	&	$\quad\beta\quad$	&	$L^3\times T$ 		&	$\kappa$		&	$a~[\fm]$	&	$m_{\pi}~[\MeV]$	&	$m_{\rho}~[\MeV]$	& $m_{\pi}L$	 &	$\#$~confs\\
	\hline
	E5	&		$5.3$		&	$32^3\times64$	& 	$0.13625$	& 	$0.0652(6)$  	& 	$437(4)$ &	971	&	 4.7 	&	500\\  
	F6	&		 	 		& 	$48^3\times96$	&	$0.13635$	& 				& 	$314(3)$ &	886	&	 5.0	&	150\\      
	F7	&		 	 		& 	$48^3\times96$	&	$0.13638$	& 				& 	$270(3)$ &	841	&	 4.3 	&	124\\       
	G8	&		 	 		& 	$64^3\times128$	&	$0.136417$	& 				& 	$194(2)$ &	781	&	 4.1	&	86\\         
	\hline
	N6	&		 $5.5$ 		& 	$48^3\times96$	&	$0.13667$	& 	0.0483(4)		& 	$342(3)$ &	917	&	 4.0	&	236\\        
	\hline
 \end{tabular} 
\label{tabsim}
\end{table}

The four-point correlation functions are computed on a subset of the
$N_f=2$ CLS (Coordinated Lattice Simulations) ensembles generated
using the plaquette gauge action for gluons~\cite{Wilson:1974sk} and
the $\mathcal{O}(a)$-improved Wilson-Clover action for
fermions~\cite{Sheikholeslami:1985ij} with the non-perturbative
parameter $c_{\rm SW}$~\cite{Jansen:1998mx}. The fermionic boundary
conditions are periodic in space and antiperiodic in time.  We
consider two different values of the lattice spacing and different
pion masses in the range from 190 to 440\;MeV. The parameters of the
ensembles used in this work are summarized in Table~\ref{tabsim}.

For each ensemble, the connected four-point correlation function is
computed at a few values of $Q_1=(n\cdot2\pi/T,0,0,0)$, the first-listed
component corresponding to the time direction, with $n=1,2,3$ on
ensemble E5; $n=1,3$ on F6 and F7; $n=1,4$ on G8 and $n=2$ on N6.  For
the (2+2) diagrams, we use $n=2$ (E5) and $n=3$ (F6).  Then, for each
value of $Q_1$, the four-point correlation function is evaluated for
many different values of $Q_2$, corresponding to different values of
$Q_2^2$ and $\nu$.

For the fully-connected diagrams, we used two source positions on
ensembles E5, F6, F7; one source position on N6; and eight source
positions on G8.  In the latter case, we used the truncated-solver
method~\cite{Blum:2012uh} for the eight sources and a computation with
exact inversions of the Dirac operator for bias correction on one
source.

To estimate the even subtracted amplitudes, we compute the
subtraction term directly at $\nu=0$, i.e., with $Q_2$ orthogonal to
$Q_1$. For the odd subtracted amplitudes, we use the approximation
$\overline{\mathcal{M}}(\nu) \approx \mathcal{M}(\nu) -
\frac{\nu}{\nu_1}\mathcal{M}(\nu_1)$, where $\nu_1$ is the smallest
available nonzero value of $\nu$. In both cases we linearly interpolate
the subtraction term in $Q_2^2$ to match the value in the unsubtracted
term.

In all tables and figures, our results for the HLbL amplitudes are
multiplied by a factor of $10^6$ for better readability.

%---------------------------------------------------------------------------------------------------
\subsection{Connected contribution to the forward light-by-light amplitudes}
%---------------------------------------------------------------------------------------------------

The results for the connected contribution to the eight amplitudes
are depicted in Figs.~\ref{fig:amps_F6_part1}--\ref{fig:amps_F6_part2} for the ensemble F6.
Additional figures for ensemble G8 can be found in appendix \ref{sec:addmat}, Fig.\ \ref{fig:amps_G8_part1}.
For F6 we show the amplitudes for two different values of the virtuality $Q_1^2$. 
We used all lattice momenta $Q_2$ up to $Q_2^2 \lesssim 4~\GeV^2$. The variable
$\nu$ is then bounded by $\nu \leq (Q_1^2Q_2^2)^{1/2}$. The four
amplitudes $\MsTT$, $\MsTL$, $\MsLT$ and , $\MsLL$ are positive as
they are related to cross sections, while the amplitudes $\MsTTa$,
$\MsTTt$, $\MsTLt$, $\MsTLa$, corresponding to interference terms, are
not sign-definite.  Since all amplitudes vanish in the limit of either 
$Q_1$ or $Q_2 \to 0$, the signal deteriorates at small $Q_1^2$ {(for
  fixed $Q_2^2$)} as can be seen by comparing the left and right
panels of Fig.~\ref{fig:amps_F6_part1}.

\afterpage{

\begin{figure}[p]

	\begin{minipage}[c]{0.49\linewidth}
	\centering 
	\includegraphics*[width=0.99\linewidth]{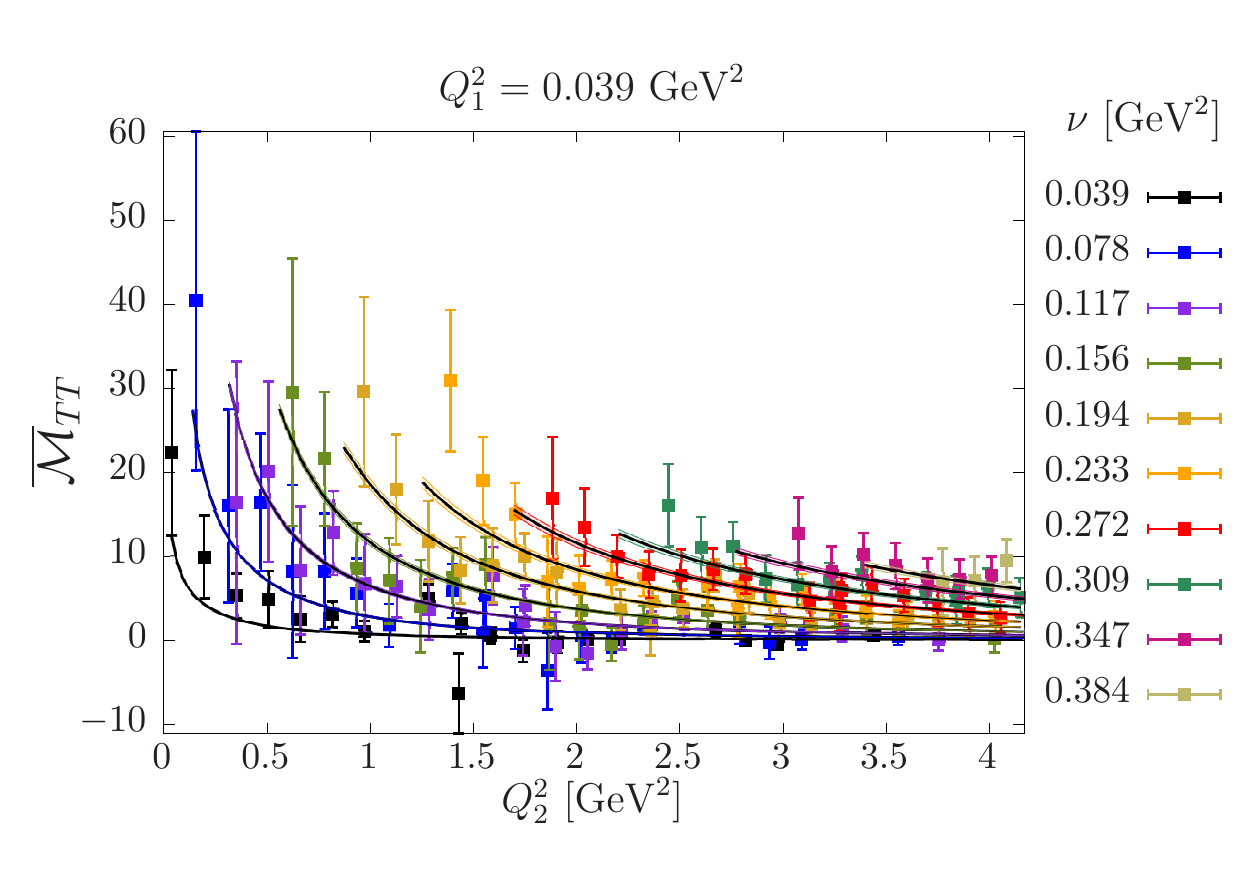}
	\end{minipage}
	\begin{minipage}[c]{0.49\linewidth}
	\centering 
	\includegraphics*[width=0.99\linewidth]{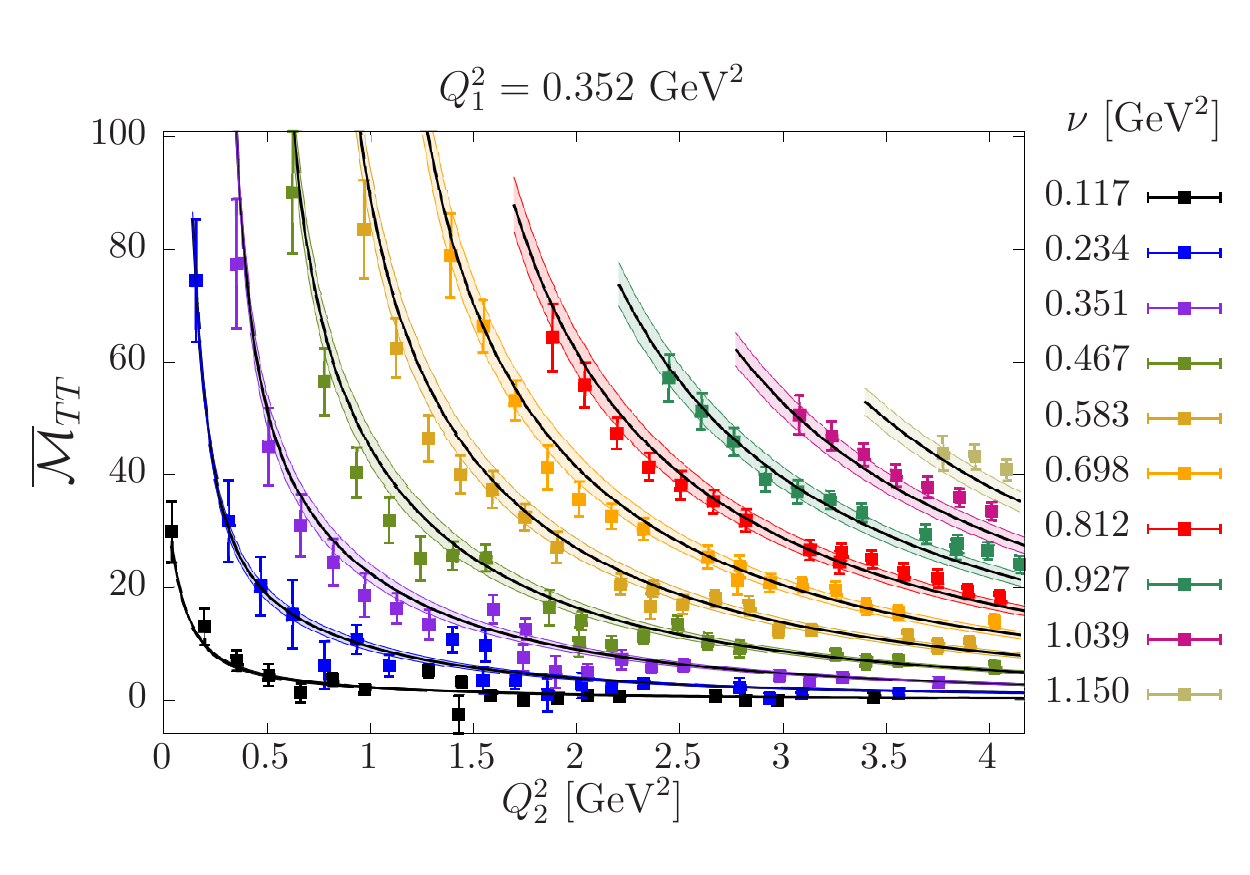}
	\end{minipage}
	
	\begin{minipage}[c]{0.49\linewidth}
	\centering 
	\includegraphics*[width=0.99\linewidth]{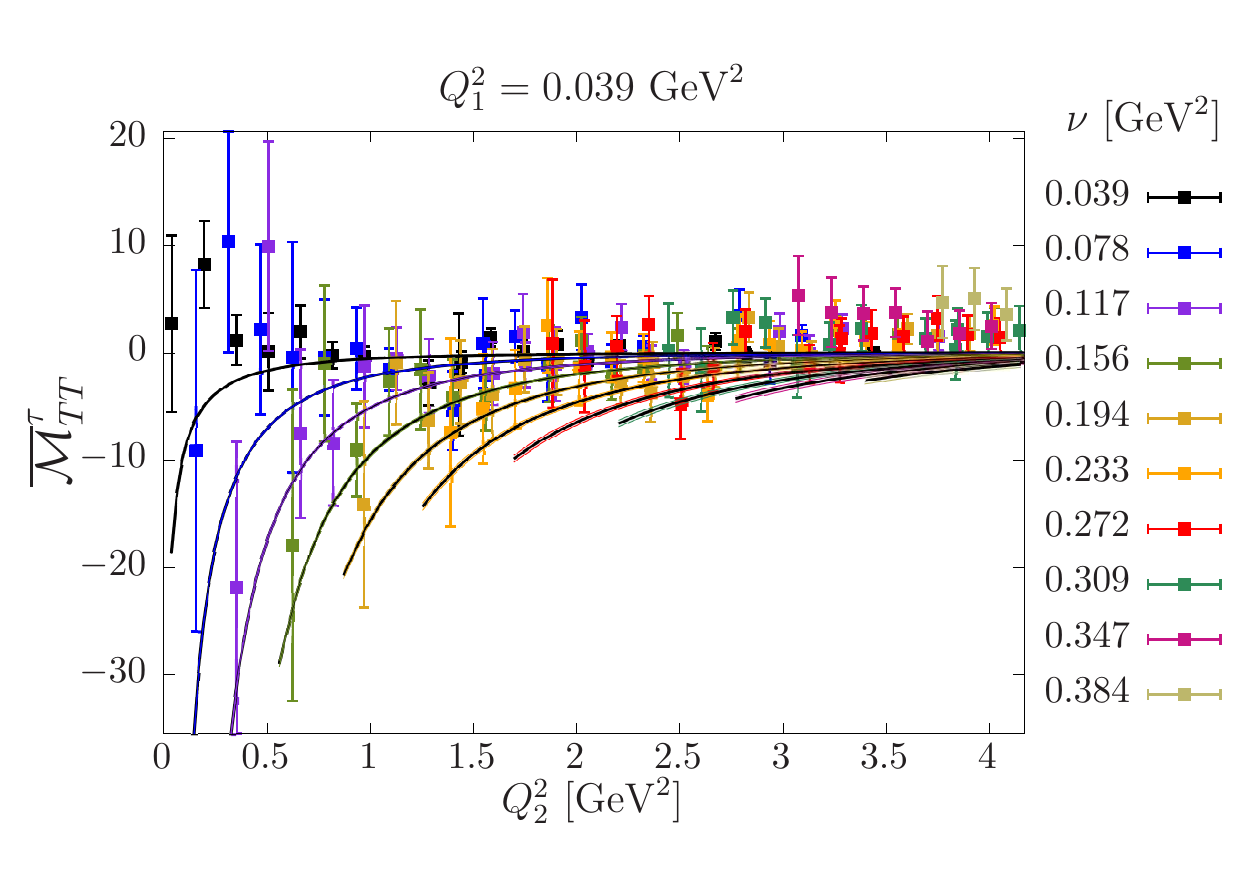}
	\end{minipage}
	\begin{minipage}[c]{0.49\linewidth}
	\centering 
	\includegraphics*[width=0.99\linewidth]{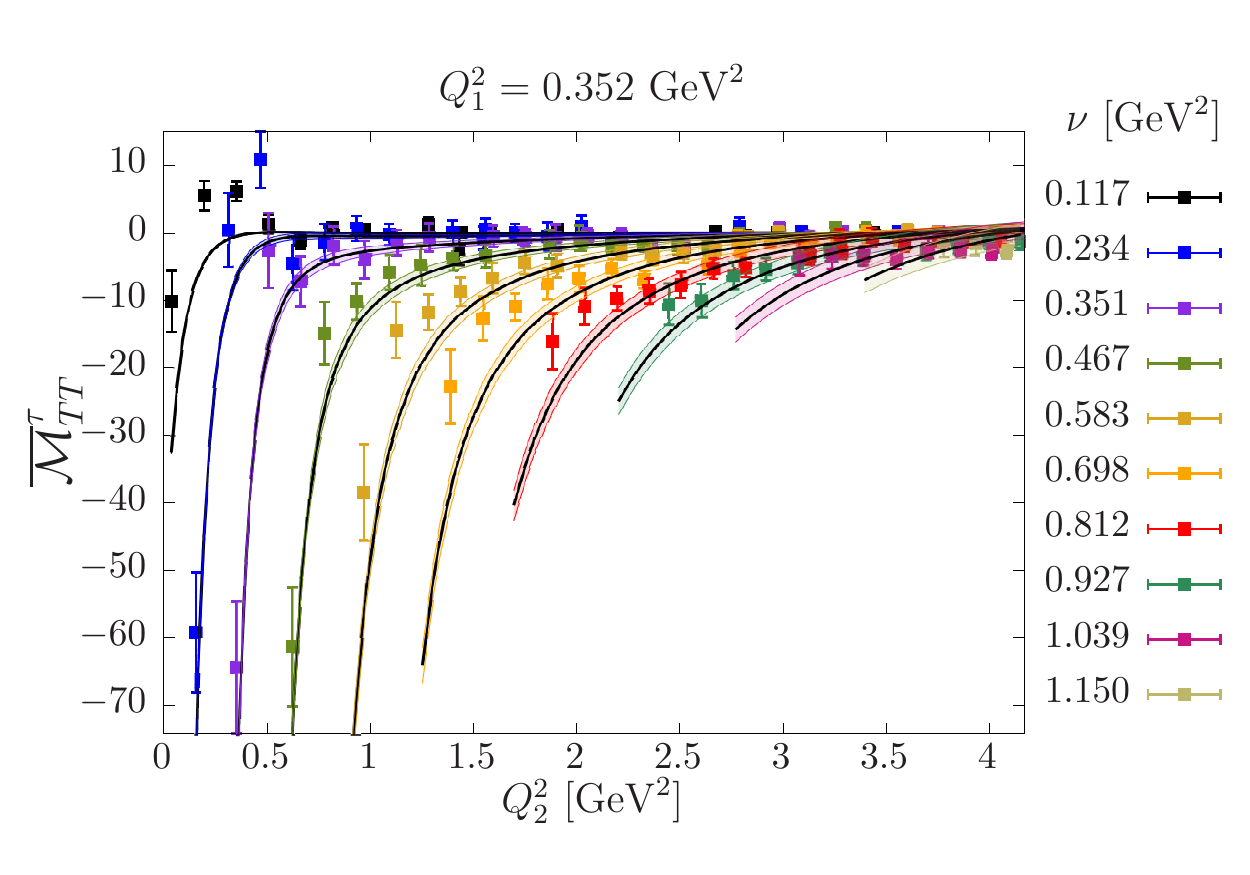}
	\end{minipage}

	\begin{minipage}[c]{0.49\linewidth}
	\centering 
	\includegraphics*[width=0.99\linewidth]{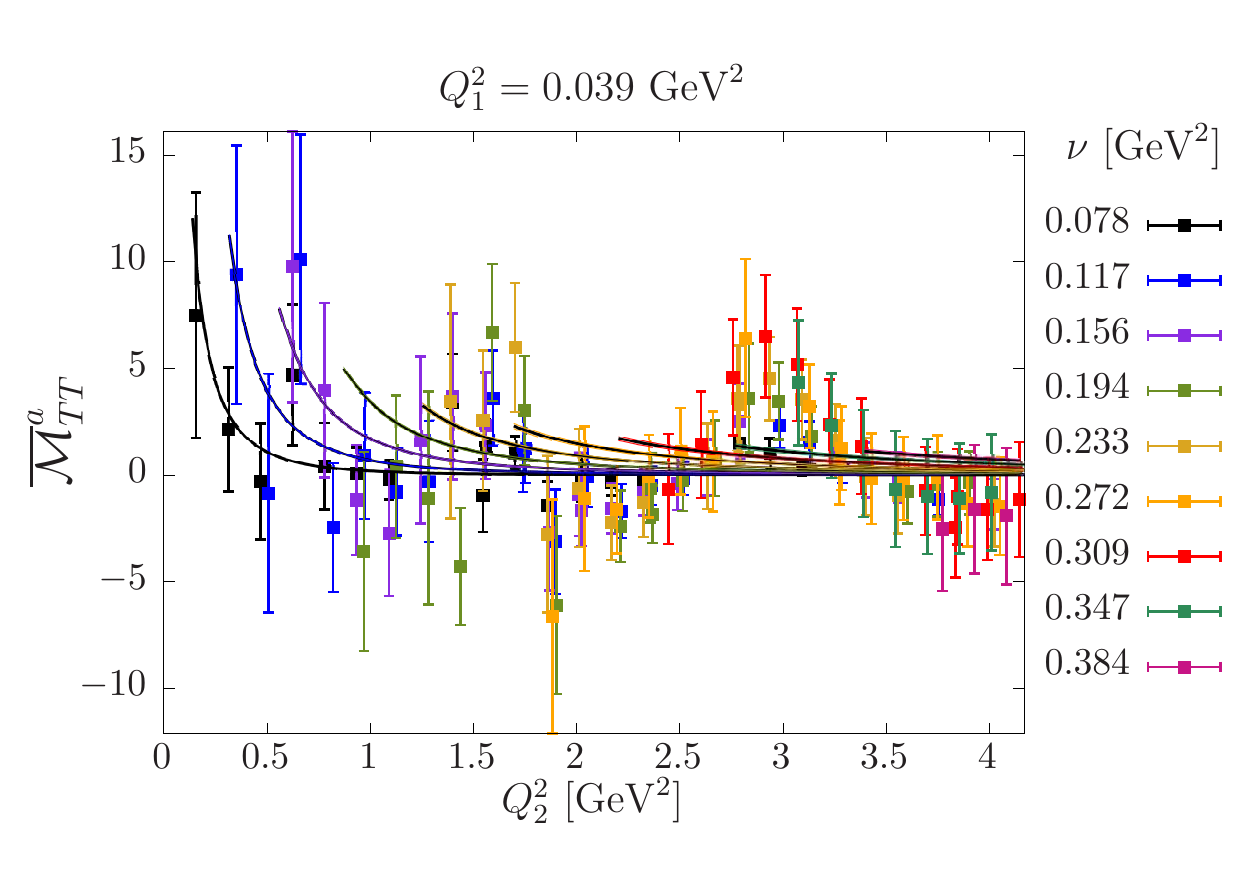}
	\end{minipage}
	\begin{minipage}[c]{0.49\linewidth}
	\centering 
	\includegraphics*[width=0.99\linewidth]{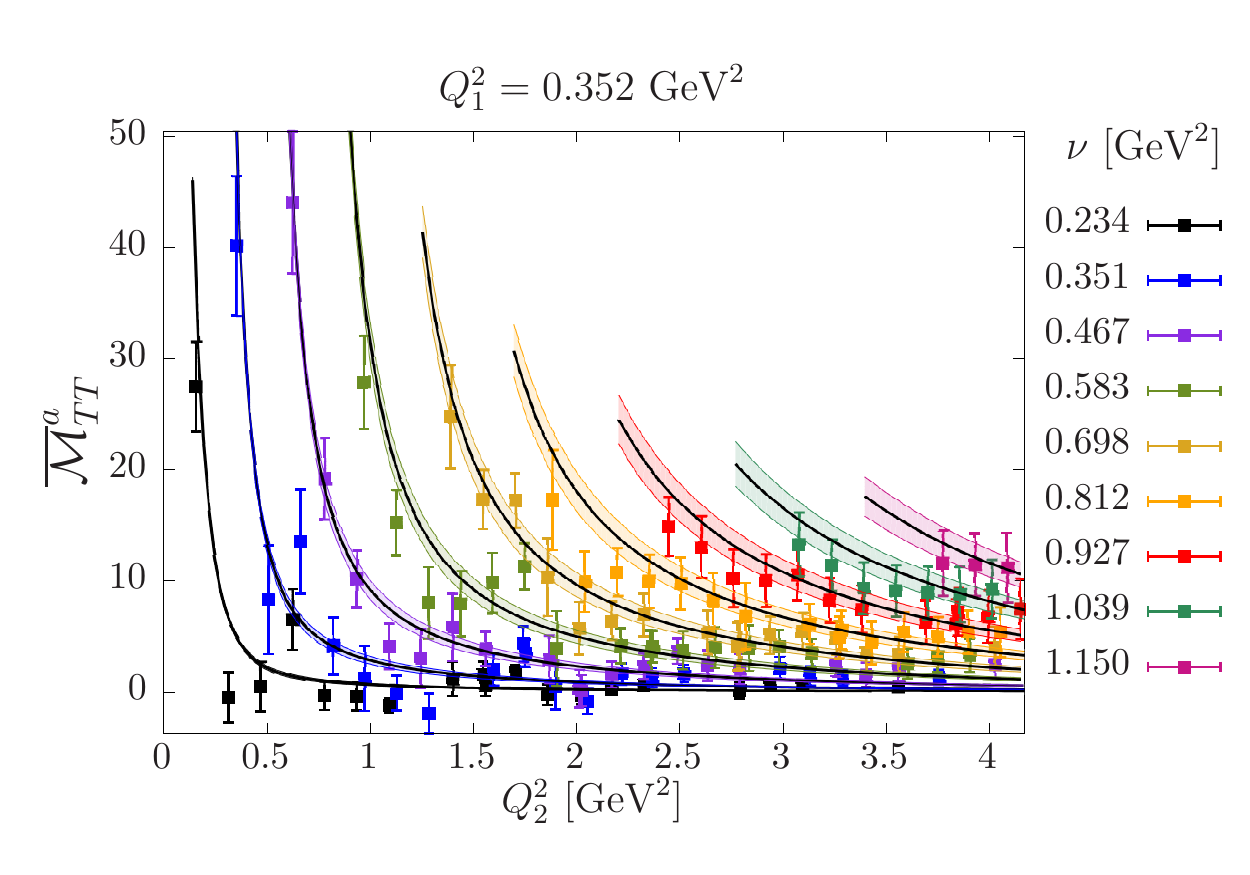}
	\end{minipage}
		
	\begin{minipage}[c]{0.49\linewidth}
	\centering 
	\includegraphics*[width=0.99\linewidth]{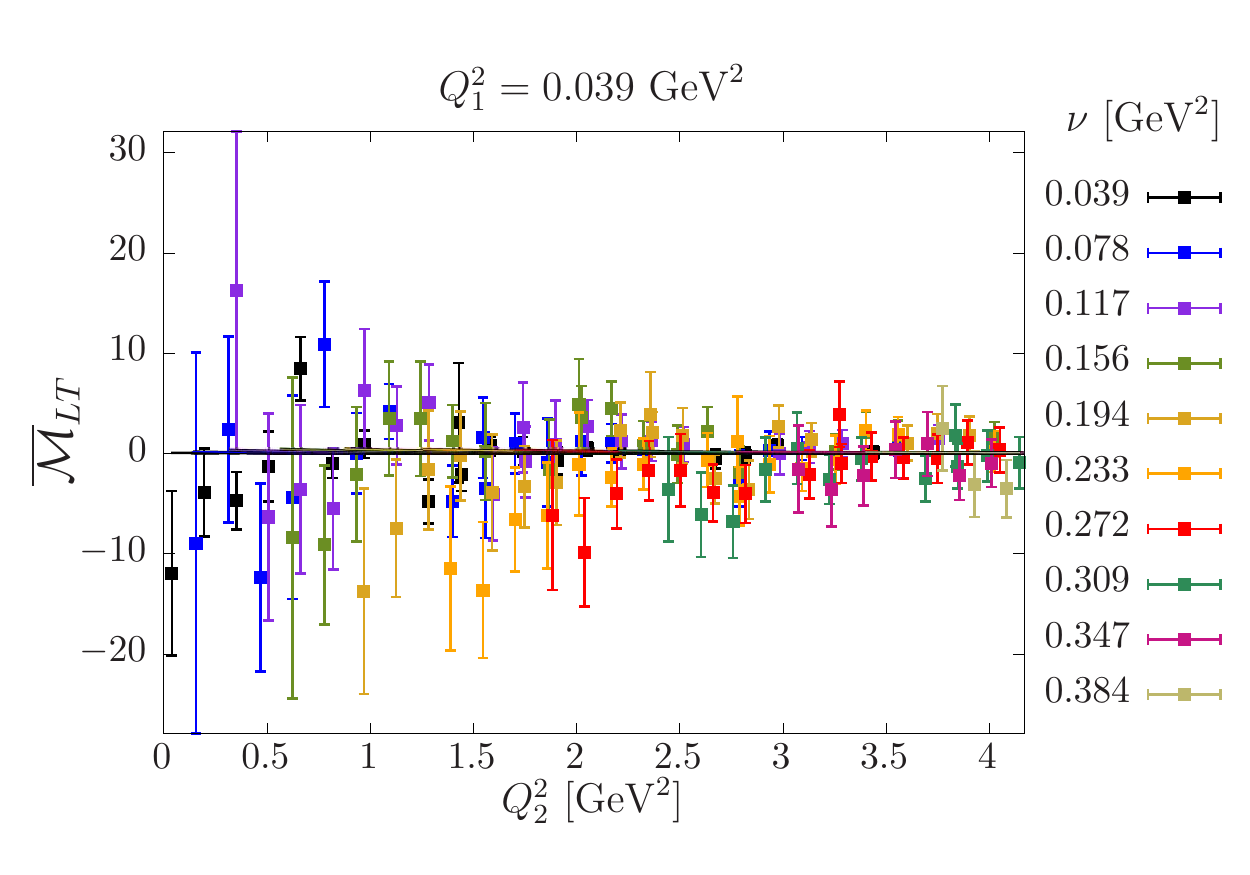}
	\end{minipage}
	\begin{minipage}[c]{0.49\linewidth}
	\centering 
	\includegraphics*[width=0.99\linewidth]{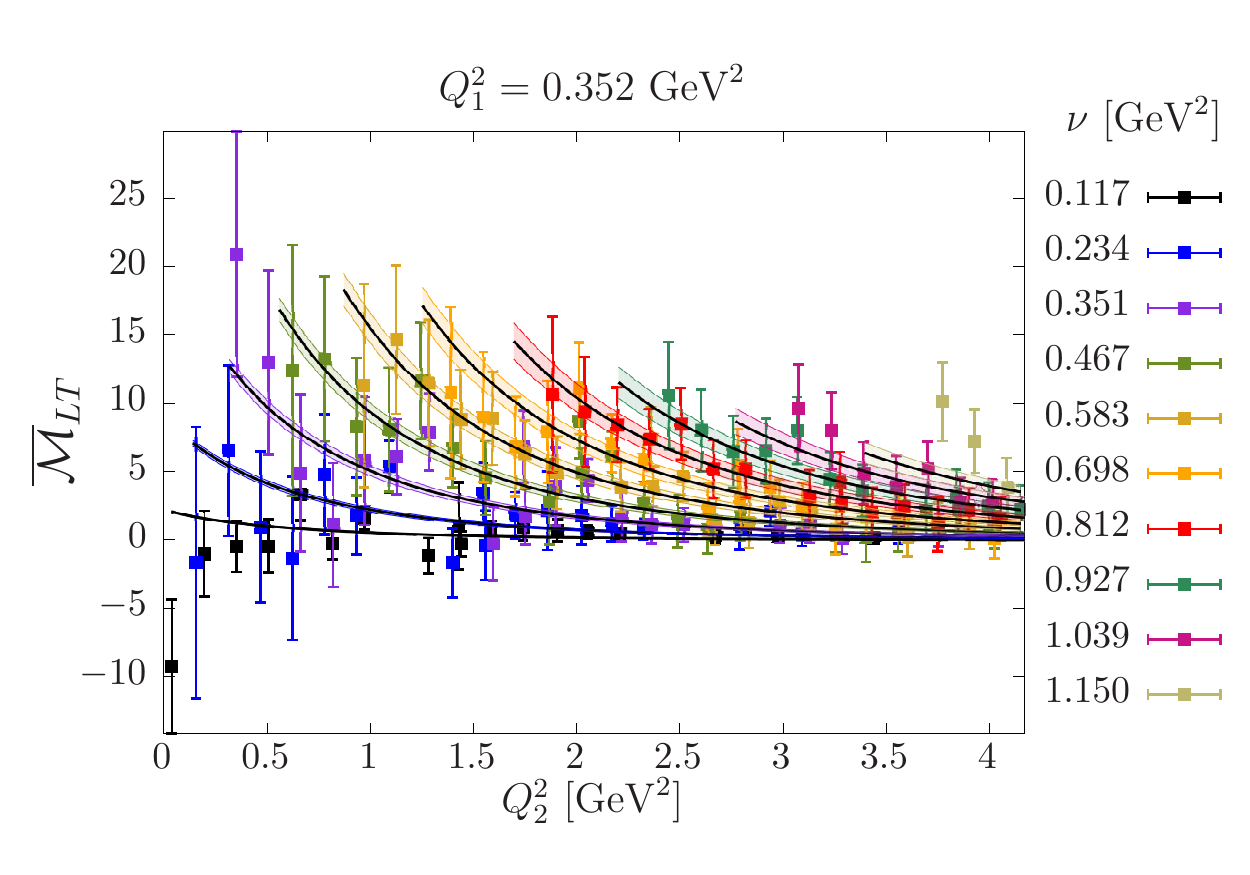}
	\end{minipage}

	\caption{Amplitudes $\MsTT$, $\MsTTt$, $\MsTTa$ and $\MsLT$ $(\times10^6)$ for the ensemble F6 and for two different values of $Q_1^2$ (left: $Q_1^2=0.039~\GeV^2$, right: $Q_1^2=0.352~\GeV^2$). The curves with error-bands represent the fit results discussed in Sec.~\ref{sec:fit}.}	
	\label{fig:amps_F6_part1}
	
\end{figure}

\clearpage}

\afterpage{

\begin{figure}[p]
	
	\begin{minipage}[c]{0.49\linewidth}
	\centering 
	\includegraphics*[width=0.99\linewidth]{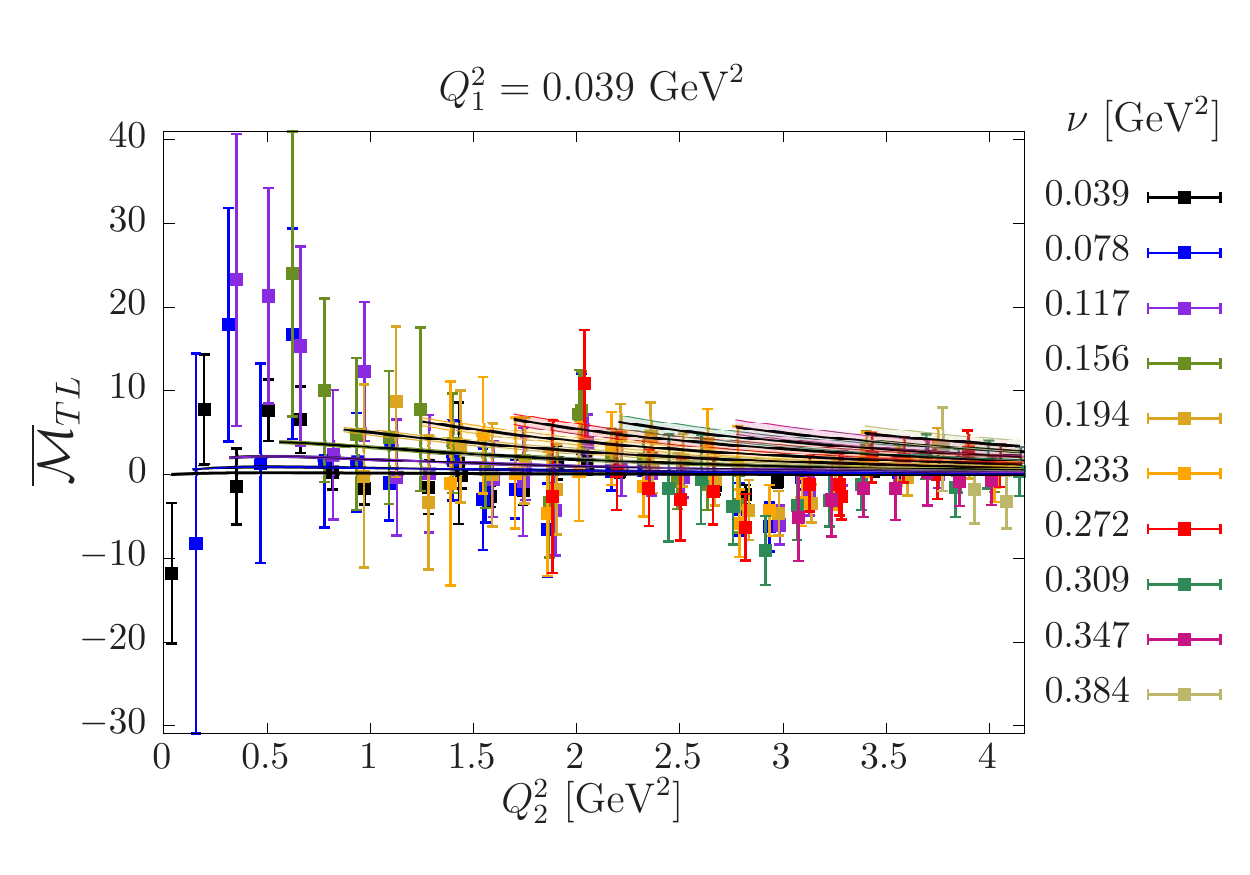}
	\end{minipage}
	\begin{minipage}[c]{0.49\linewidth}
	\centering 
	\includegraphics*[width=0.99\linewidth]{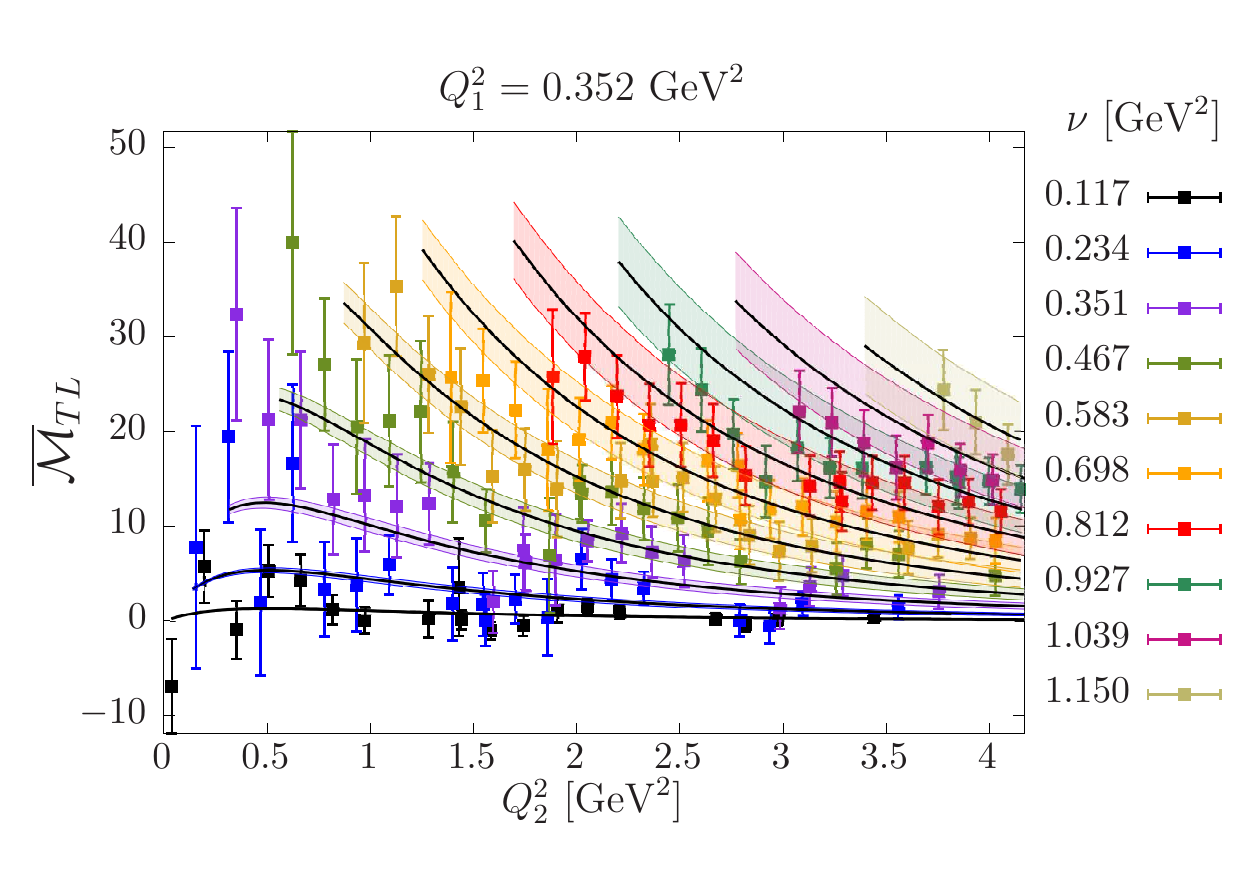}
	\end{minipage}	
	
	\begin{minipage}[c]{0.49\linewidth}
	\centering 
	\includegraphics*[width=0.99\linewidth]{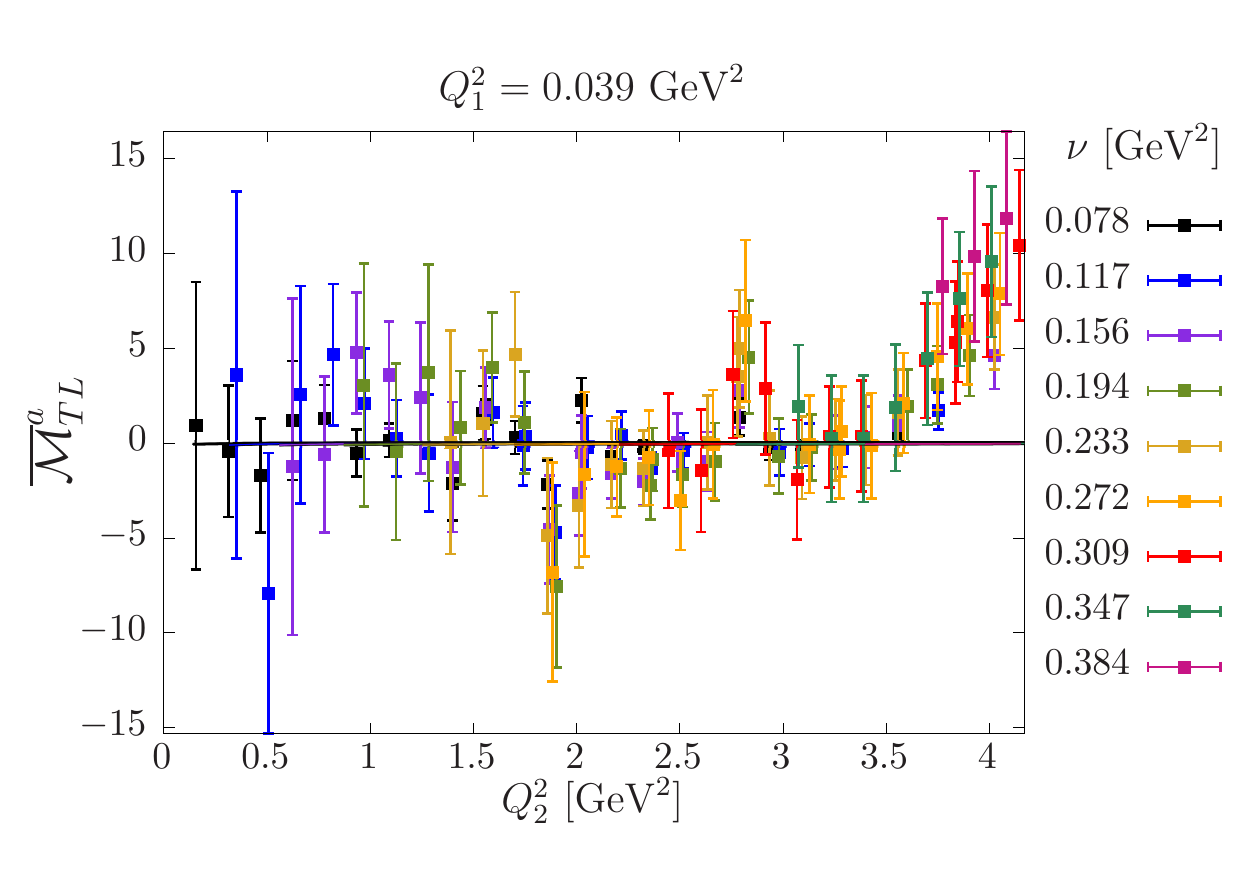}
	\end{minipage}
	\begin{minipage}[c]{0.49\linewidth}
	\centering 
	\includegraphics*[width=0.99\linewidth]{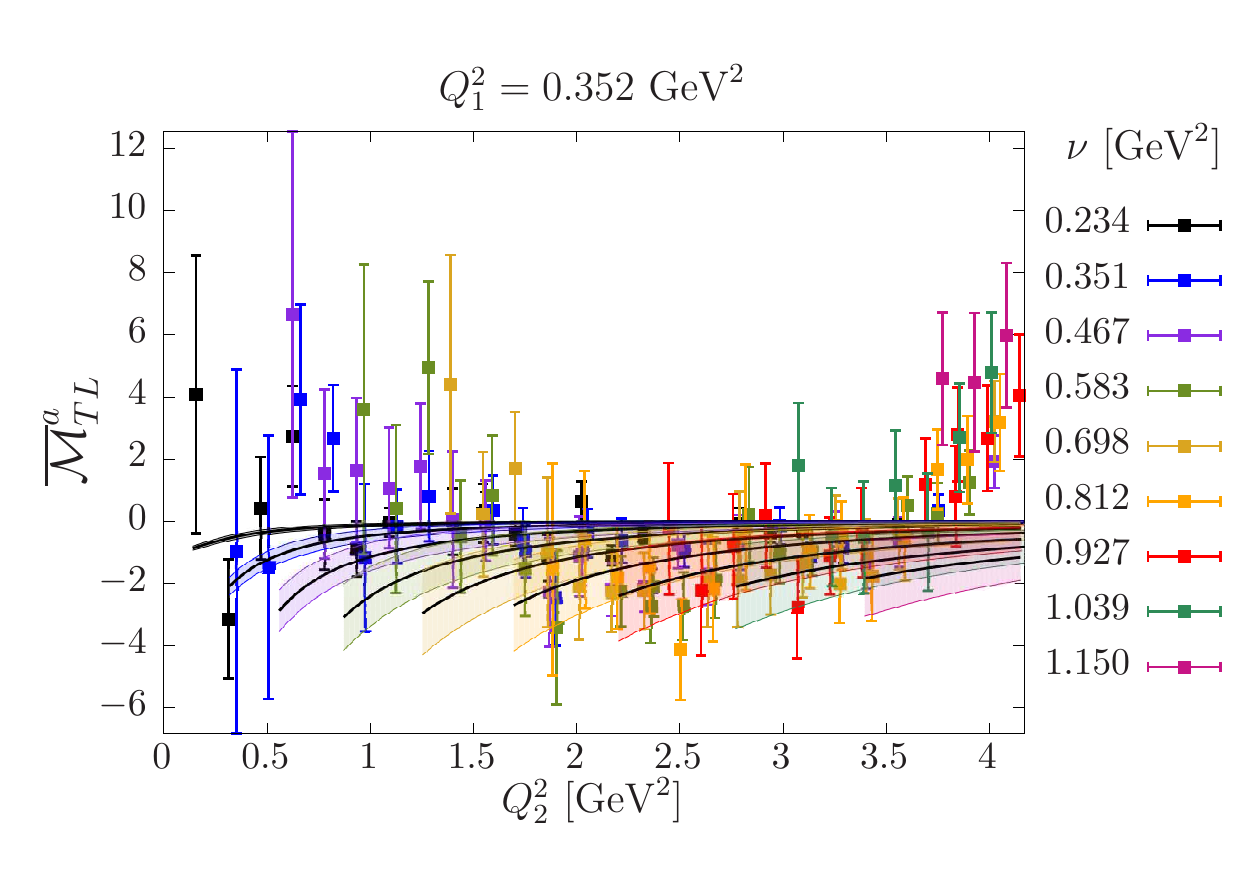}
	\end{minipage}
	
	\begin{minipage}[c]{0.49\linewidth}
	\centering 
	\includegraphics*[width=0.99\linewidth]{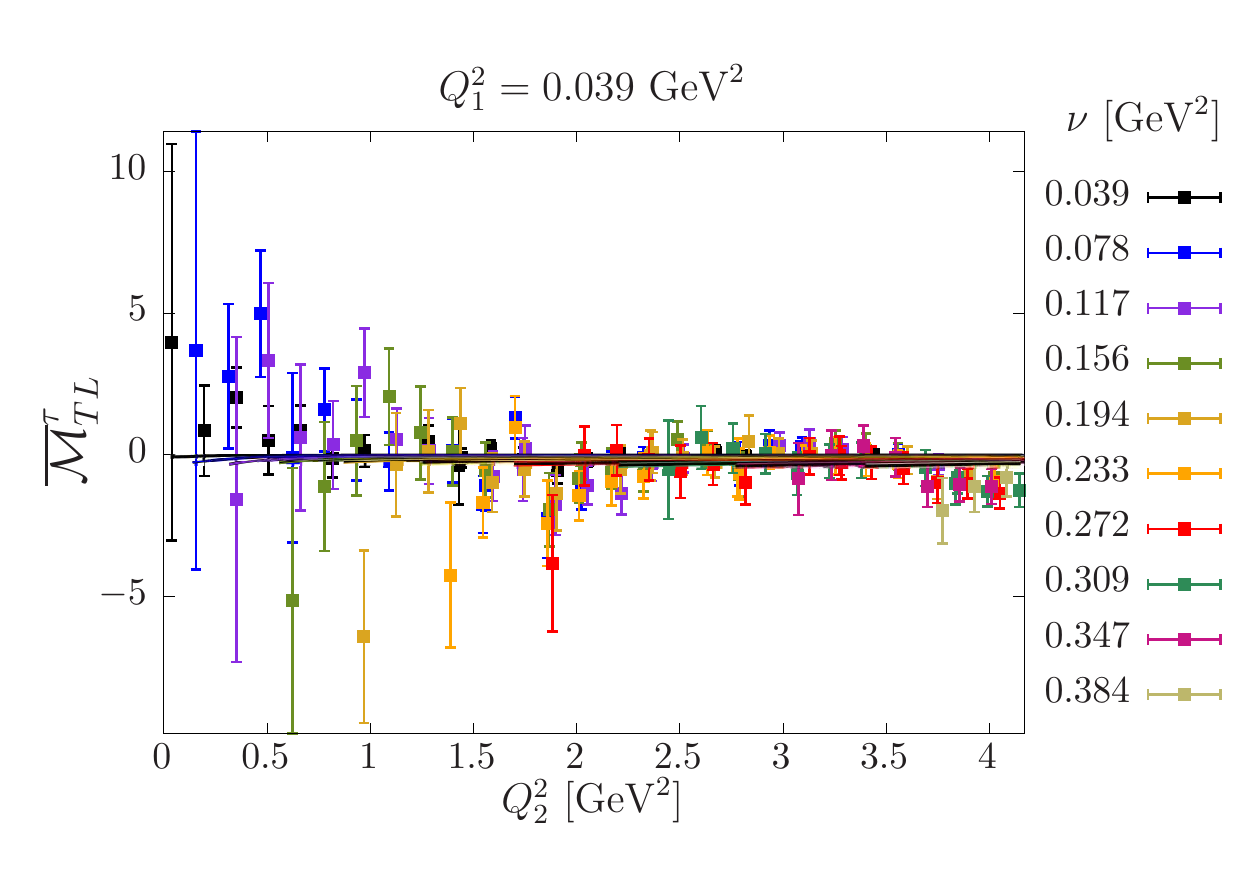}
	\end{minipage}
	\begin{minipage}[c]{0.49\linewidth}
	\centering 
	\includegraphics*[width=0.99\linewidth]{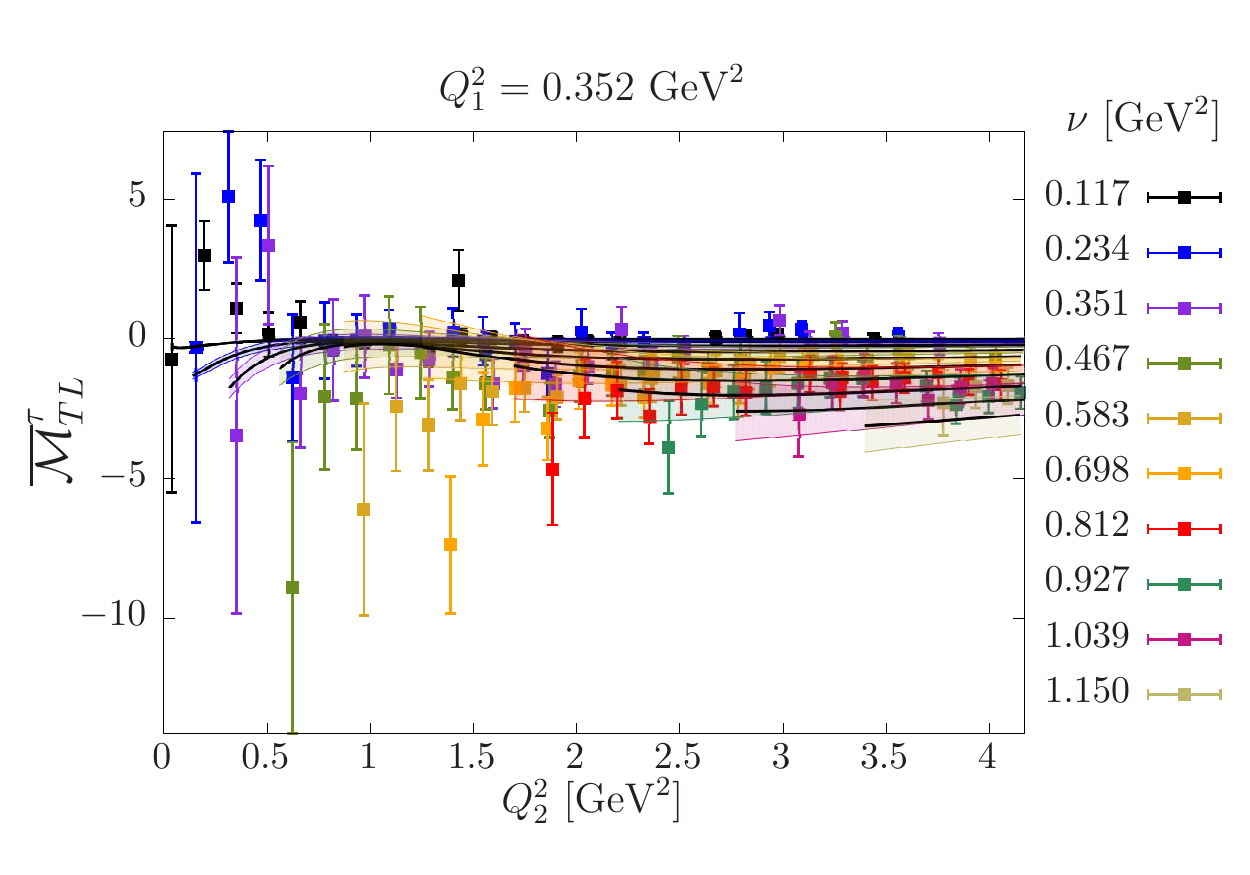}
	\end{minipage}
		
	\begin{minipage}[c]{0.49\linewidth}
	\centering 
	\includegraphics*[width=0.99\linewidth]{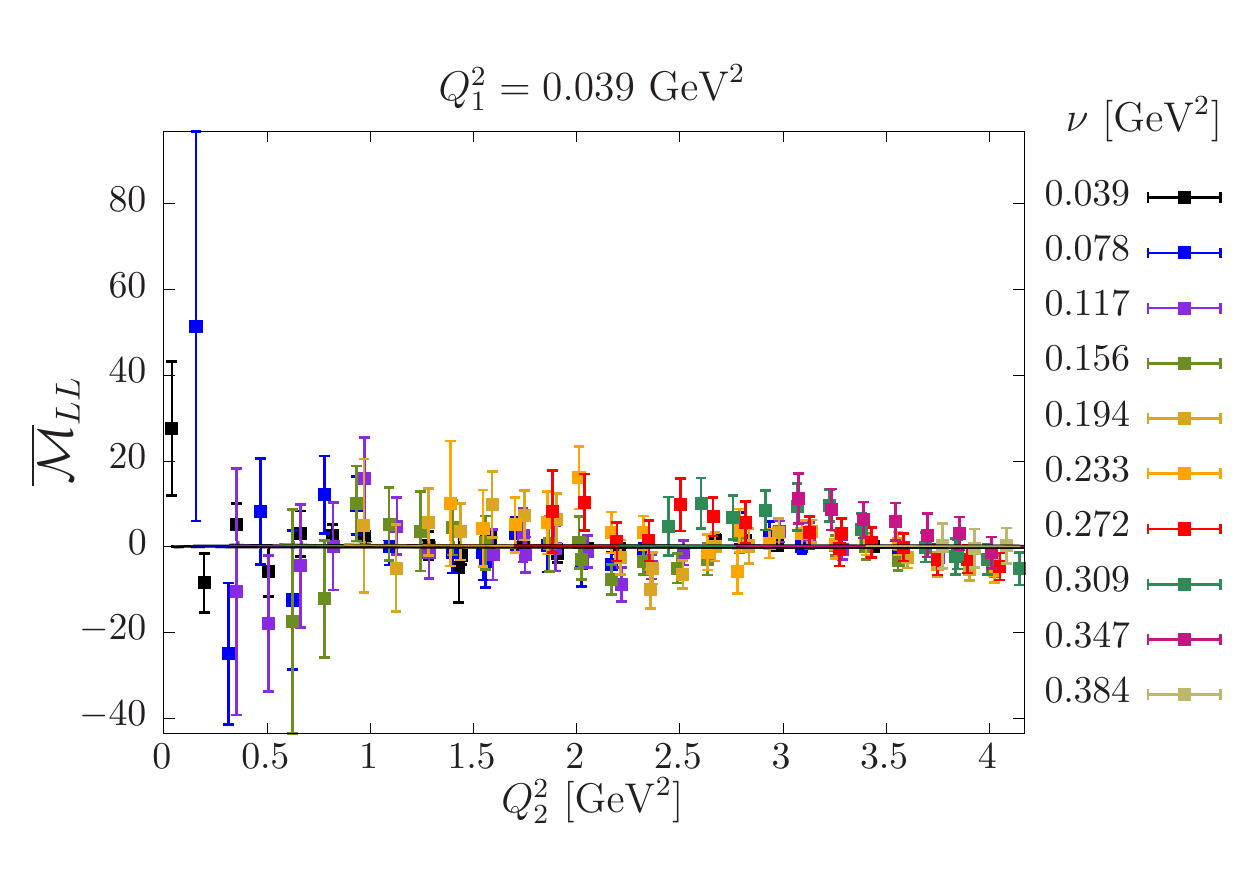}
	\end{minipage}
	\begin{minipage}[c]{0.49\linewidth}
	\centering 
	\includegraphics*[width=0.99\linewidth]{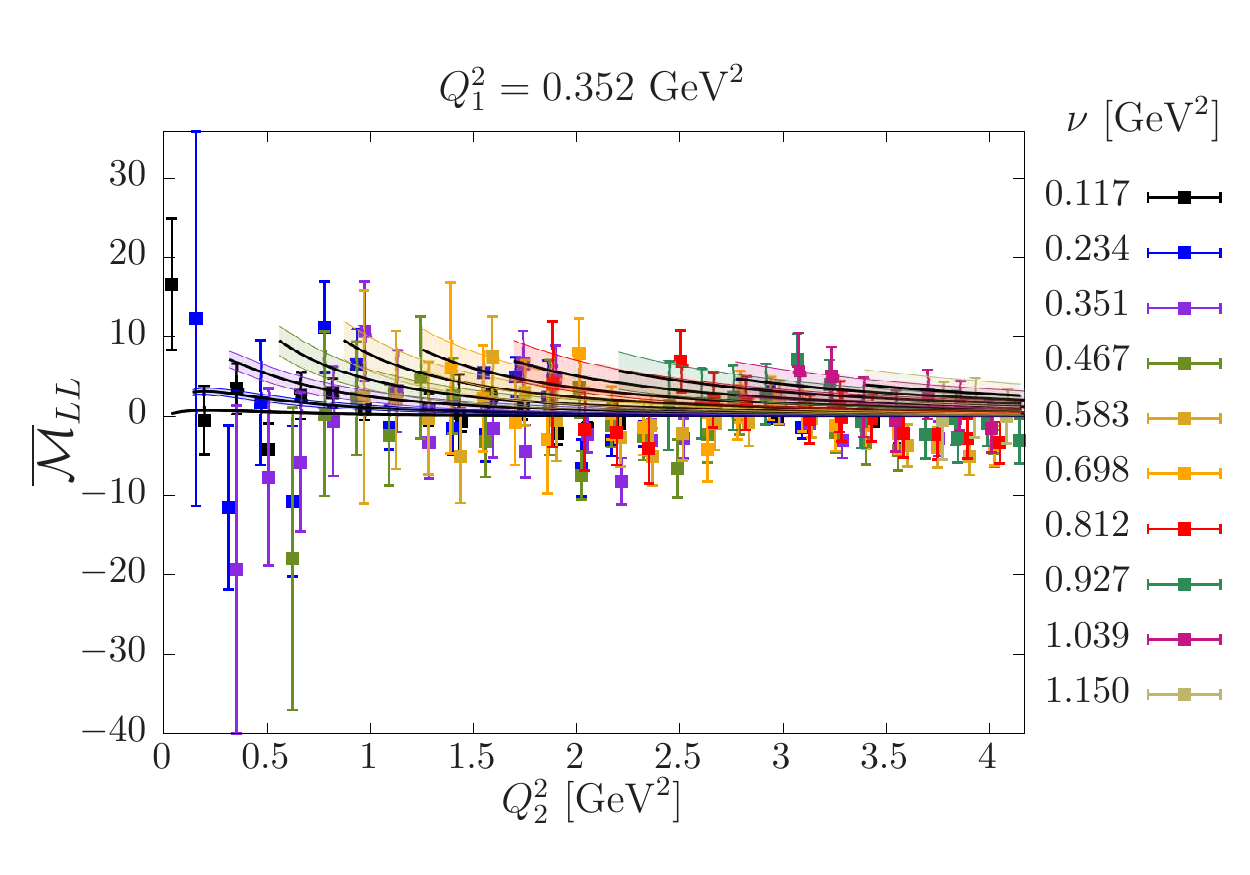}
	\end{minipage}
	
	\caption{Amplitudes $\MsTL$, $\MsTLa$, $\MsTLt$ and $\MsLL$ $(\times10^6)$ for the ensemble F6 and for two different values of $Q_1^2$ (left: $Q_1^2=0.039~\GeV^2$, right: $Q_1^2=0.352~\GeV^2$). The curves with error-bands represent the fit results discussed in Sec.~\ref{sec:fit}.}		
	\label{fig:amps_F6_part2}
\end{figure}

\clearpage}

%---------------------------------------------------------------------------------------------------
\subsection{Disconnected contribution to the forward light-by-light amplitudes}
%---------------------------------------------------------------------------------------------------

\begin{figure}[t]
  \centering
\begin{minipage}{0.428\textwidth}
\vspace{-5.4cm}
    \includegraphics[width=\textwidth]{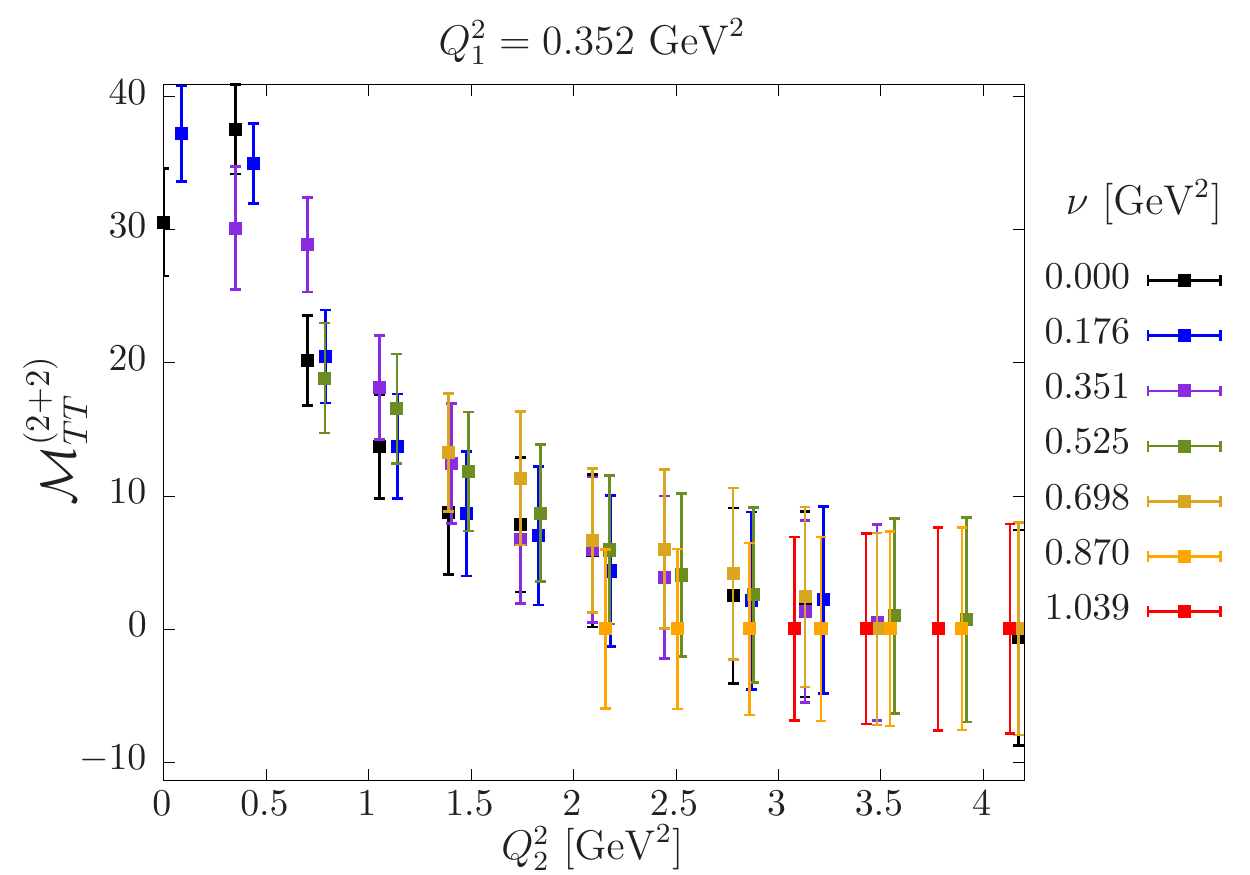}
\end{minipage}
  \includegraphics[width=0.495\textwidth]{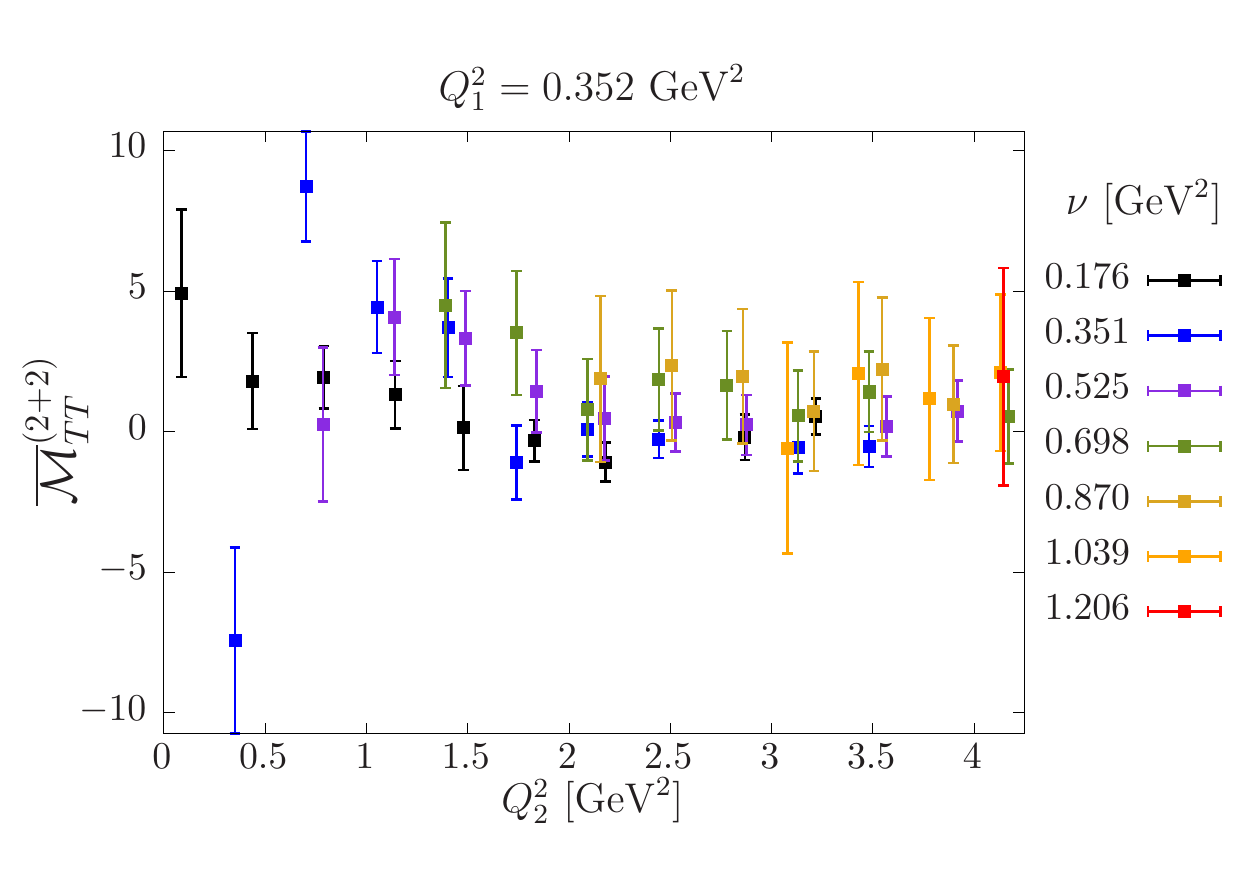}
  \caption{Contribution $(\times10^6)$ from $(2+2)$-disconnected diagrams to the
    forward scattering amplitude $\mathcal{M}_{TT}$ on ensemble E5,
    without (left) and with (right) subtraction of the value at
    $\nu=0$.}
  \label{fig:22_TT_sub}
\end{figure}

We now come to our results for the (2+2) disconnected diagram contribution to the eight subtracted 
amplitudes. We obtain this contribution with a reasonable statistical precision;
however, some of the amplitudes are significantly different
from zero when $Q^2_2=0$, as shown in the left panel of Fig.~\ref{fig:22_TT_sub}. 
In infinite volume, the Euclidean four-point function
should vanish at this kinematic point, since a conserved current can be written as the divergence of a tensor field,
$J_\mu(x) = \partial_\nu(x_\mu J_\nu(x))$, so that $\int d^4x \; J_\mu(x)$ is a pure boundary term,
which vanishes in the presence of a mass gap.
Therefore this is a sign of significant finite-volume effects. The bulk of the effect may be removed
when subtracting the amplitude at $\nu=0$, but some of it may remain. Figure~\ref{fig:22_TT_sub} also
shows that due to correlations, the subtraction significantly reduces the statistical uncertainty.
The full set of subtracted amplitudes on ensemble F6 is shown in Fig.~\ref{fig:22_F6}.

\afterpage{

\begin{figure}[p]
	
	\begin{minipage}[c]{0.49\linewidth}
	\centering 
	\includegraphics*[width=0.99\linewidth]{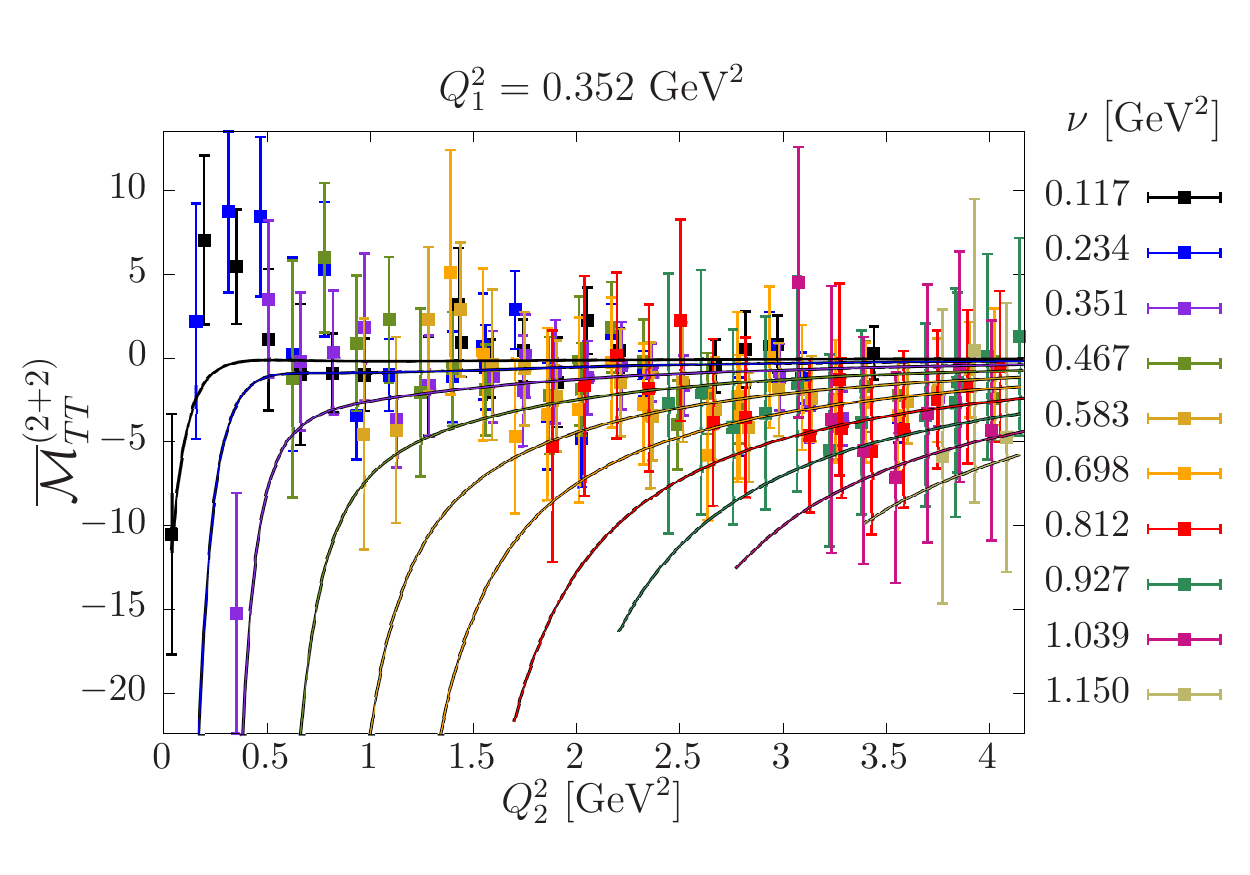}
	\end{minipage}
	\begin{minipage}[c]{0.49\linewidth}
	\centering 
	\includegraphics*[width=0.99\linewidth]{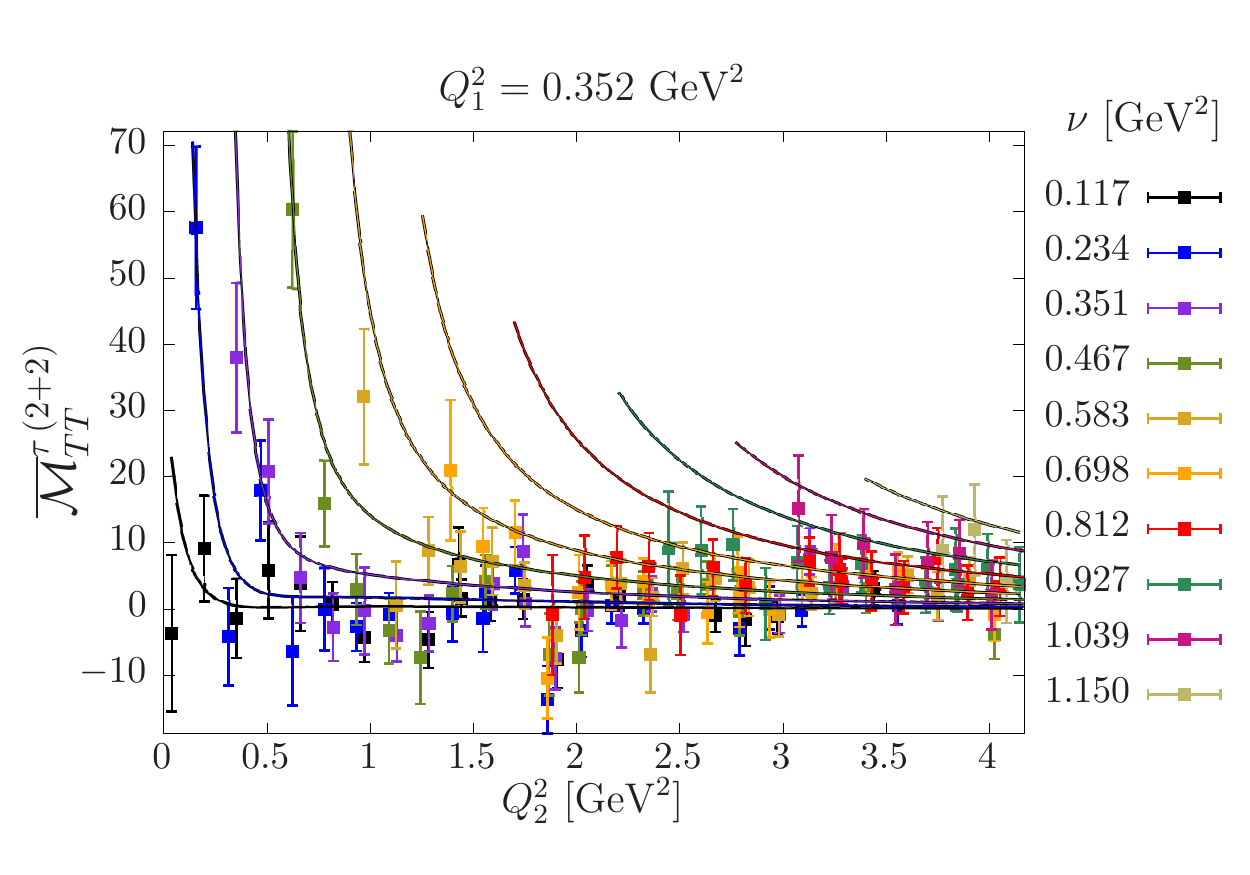}
	\end{minipage}	
	
	\begin{minipage}[c]{0.49\linewidth}
	\centering 
	\includegraphics*[width=0.99\linewidth]{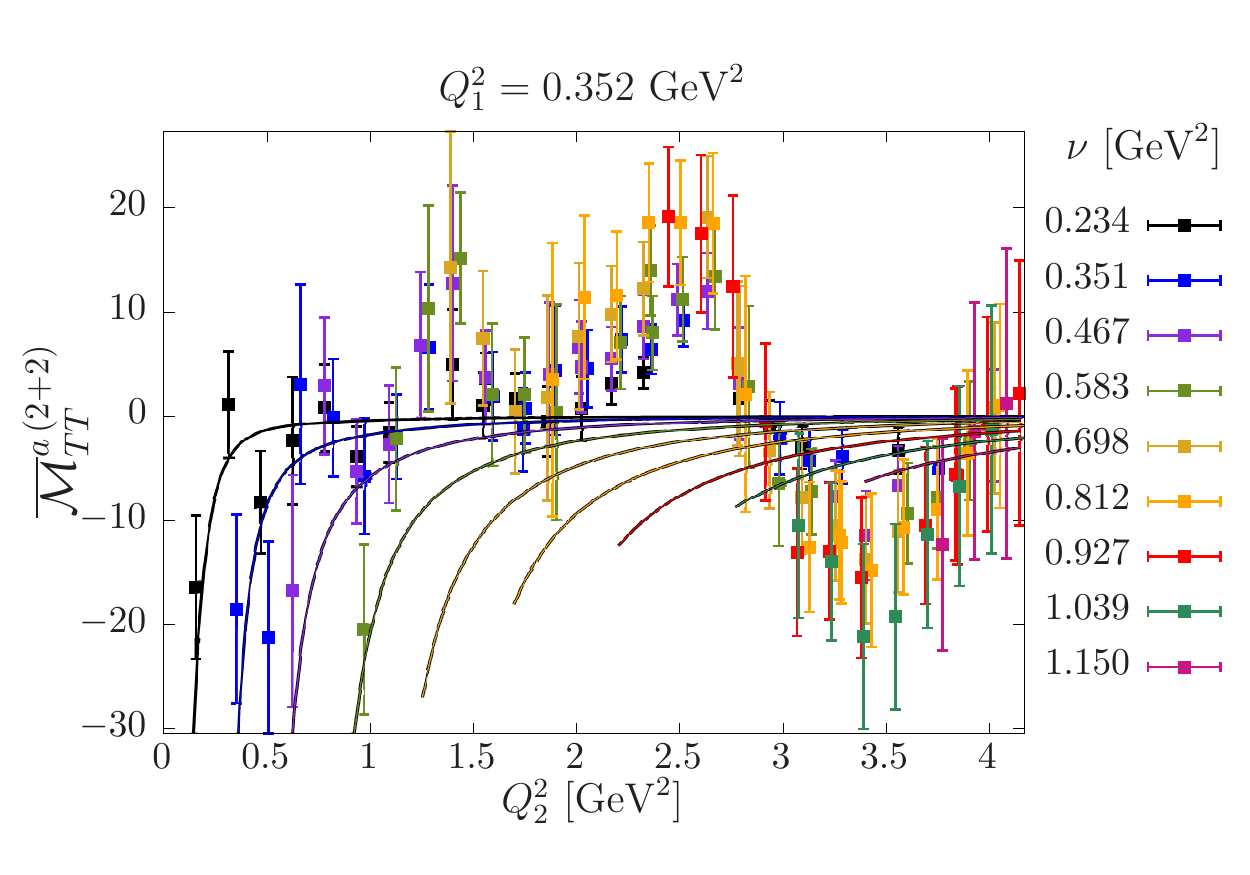}
	\end{minipage}
	\begin{minipage}[c]{0.49\linewidth}
	\centering 
	\includegraphics*[width=0.99\linewidth]{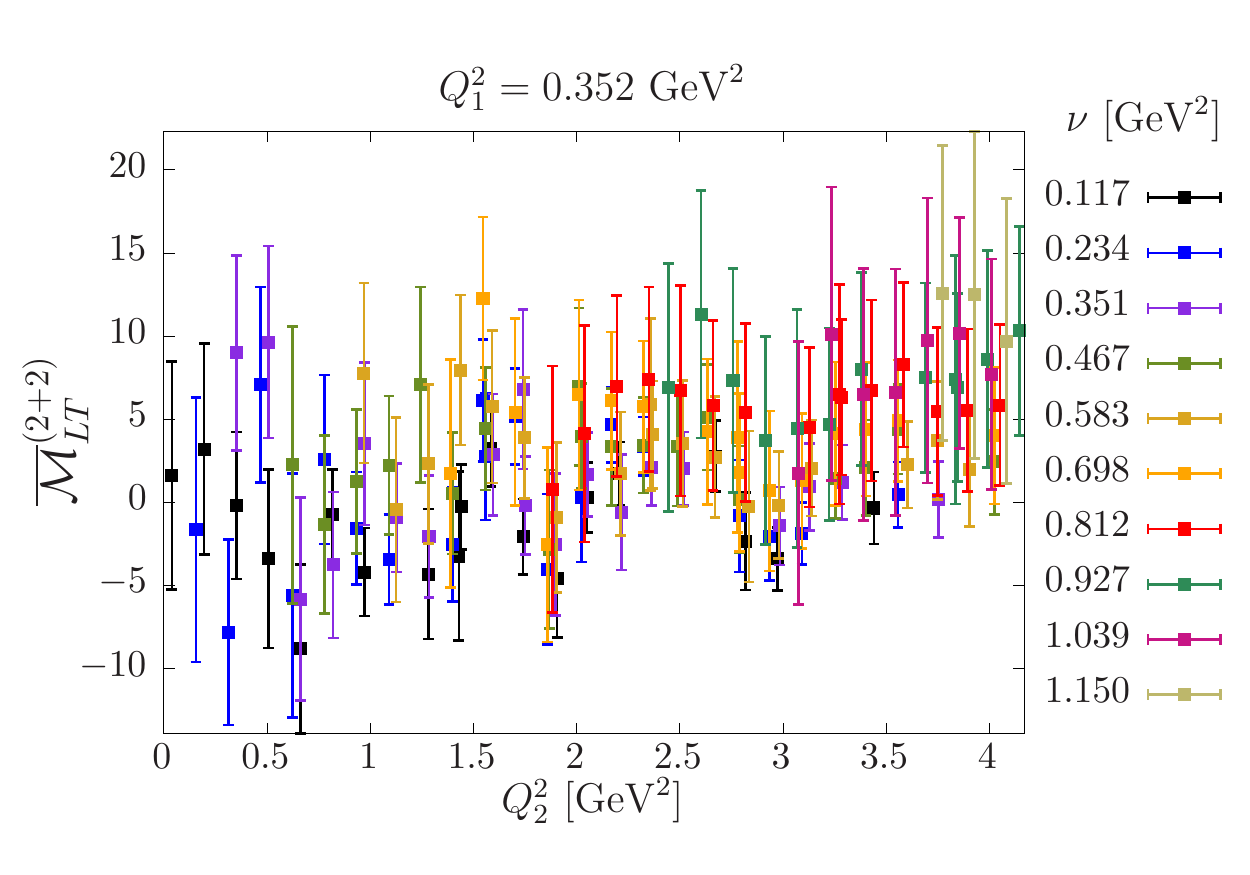}
	\end{minipage}
	
	\begin{minipage}[c]{0.49\linewidth}
	\centering 
	\includegraphics*[width=0.99\linewidth]{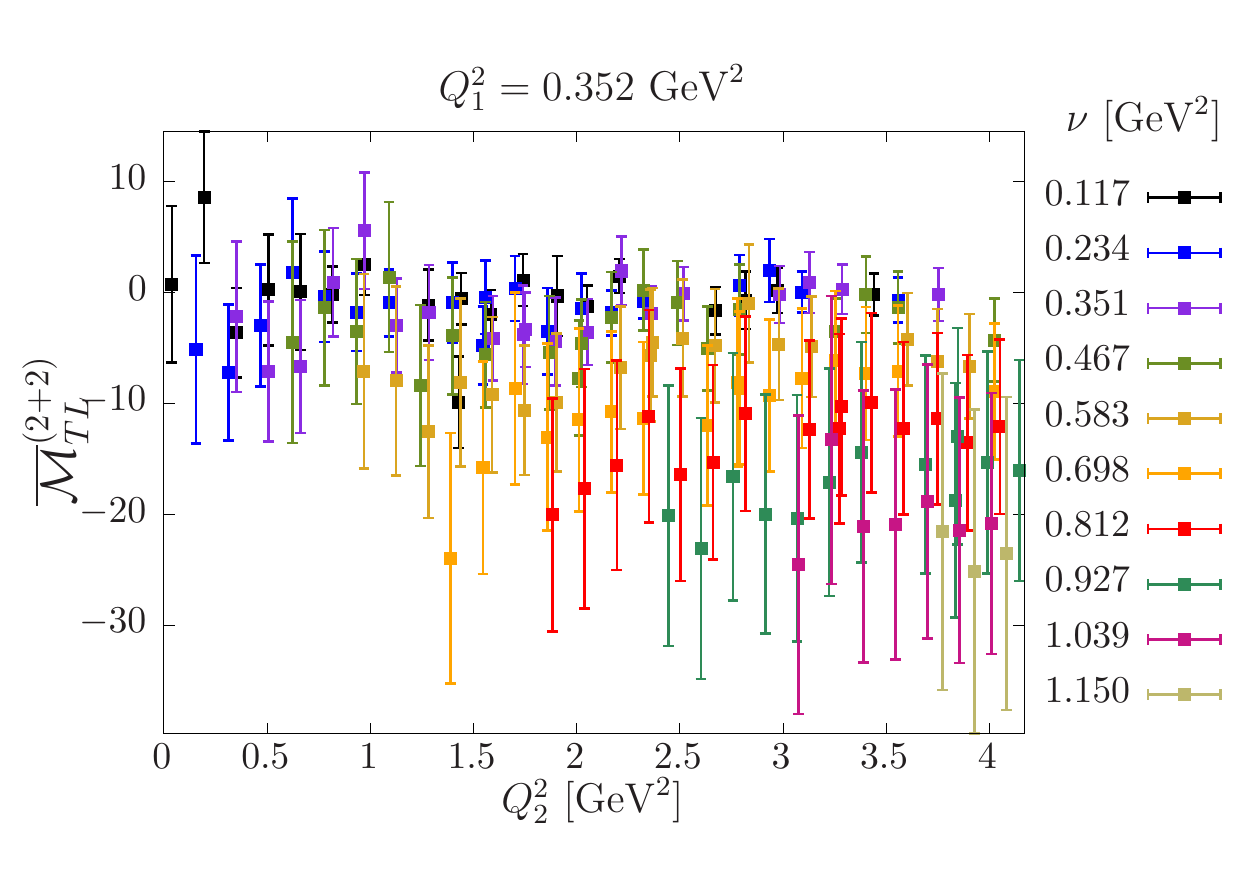}
	\end{minipage}
	\begin{minipage}[c]{0.49\linewidth}
	\centering 
	\includegraphics*[width=0.99\linewidth]{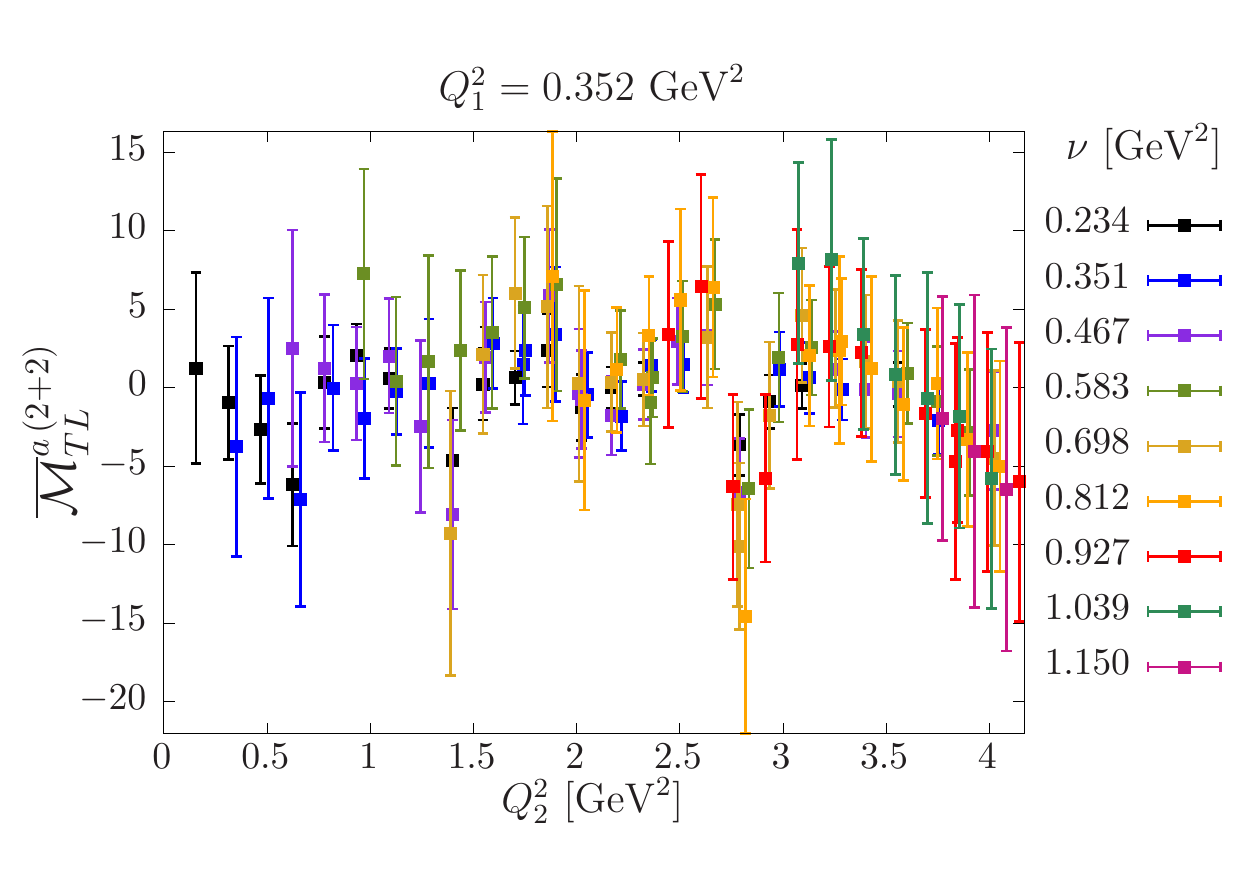}
	\end{minipage}
	
	\begin{minipage}[c]{0.49\linewidth}
	\centering 
	\includegraphics*[width=0.99\linewidth]{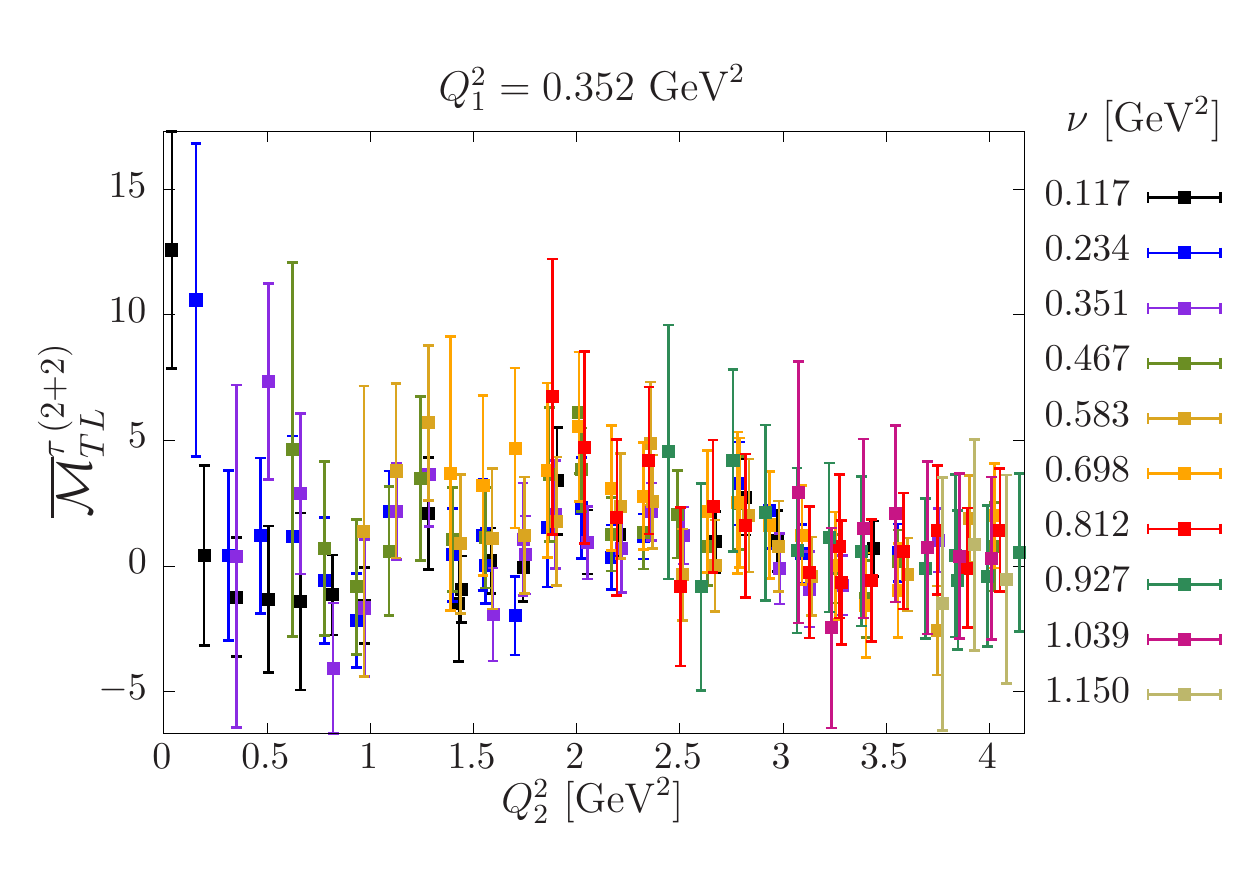}
	\end{minipage}
	\begin{minipage}[c]{0.49\linewidth}
	\centering 
	\includegraphics*[width=0.99\linewidth]{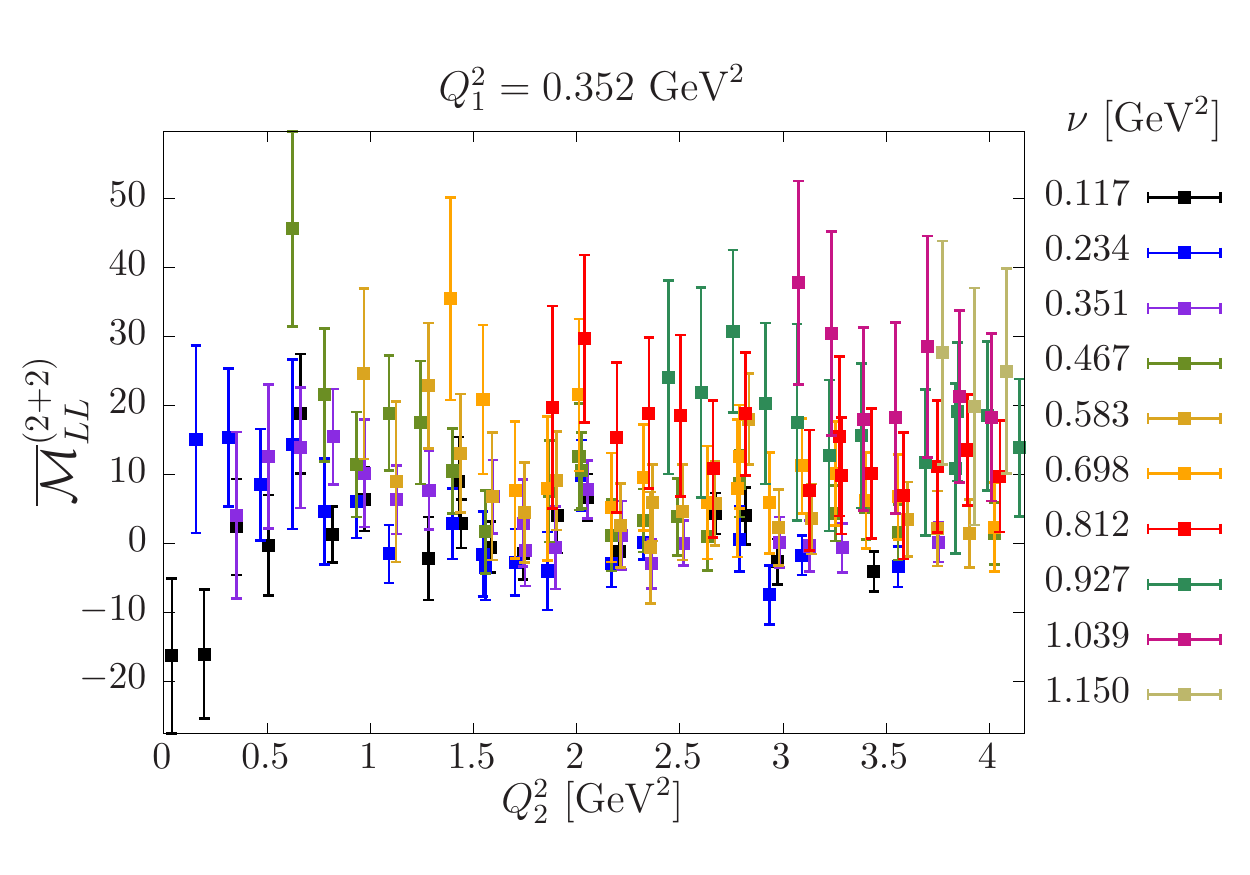}
	\end{minipage}

    \caption{Contribution $(\times10^6)$ from $(2+2)$-disconnected diagrams to the
    eight subtracted forward scattering amplitudes on ensemble F6 with one fixed virtuality $Q_1^2=0.352\,\GeV^2$.}
  \label{fig:22_F6}
\end{figure}

\clearpage}

%%%%%%%%%%%%%%%%%%%%%%%%%%%%%%%%%%%%%%%
\section{Empirical parametrization of the hadronic $\gamma^*\gamma^*$-fusion cross section}
\label{sec:model}
%%%%%%%%%%%%%%%%%%%%%%%%%%%%%%%%%%%%%%%

%-----------------------------------------------------------------------
\subsection{Model description and particle content}
%-----------------------------------------------------------------------

\begin{table}[t] 
\begin{center} 
\caption{Particle multiplets and physical values for the mass and two-photon width as quoted by the PDG~\cite{Olive:2016xmw}, 
as well as by~\cite{Dai:2014zta} for the two-photon width of the $f_2(1270)$ meson and~\cite{Larin:2010kq} for the $\pi^0$ width. 
In the case of the axial-vector mesons, the indicated width is the effective width defined in Eq.\ (\ref{eq:GammaGGt}) 
and obtained phenomenologically in~\cite{Danilkin:2016hnh}.
A cross indicates an absent or imprecise value in the PDG. An asterisk means that we use the isoscalar result 
divided by a factor 25/9 as explained in section~\ref{sec:34ov9}. \label{tab:particles} }
    \vskip 0.1in
\renewcommand{\arraystretch}{1.3}
\begin{tabular}{l@{\hskip 0.2in}lcc@{\hskip 0.2in}lcc@{\hskip 0.2in}lcc}
\hline 
		& 	\multicolumn{3}{c}{Isovector}			&	\multicolumn{3}{c}{Isoscalar}	&	\multicolumn{3}{c}{Isoscalar}		 \\ 
\cline{2-4}	\cline{5-7} \cline{8-10}
		& name			& $m$~[MeV]		& 	$\GammaGG$~[keV] 			& name	& $m$~[MeV]	& $\GammaGG$~[keV] & name	& $m$~[MeV]	& $\GammaGG$~[keV] \\
\hline 
 $0^{-+}$	&	$\pi$	&	134.98	&	$0.0078(2)$	& 	$\eta^{\prime}$		&	$957.78(6)$	&	$4.35(25)$		&	$\eta$	&	$547.86(2)$	&	$0.515(18)$\\  
\hline 
 $0^{++}$	&	$a_0(980)$	&	$980(20)$	&	$0.30(10)$	& 	$f_0(600)$ &	$\times$  	& 	$\times$  	&	$f_0(980)$ &	$990(20)$	& $0.31(5)$		\\  
\hline 
 $1^{++}$	&	$a_1(1260)$	& $1230(40)$ 		&	$1.26^{*}$  						&	$f_1(1285)$	& $1281.8(0.6)$  &	$3.5(0.8)$	&	$f_1(1420)$	&	$1426.4(0.9)$	&	$3.2(0.9)$   \\
\hline 
 $2^{++}$	&	$a_2(1320)$	& $1318.3(0.6)$	&	$1.00(6)$				& 	$f_2(1270)$	& $1275.5(0.8)$	&	$2.93 (40)$	&	$f_2^{\prime}(1525) $ 	&	$1525(5)$	&	$0.081(9)$ \\  
\hline
\end{tabular} 
\end{center}
\end{table} 

In this section, we describe how we model the hadronic
$\gamma^*\gamma^*$-fusion cross section.  We represent it as a sum of
contributions from charge-conjugation even mesonic resonances produced
in the $s$-channel. Specifically, we include 
the pseudoscalar ($J^{\rm PC} = 0^{-+}$), scalar ($J^{\rm PC} = 0^{++}$), 
axial-vector ($J^{\rm PC} = 1^{++}$) and tensor ($J^{\rm PC} = 2^{++}$) mesons. 
Table~\ref{tab:particles} lists the most relevant light mesons with these quantum numbers.
In our implementation, we limit ourselves to the lightest state in each symmetry channel.
The assumption that those states are sufficient to saturate the sum rules is
motivated by the fact that, at small energies, higher mass
singularities are suppressed in Eq.~(\ref{eq:sumrules}). 
Moreover, we have revised the model used in~\cite{Green:2015sra}
to better account for the fact that we perform fits to the fully-connected diagrams. 
Rather than including isovector and isoscalar
mesons, we consider only isovector mesons, enhanced by a factor $34/9$:
we refer the reader to section \ref{sec:34ov9} for a justification of this approximation, 
which we expect to be superior. The procedure mostly 
modifies the contribution of the pseudoscalar sector, due to the large mass difference between 
the pion and the $\eta'$ meson.
Also, since lattice simulations are performed using $N_f=2$ dynamical quarks, 
we do not include the $\eta$ meson.
Finally, we include the Born approximation to the $\gamma^*\gamma^* \to
\pi^+\pi^{-}$ cross section using scalar QED, as described in
Ref.~\cite{Pascalutsa:2012pr},  using a monopole vector form factor,
the monopole mass being set to the $\rho$ meson mass.  Explicit formulae for cross
sections used in our model are given in Appendix~\ref{app:CS}. The
individual contributions  to the eight amplitudes from each channel are
summarized in Table~\ref{tab:rel_contrib}.

\begin{table}[t] 
\begin{center} 
\caption{List of individual contributions to each of the eight helicity amplitudes. 
A cross indicates the absence of a contribution in the given channel.
The relevant cross sections for each channel are given in Appendix~\ref{app:CS}. } \label{tab:rel_contrib}
    \vskip 0.1in
\renewcommand{\arraystretch}{1.3}
\begin{tabular}{l@{\hskip 0.3in}c@{\hskip 0.3in}c@{\hskip 0.3in}c@{\hskip 0.3in}c@{\hskip 0.3in}c@{\hskip 0.3in}c@{\hskip 0.3in}c@{\hskip 0.3in}c}
\hline 
			& $\MTT$	&	$\MTTt$	&	$\MTTa$	&	$\MTL$	&	$\MLT$	&	$\MTLt$	&	$\MTLa$ 	&	$\MLL$\\
\hline 
Pseudoscalar	&	$\sigma_{0}/2$	&	$-\sigma_{0}$	&	$\sigma_{0}/2$	&	$\times$		&	$\times$		&	$\times$		&	$\times$		&	$\times$ \\
Scalar		&	$\sigma_{0}/2$	&	$\sigma_{0}$	&	$\sigma_{0}/2$	&	$\times$		&	$\times$		&	$\tau_{TL}$	&	$\tau_{TL}$	&	$\sigma_{LL}$\\
Axial			&	$\sigma_{0}/2$	&	$-\sigma_{0}$	&	$\sigma_{0}/2 $	&	$\sigma_{TL}$	&	$\sigma_{LT}$	&	$\tau_{TL}$	&	$-\tau_{TL}$ 	&	$\times$\\
Tensor		& 	$\frac{\sigma_{0}+\sigma_{2}}{2}$	& 	$\sigma_0$	& 	$\frac{\sigma_{0}-\sigma_{2}}{2}$	& 	$\sigma_{TL}$	& 	$\sigma_{LT}$	& 	$\tau_{TL}$	& 	$\tau_{TL}^a$	& 	$\sigma_{LL}$	\\
Scalar QED	&	 $\sigma_{TT}$	&	 $\tau_{TT}$	&	$\tau_{TT}^a$	&	 $ \sigma_{TL}$		&	 $\sigma_{LT}$	&	 $ \tau_{TL}$	&	 $ \tau_{TL}^a$	&	$\sigma_{LL}$	\\
\hline
\end{tabular} 
\end{center}
\end{table} 

%-----------------------------------------------------------------------
\subsection{Assumptions on masses and resonances}
%-----------------------------------------------------------------------

Our lattice simulations are performed at larger-than-physical quark
masses. For each ensemble, the pion and $\rho$ meson masses are
determined from the pseudoscalar and vector two-point correlation
functions respectively; see Table~\ref{tabsim} for the obtained
values. To obtain an estimate of the lowest-lying meson mass $m_X$ in
every other symmetry channel, we assume that $m_X$ admits a constant
additive shift relative to its physical value $m_X^{\phys}$.  The
shift $\delta m$ is determined from the difference between the $\rho$
mass computed on the lattice and its experimental value,
\begin{equation}\label{eq:deltam}
m_X = m_X^{\phys} + \delta m, \qquad \delta m = m_{\rho}^{\lat} - m_{\rho}^{\phys} \,. 
\end{equation}
In section \ref{sec:fit}, we will test the sensitivity of our results to variations of $\delta m $ by a factor of two.
As for resonances, we assume that their contributions are well approximated by Breit-Wigner distributions
and use the following formal substitution in the cross sections given in Appendix~\ref{app:CS},
\begin{equation}
\delta(s-m_X^2) \ \leftrightarrow \ \frac{m_X}{\pi} \frac{\Gamma_X}{ (s-m_X^2)^2 + m_X^2 \Gamma_X^2} \,,
\end{equation}
where $m_X$ and $\Gamma_X$  are the mass and the total width of the particle respectively. However, the remaining part of the cross section is still evaluated at $s=m_X^2$. For the (very narrow) pseudoscalar mesons, one can perform the integration explicitly and obtain the following contribution to the sum rules (using $\delta(\nu-\nu_P) = 2 \delta(s-s_P)$, where $\nu_P = \frac{1}{2}(m_P^2+Q_1^2+Q_2^2)$)~:
\begin{align}
\frac{4\nu^2}{\pi} \int_{\nu_0}^\infty \!\! d \nu^\prime \, \frac{ \sqrt{X^\prime}  \sigma_{0}(\nu^\prime) }{ \nu^\prime ( \nu^{\prime \, 2} - \nu^2-i\epsilon )} &= 64 \pi \ \frac{ \GammaGG }{ m_P } \ \frac{ \nu^2 X_P }{ m_P^2\, \nu_P (\nu_P^2 - \nu^2) }  \left[ \frac{  F_{{\cal P} \gamma^\ast \gamma^\ast}(Q_1^2, Q_2^2) }{ F_{{\cal P} \gamma^\ast \gamma^\ast}(0, 0)  } \right]^2 \\
& = 16 \pi^2 \alpha^2 \ \frac{ \nu^2 X_P }{\nu_P (\nu_P^2 - \nu^2) }  \left[   F_{{\cal P} \gamma^\ast \gamma^\ast}(Q_1^2, Q_2^2)  \right]^2 \,,
\end{align}
in the even case, and
\begin{align}
\frac{4\nu^3}{\pi} \int_{\nu_0}^\infty \!\! d \nu^\prime \, \frac{ \sqrt{X^\prime}  \sigma_{0}(\nu^\prime) }{ \nu^{\prime 2} ( \nu^{\prime \, 2} - \nu^2-i\epsilon )} &
=  16 \pi^2 \alpha^2 \ \frac{ \nu^3 X_P}{ \nu_P^2 (\nu_P^2 - \nu^2) }  \left[   F_{{\cal P} \gamma^\ast \gamma^\ast}(Q_1^2, Q_2^2)  \right]^2 \,,
\end{align}
in the odd case, where $X_P \equiv \nu_P^2 - Q_1^2 Q_2^2 $.

%--------------------------------------------------------------
\subsection{Parametrization of the form factors}
%--------------------------------------------------------------

In this subsection, we briefly review the available information on the transition form factors
of the exchanged mesons in the hadronic model, and present the parametrization we use in fitting the lattice HLbL amplitudes.
While detailed information is available in the case of the pion from lattice QCD, no experimental data is presently available at doubly virtual 
kinematics in any channel. In these cases, a monopole or dipole ansatz, in which the $Q_1^2$ and $Q_2^2$ dependence factorizes,
is made to describe the photon-virtuality dependence, even though such an ansatz might not have the asymptotic
behavior predicted by the operator-product expansion.
Our motivation is that this type of parametrization is used in model calculations of $a_\mu^{\rm HLbL}$.
Also, given our goal of performing fits to  the HLbL amplitudes computed on the lattice, 
the number of free parameters characterizing the transition form factors should be 
commensurate with the precision of the lattice data.

\subsubsection{Pseudoscalar mesons}

For pseudoscalar mesons, experimental data are available when at
least one photon is on-shell, and in this case a good parametrization
of the data is obtained using a monopole form factor
\cite{Behrend:1990sr,Gronberg:1997fj,Aubert:2009mc,Uehara:2012ag}. However,
as shown in Ref.~\cite{Gerardin:2016cqj}, a monopole form factor
failed to reproduce the lattice data in the doubly-virtual case,
in contrast to the LMD+V model. Furthermore, the LMD+V model is
compatible with the Brodsky-Lepage behavior~\cite{Lepage:1979zb,Lepage:1980fj,Brodsky:1981rp} in the singly-virtual
case and with the operator-product expansion (OPE) prediction~\cite{Nesterenko:1982dn,Novikov:1983jt} in the 
$Q_1^2=Q_2^2$ doubly-virtual case. We
therefore use this model for the pion transition form factor, of which the 
parameters were determined in Ref.~\cite{Gerardin:2016cqj} for each ensemble listed in Table \ref{tabsim}.  

\subsubsection{Scalar mesons}
Scalar mesons can be produced by two transverse (T) or two longitudinal
 (L) photons. Correspondingly, the amplitude is parametrized by two form factors,
$F^T_{{\cal S} \gamma^\ast \gamma^\ast}$ and $F^L_{{\cal S}  \gamma^\ast \gamma^\ast}$. 
Only the first one has been measured experimentally: this was done for the $f_0(980)$ meson in the
region $Q^2 < 30 {\rm \,GeV}^2$ by the Belle Collaboration~\cite{Masuda:2015yoh},
and the results are compatible with a monopole form factor with a monopole mass $M_S =0.800(50)~\GeV$. 
Therefore, we assume the form
\begin{equation}
\frac{  F^T_{{\cal S} \gamma^\ast \gamma^\ast}(Q_1^2, Q_2^2) }{ F^T_{{\cal S} \gamma^\ast \gamma^\ast}(0, 0)  } = \frac{1}{( 1 + Q_1^2/M_S^2)(1 + Q_2^2/M_S^2)} \,.
\end{equation}
For simplicity, we also assume that the transverse and longitudinal form factors are equal 
(the longitudinal one is only relevant for the amplitudes $M^{a}_{TL}$, $M^{\tau}_{TL}$ and $M_{LL}$),
\begin{equation}
{ F^L_{{\cal S} \gamma^\ast \gamma^\ast}(Q_1^2, Q_2^2) = - F^T_{{\cal S} \gamma^\ast \gamma^\ast}(Q_1^2, Q_2^2) \,. }
\end{equation}
The normalization is obtained from the experimentally measured two-photon decay width 
$\Gamma_{\gamma\gamma}$ given by (see Table~\ref{tab:particles})
\begin{equation}
\GammaGG = \frac{\pi \alpha^2}{4} m_S \left[ F^T_{{\cal S} \gamma^\ast \gamma^\ast}(0, 0) \right]^2 \,,
\end{equation}
while the monopole mass $M_S$ will be treated as a free parameter.
%
%%%%%%%%%%%%%%%%%%%%%%%%%%%%%%%%%%%%
\subsubsection{Axial mesons}
%%%%%%%%%%%%%%%%%%%%%%%%%%%%%%%%%%%%
For axial mesons, we have two form factors, $F^{(0)}_{{\cal A} \gamma^\ast \gamma^\ast}$ and $F^{(1)}_{{\cal A} \gamma^\ast \gamma^\ast}$, 
corresponding to the two helicity states of the meson. We use the same parametrization as in Ref.~\cite{Pascalutsa:2012pr}, inspired by quark models,
\begin{subequations}
\begin{align}
F^{(0)}_{{\cal A} \gamma^\ast \gamma^\ast}(Q_1^2, Q_2^2) &= m_A^2 A(Q_1^2, Q_2^2) \,, \\
F^{(1)}_{{\cal A} \gamma^\ast \gamma^\ast}(Q_1^2, Q_2^2) &= - \frac{\nu}{X} \left( \nu + Q_2^2\right) \, m_A^2 A(Q_1^2, Q_2^2) \,, \\
F^{(1)}_{{\cal A} \gamma^\ast \gamma^\ast}(Q_2^2, Q_1^2) &= - \frac{\nu}{X} \left(\nu + Q_1^2\right) \, m_A^2 A(Q_1^2, Q_2^2) \,,
\end{align}
\end{subequations}
in which $2\nu = m_A^2 + Q_1^2 + Q_2^2$ with $m_A$ the meson mass,  
\begin{equation}
\frac{  A(Q_1^2, 0) }{ A(0, 0)  }  = \frac{1}{( 1 + Q_1^2/M_A^2)^2} \,,
\label{eq:axialFF}
\end{equation}
and assuming factorization such that $A(Q_1^2, Q_2^2)= A(Q_1^2, 0)A(0,Q_2^2)/A(0, 0)   = A(Q_2^2, Q_1^2)$. In particular, the form factor $F^{(1)}_{{\cal A} \gamma^\ast \gamma^\ast}$ is not symmetric in the photon virtualities $Q_1^2,Q_2^2$. These form factors have been measured by the L3 Collaboration for one real and one virtual photon in the region $Q^2 < 5~\GeV^2$~\cite{Achard:2001uu,Achard:2007hm} for the isoscalar resonance. Using the previous parametrization, the authors obtain the dipole mass $M_A = 1040(78)~\MeV$ for the $f_1(1285)$ meson. We obtain the normalization of the form factors from the 
values given in~\cite{Danilkin:2016hnh} for the effective two-photon width, defined as
\begin{equation}\label{eq:GammaGGt}
\GammaGGt \equiv \lim_{Q_1^2 \to 0} \frac{m_A^2}{Q_1^2} \frac{1}{2} \Gamma(\mathcal{A} \to \gamma_L^* \gamma_T) 
= \frac{\pi \alpha^2}{4} \frac{m_A}{3} \left[ F^{(1)}_{{\cal A} \gamma^\ast \gamma^\ast}(0, 0) \right]^2 \,, 
\end{equation}
and we will consider $M_A$ as a free parameter in our fits. 
\begin{table}[t] 
\caption{Tensor form factor normalizations for the isoscalar meson $f_2(1270)$. For helicities $\Lambda=2$ and $\Lambda=(0,T)$ the normalization is obtained using Eq.~(\ref{eq:tensorGG}) and the measured two-photon decay width. For helicities $\Lambda=1$  and $\Lambda=(0,L)$ the results are extracted from Ref.~\cite{Danilkin:2016hnh}.} \label{tab:tensorFF}
\begin{center} 
\renewcommand{\arraystretch}{1.3}
\begin{tabular}{l@{\hskip 0.2in}c@{\hskip 0.2in}c@{\hskip 0.2in}c@{\hskip 0.2in}c@{\hskip 0.2in}c}
\hline 
  & $\Lambda=2$ & 	$\Lambda=(0,T)$  & $\Lambda=1$  & $\Lambda=(0,L)$\\
\hline 
$F^{(\Lambda)}_{{\cal T} \gamma^\ast \gamma^\ast}(0, 0)$	&	$0.500 \pm 0.034$ 	&	$0.095 \pm 0.011$ &	$0.24 \pm 0.05$ &	$-0.90 \pm 0.30$ 	\\  
\hline
\end{tabular} 
\end{center}
\end{table} 
%
%
%%%%%%%%%%%%%%%%%%%%%%%%%%%%%%%%%%%%
\subsubsection{Tensor mesons}
%%%%%%%%%%%%%%%%%%%%%%%%%%%%%%%%%%%%
We now turn our attention to the tensor mesons. The singly-virtual form factors of the isoscalar resonance $f_2$ for helicities $\Lambda=2,1,(0,T)$ have also been measured experimentally in the region \mbox{$Q^2 < 30~\GeV^2$}  by the Belle Collaboration~\cite{Masuda:2015yoh}, where the data are compatible with a dipole form factor~\cite{Danilkin:2016hnh}. Therefore, we use the following parametrization for all helicities $\Lambda = (0,T), (0,L), 1, 2$,
\begin{equation}
\frac{  F^{(\Lambda)}_{{\cal T} \gamma^\ast \gamma^\ast}(Q_1^2, Q_2^2) }{ F^{(\Lambda)}_{{\cal T} \gamma^\ast \gamma^\ast}(0, 0)  }  =   \frac{1}{( 1 + Q_1^2/M_{T,(\Lambda)}^2)^2(1 + Q_2^2/M_{T,(\Lambda)}^2)^2} \,,
\end{equation}
where we allow for a different dipole mass for each helicity. 
The normalization of the transverse form factors is computed from the experimentally measured 
two-photons widths~\cite{Olive:2016xmw}, $\Gamma_{\gamma\gamma} = \GammaGG^{(0)} + \GammaGG^{(2)}$,
assuming that the ratio of helicity 2 to helicity 0 decays is $r = 91.3~\%$ (see Ref.\ \cite{Dai:2014lza}):
\begin{align}
\GammaGG^{(0)} &= {\pi \alpha^2} \, m_T   \, \frac{2}{15} \,  \left[ F^{(0, T)}_{{\cal T} \gamma^\ast \gamma^\ast}(0,0) \right]^2 \,, \nonumber \\
\GammaGG^{(2)}  &= \frac{\pi \alpha^2}{4}  \, m_T \, \frac{1}{5} \, \left[ F^{(2)}_{{\cal T} \gamma^\ast \gamma^\ast} (0,0) \right]^2 \,.
\label{eq:tensorGG}
\end{align}
 In Ref.~\cite{Danilkin:2016hnh}, the authors obtain the normalization of the two other form factors by saturating two different sum rules involving one real and one virtual photon; their results are summarized in Table~\ref{tab:tensorFF}.\\

Finally, based on large-$N$ arguments reviewed in section \ref{sec:34ov9}, 
we assume the following relationship between the two-photon decay widths of the isoscalar and isovector mesons,
\begin{equation}
\GammaGG(f_X) = \frac{25}{9} \, \GammaGG(a_X) \,.
\end{equation}
In particular, we observe that this approximation works well for the tensor meson,
where the  two-photon decay widths have been measured both for the isovector and isoscalar mesons
 (see Table~\ref{tab:particles}).
%-------------------------------------------------------------------------------------
\section{Fitting the $\gamma^*\gamma^*\to\,$hadrons model  to the lattice HLbL amplitudes}
\label{sec:fit}
%-------------------------------------------------------------------------------------

%-------------------------------------------
\subsection{Preliminary checks}
%-------------------------------------------

In this section, we fit simultaneously the eight forward light-by-light amplitudes using the phenomenological model described in Sec.~\ref{sec:model}. 
We have checked that we can reproduce the results given in Refs.~\cite{Pascalutsa:2012pr,Danilkin:2016hnh} in the limit where only one photon is virtual to the quoted accuracy~\footnote{In the second paper, the authors worked in the narrow width approximation.} (Tables~I and II of \cite{Pascalutsa:2012pr} and Table~III and IV of \cite{Danilkin:2016hnh}). Moreover, fits have been checked using two different routines: the Minuit package from CERN~\cite{Minuit}
 and the GSL library~\cite{GSL}. 

%-------------------------------------------------------------------
\subsection{Fit of the eight helicity amplitudes}
%-------------------------------------------------------------------

It appears that the five subtracted amplitudes $\MsTT$, $\MsTTt$, $\MsTTa$,  $\MsTL$ and  $\MsLT$ are statistically more precise than the three other amplitudes $\MsTLt$, $\MsTLa$ and $\MsLL$. Moreover, these last three amplitudes also depend on the longitudinal scalar form factor and on the tensor form factor with helicity $\Lambda=(0,L)$ which are unknown from experiment and for which we use values from phenomenology (see Table~\ref{tab:tensorFF}).
As shown in the last row of Table~\ref{tab:contrib}, the contribution from scalar QED is always small and therefore we do not try to fit the associated monopole mass which is explicitly set to the rho mass computed on the lattice.
We therefore have six fit parameters, which correspond to the monopole and dipole masses of the scalar ($M_S$), axial ($M_A$) and tensor ($M^{(2)}_{T}, M^{(0,T)}_{T}, M^{(1)}_{T}, M^{(0,L)}_{T}$) mesons. The results are given in Table~\ref{tab:fit}, and the corresponding plots for the ensemble F6  are shown in Figs.~(\ref{fig:amps_F6_part1} and \ref{fig:amps_F6_part2}; additional plots for G8 are shown in appendix \ref{sec:addmat}, \ref{fig:amps_G8_part1}). The quoted error on the fit parameters is only statistical and estimated using the jackknife method. The quoted $\chi^2$ correspond to uncorrelated fits. The $\chi^2$ per degree of freedom
 are slightly above unity, with the exception of the value for ensemble E5. Here we attribute its large value to the fact that the statistical errors
are smallest on E5  and that finite-volume effects could be significant for this ensemble. Given that lattice artifacts and finite-size effects are not taken into account by the $\chi^2$, we consider the obtained description of the data on the other ensembles to be satisfactory.

In Table~\ref{tab:contrib}, we show the relative contribution of each channel to the different amplitudes at $Q_1^2 = 0.352~\GeV^2$, $\nu = 0.467~\GeV^2$ and for two values of $Q_2^2$. The amplitudes $\MsTTa$, $\MsTTt$,  $\MsTLt$ and  $\MsTLa$ involve interference cross sections and are not sign-definite: we observe large cancellations between the different contributions. The latter help to stabilize the fit due to the enhanced sensitivity to the relative size of these contributions. In particular, fitting only the amplitudes $\MsTT$,  $\MsTL$ and  $\MsLT$ leads to unstable fits. 
Figures~\ref{fig:nu_G8_part1} and \ref{fig:nu_G8_part2}, in addition to displaying
 the $\nu$-dependence of the amplitudes for two sets of values of $(Q_1^2,Q_2^2)$, show the contributions of the individual mesons.
The pseudoscalar and tensor mesons give the dominant contribution to the amplitudes $\MsTT$, $\MsTTt$ and $\MsTTa$, which involve two transverse photons.
As stated above, the scalar QED contribution is always small, except for $\MsLL$.
The axial form factor is mainly constrained from $\MsTL$, $\MsLT$ where the axial and tensor mesons make the dominant contribution; this is clearly visible from Figs.~\ref{fig:nu_G8_part1} and \ref{fig:nu_G8_part2}. It also contributes significantly to the  amplitudes $\MsTLa$ and $\MsTLt$, which involve one transverse and one longitudinal photon.
On the other hand, the axial meson does not contribute significantly to the amplitudes $\MsTT$, $\MsTTt$ and $\MsTTa$ involving two transverse photons, especially at low virtualities. This suppression is expected since axial mesons have vanishing contribution when at least one photon is real according to the Landau-Yang theorem~\cite{Landau:1948kw,Yang:1950rg}. 
Finally, the tensor meson contributes significantly to all amplitudes.
\begin{table}[t]
\caption{Results of the simultaneous fit to the eight subtracted amplitudes
 $\MsTT$, $\MsTTt$, $\MsTTa$,  $\MsTL$, $\MsLT$, $\MsTLa$, $\MsTLt$ and $\MsLL$ for the five lattice ensembles. The six mass parameters are given
in units of GeV.}
\vskip 0.1in
\begin{tabular}{l@{\hskip 01em}c@{\hskip 01em}c@{\hskip 01em}c@{\hskip 01em}c@{\hskip 01em}c@{\hskip 01em}c@{\hskip 01em}c}
	\hline
		&	$M_{S}$	&	$M_{A}$	&	$M^{(2)}_{T}$	&	$M^{(0,T)}_{T}$ &	$M^{(1)}_{T}$	& $M^{(0,L)}_{T}$ & $\chi^2/\dof$ \\
	\hline
	E5  	&	1.38(11)	&	1.26(10)	&	1.93(3)	&	2.24(5)	 &	2.36(4)	&	0.60(10)	& 4.22	\\
	F6  	&	1.12(14)	&	1.44(5)	&	1.66(9)	&	2.17(5)	 &	1.85(14)	&	0.89(28)	& 1.15	\\
	F7  	&	1.04(18)	&	1.29(8)	&	1.61(12)	&	2.08(7)	 &	2.03(7)	&	0.57(16)	& 1.19	\\
	G8  	&	1.07(10)	&	1.36(5)	&	1.37(24)	&	2.03(6)	 &	1.63(13)	&	0.73(14)	& 1.13	\\
	N6  	&	0.86(37)	&	1.59(3)	&	1.72(17)	&	2.19(4)	 &	1.72(18)	&	0.51(8)	& 1.35	\\
	\hline
 \end{tabular} 
\label{tab:fit}
\end{table}

\begin{table}[t!]
\caption{Relative contributions in \% of each particle to the different amplitudes for the ensemble F7 at $Q_1^2 = 0.352~\GeV^2$, $\nu = 0.467~\GeV^2$ and for two values of $Q_2^2$. For each $Q_2^2$ value, the normalization is such that the absolute values of the entries in a given column add up to 100.}
\vskip 0.1in
\begin{tabular}{l@{\hskip 02em}c@{\hskip 02em}c@{\hskip 02em}c@{\hskip 02em}c@{\hskip 02em}c@{\hskip 02em}c@{\hskip 02em}c@{\hskip 02em}c@{\hskip 02em}c}
        \hline
        &       $Q_2^2~[\GeV^2]$                &       $\MsTT$ &       $\MsTTt$        &       $\MsTTa$                &       $\MsTL$ &       $\MsLT$ &       $\MsTLa$        &       $\MsTLt$        &       $\MsLL$          \\
        \hline
        \multirow{ 2}{*}{$0^{-+}$}              &       1.0     &       35      &       $-56$   &       68      &       $\times$        &       $\times$        &       $\times$        &       $\times$        &       $\times$ \\
                                                        &       3.0     &       30      &       $-38$   &       61      &       $\times$        &       $\times$        &       $\times$        &       $\times$        &       $\times$ \\
        \hline
        \multirow{ 2}{*}{$0^{++}$}              &       1.0     &       7       &       11      &       8       &       $\times$        &       $\times$        &       23              &       14      &       42      \\
                                                        &       3.0     &       5       &       6       &       8       &       $\times$        &       $\times$        &       19              &       9       &       50      \\
        \hline
        \multirow{ 2}{*}{$1^{++}$}              &       1.0     &       2       &       $-2$    &       1       &       43              &       57              &       $-43$           &       32      &       $\times$\\
                                                        &       3.0     &       8       &       $-11$   &       11      &       21              &       49              &       $-40$           &       23      &       $\times$\\
        \hline
        \multirow{ 2}{*}{$2^{++}$}              &       1.0     &       53      &       25      &       $-20$   &       56              &       42              &       19              &     $-47$     &       25\\
                                                        &       3.0     &       56      &       44      &       19      &       79              &       51              &       $-38$           &       $-67$   &       40\\
        \hline
        \multirow{ 2}{*}{Scalar QED}    &       1.0     &       4       &       5       &       3       &       1               &       $<1$            &       $-15$           &       $-7$    &       33      \\
                                                        &       3.0     &       1       &       1       &       1       &       $<1$            &       $<1$            &       $-3$            &       $-1$    &       10\\
        \hline
 \end{tabular}
\label{tab:contrib}
\end{table}

\afterpage{

\begin{figure}[p]

	\begin{minipage}[c]{0.49\linewidth}
	\centering 
	\includegraphics*[width=0.89\linewidth]{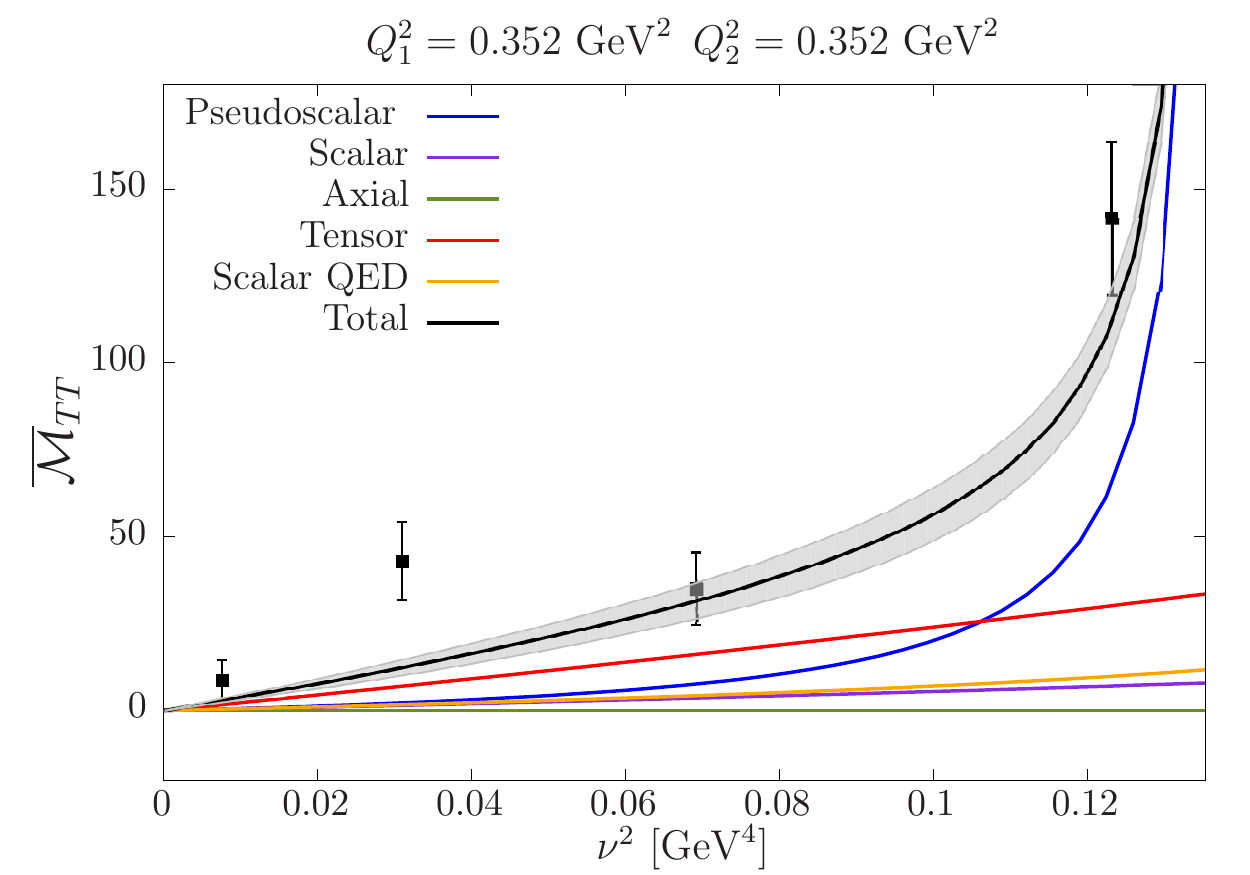}
	\end{minipage}
	\begin{minipage}[c]{0.49\linewidth}
	\centering 
	\includegraphics*[width=0.89\linewidth]{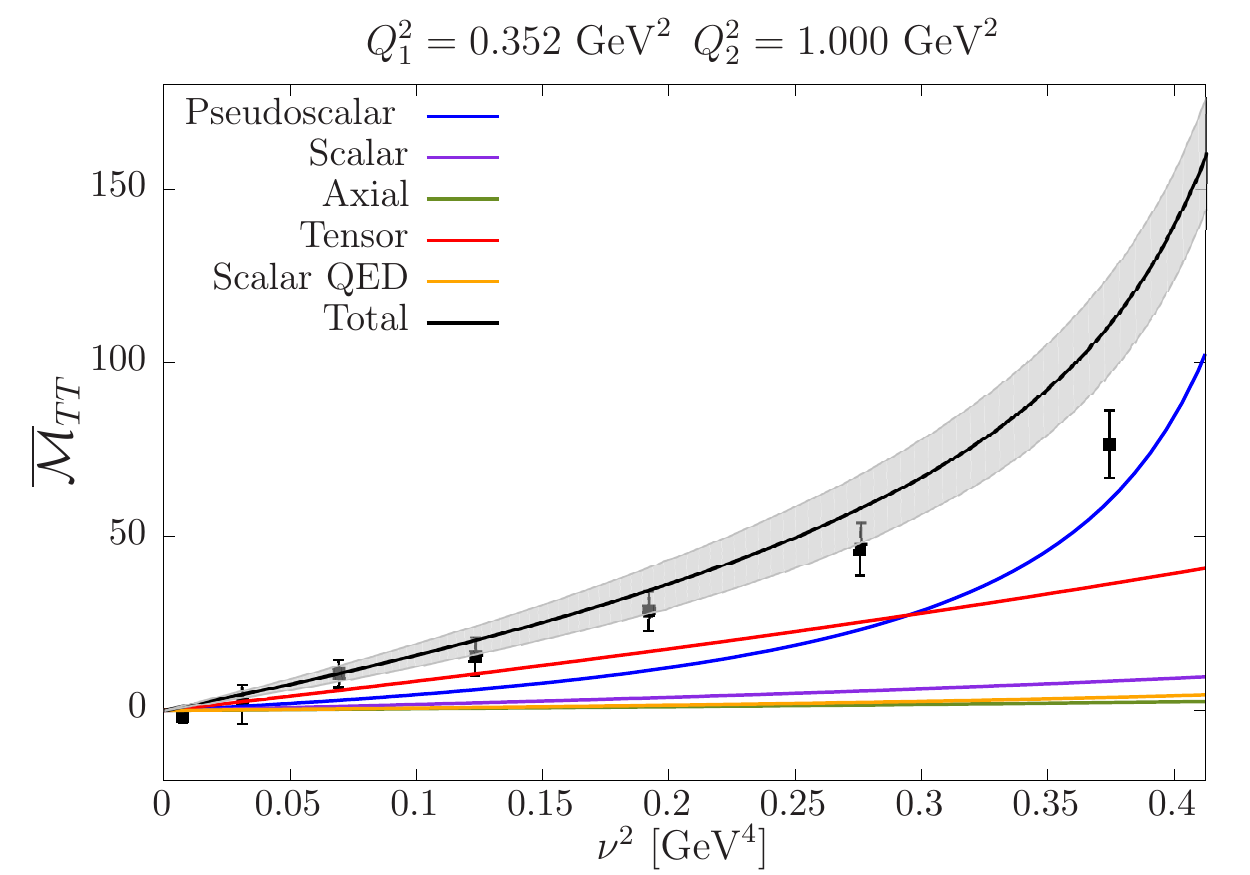}
	\end{minipage}
		
	\vskip 0.15in
	\begin{minipage}[c]{0.49\linewidth}
	\centering 
	\includegraphics*[width=0.89\linewidth]{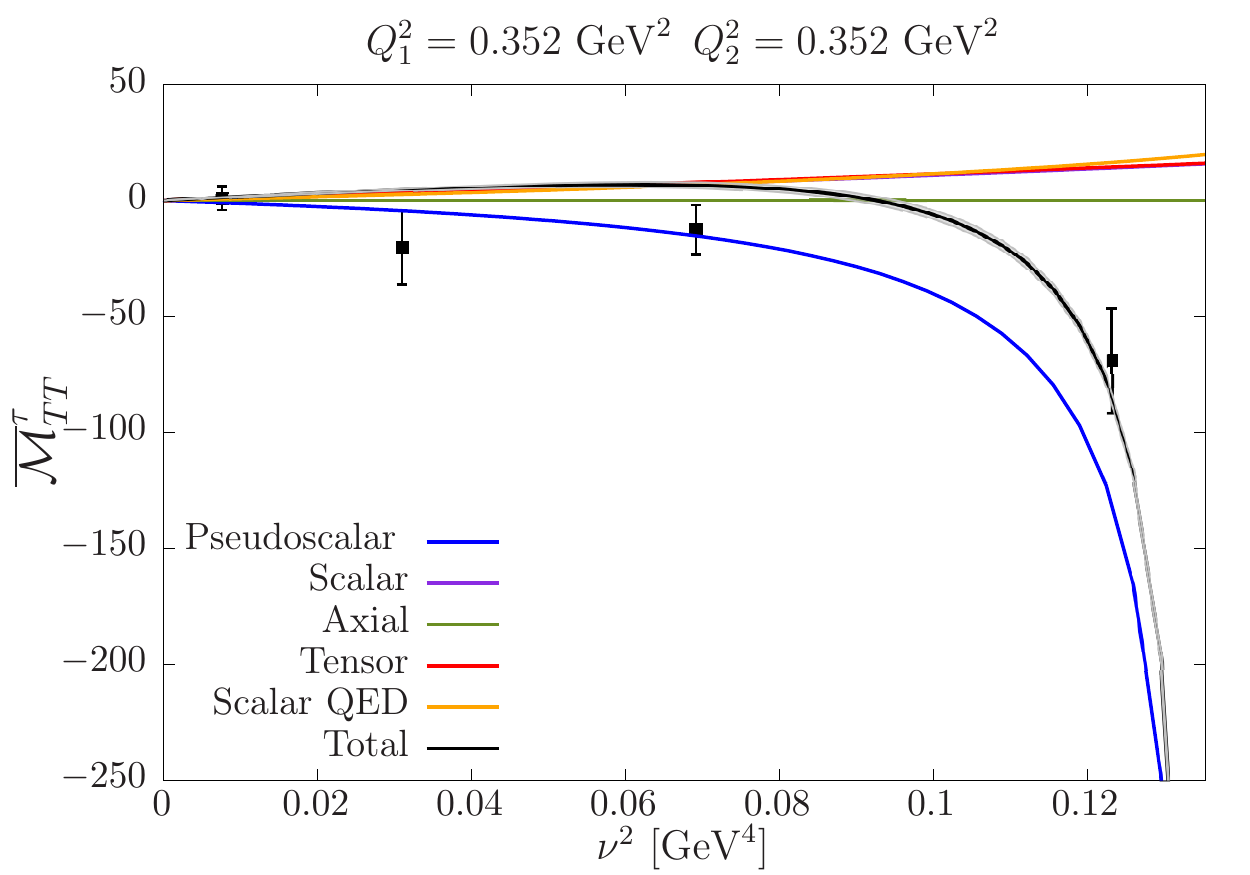}
	\end{minipage}
	\begin{minipage}[c]{0.49\linewidth}
	\centering 
	\includegraphics*[width=0.89\linewidth]{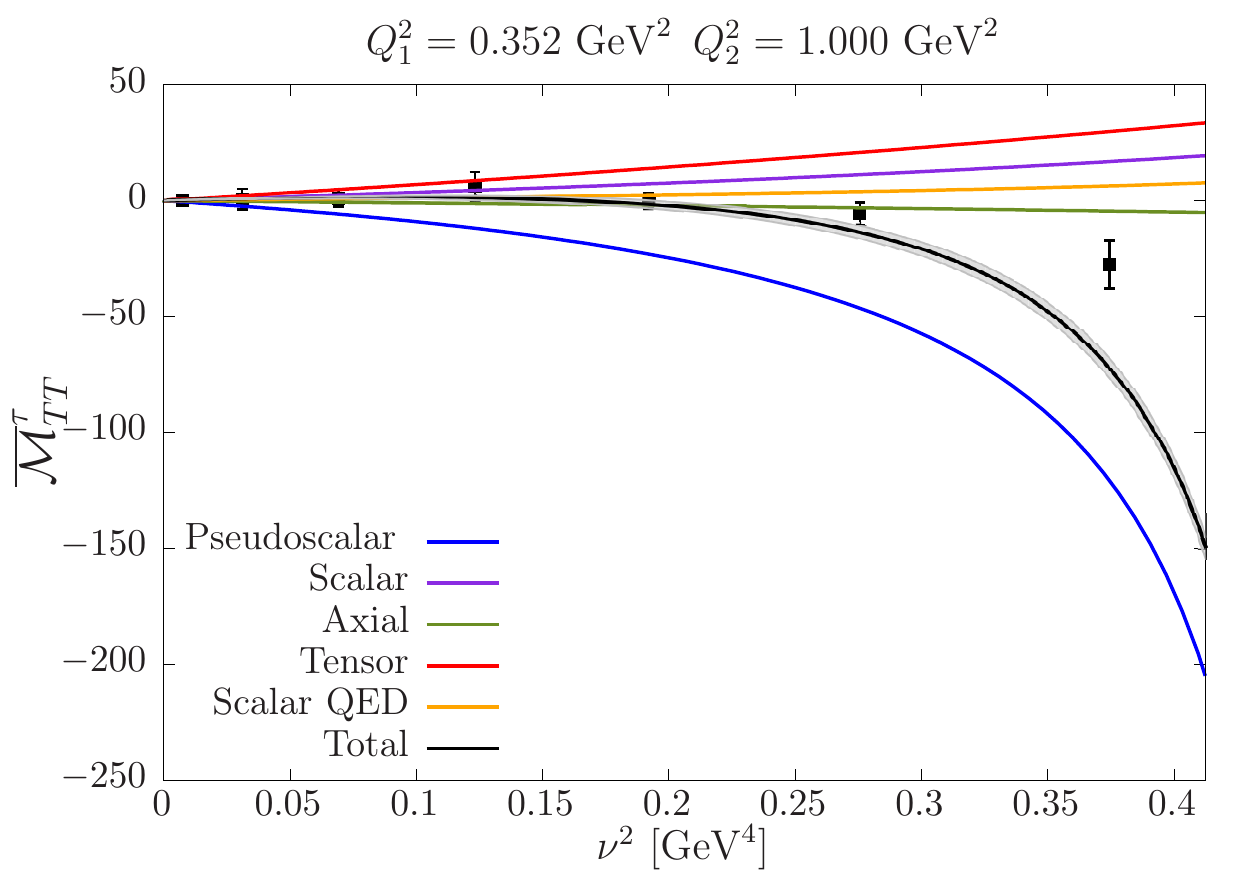}
	\end{minipage}
	
	\vskip 0.15in
	\begin{minipage}[c]{0.49\linewidth}
	\centering 
	\includegraphics*[width=0.89\linewidth]{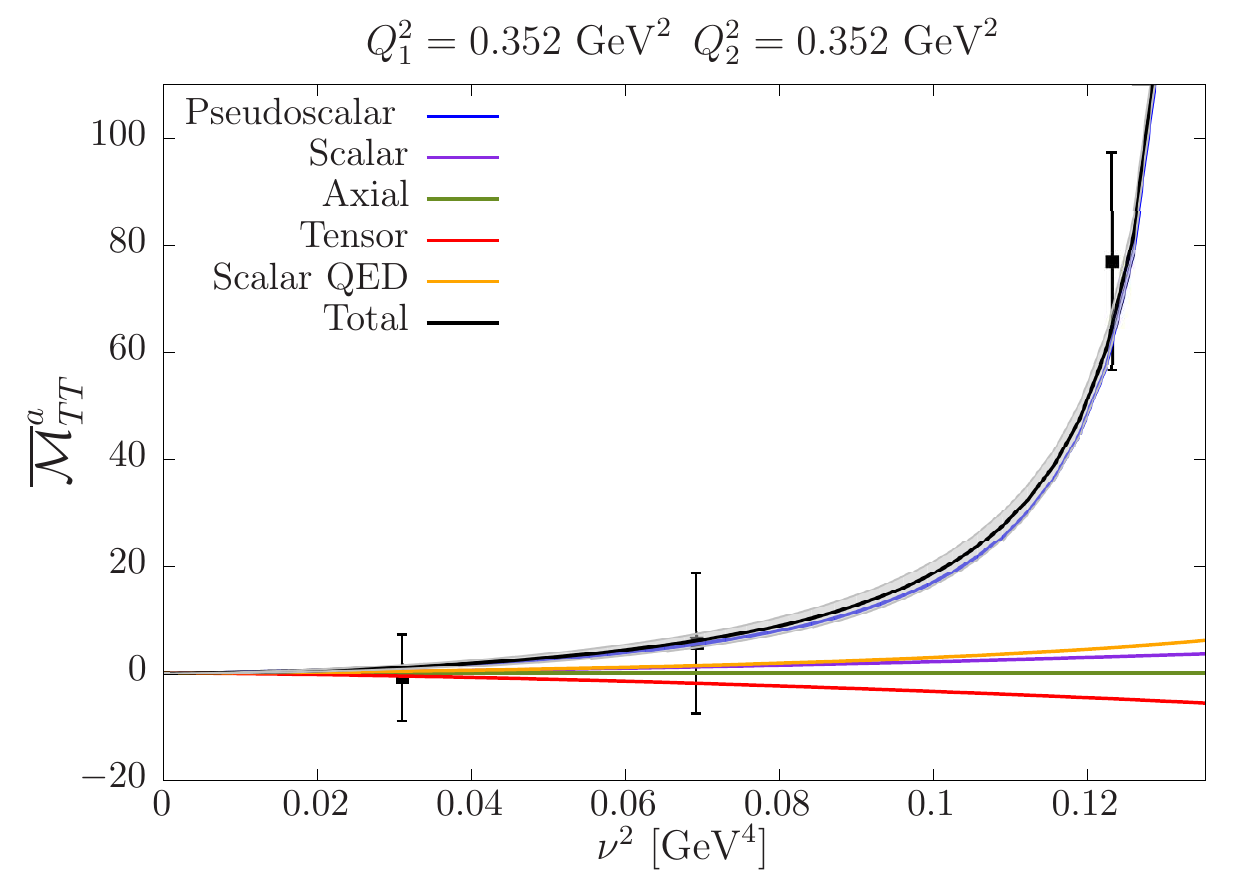}
	\end{minipage}
	\begin{minipage}[c]{0.49\linewidth}
	\centering 
	\includegraphics*[width=0.89\linewidth]{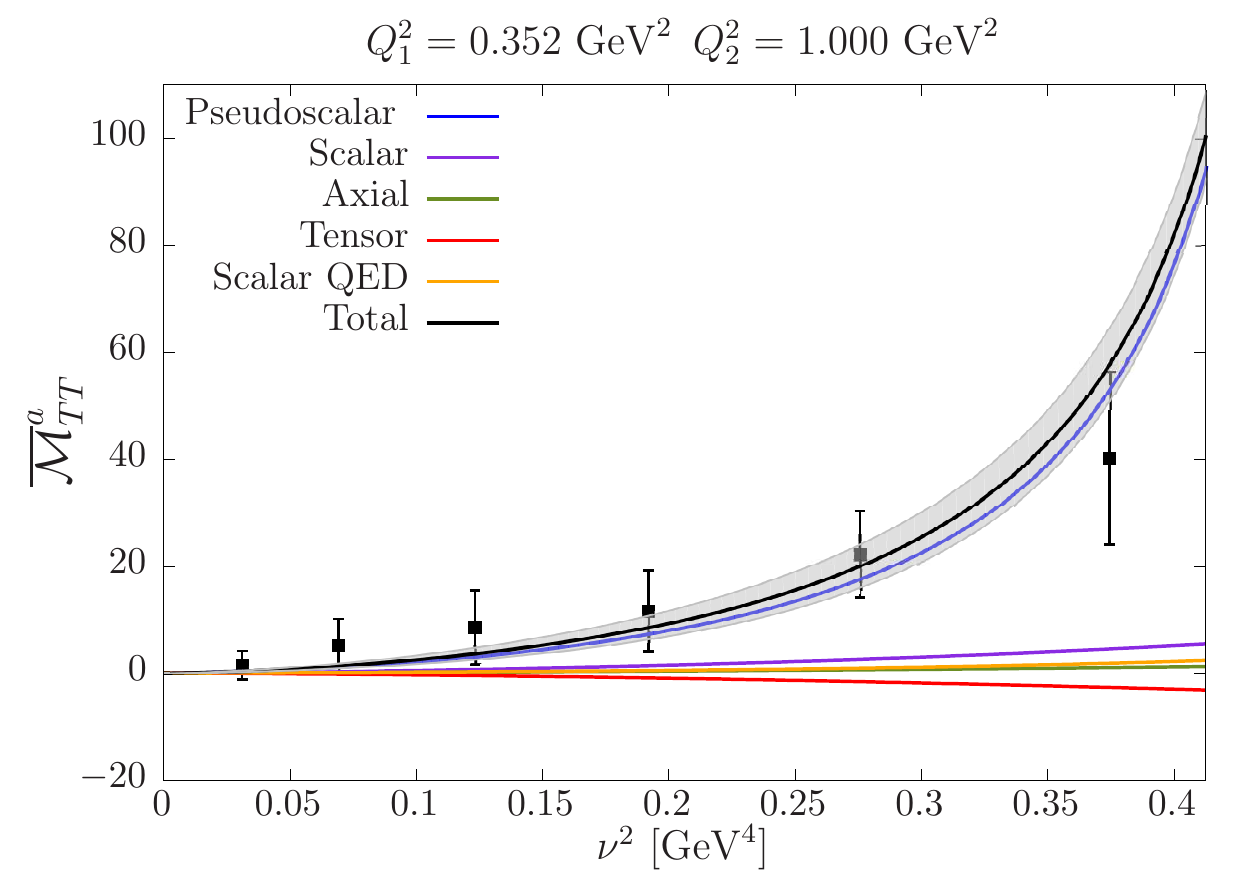}
	\end{minipage}

	\vskip 0.15in
	\begin{minipage}[c]{0.49\linewidth}
	\centering 
	\includegraphics*[width=0.89\linewidth]{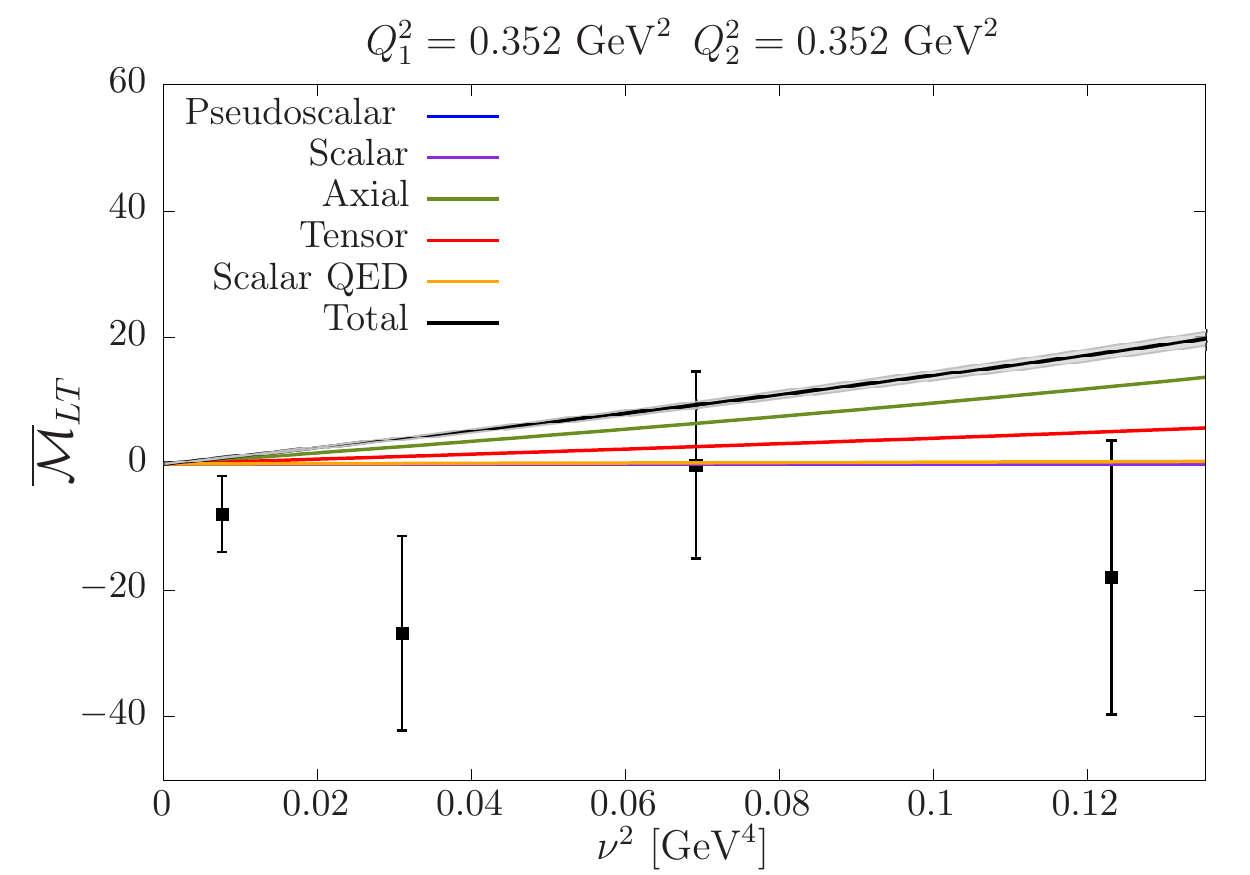}
	\end{minipage}
	\begin{minipage}[c]{0.49\linewidth}
	\centering 
	\includegraphics*[width=0.89\linewidth]{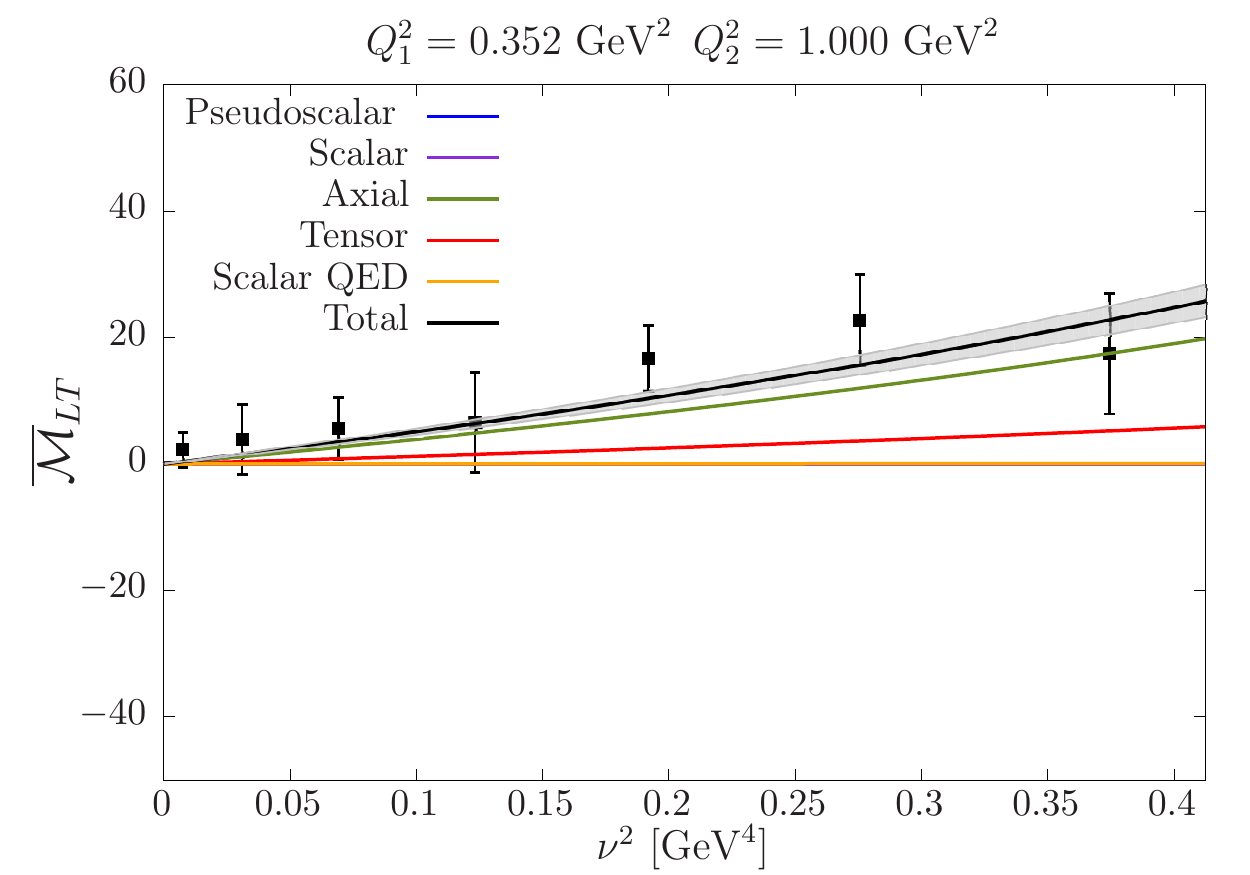}
	\end{minipage}

	\caption{The dependence of the amplitudes $\overline{\cal M}_{TT}$, $\overline{\cal M}_{TT}^\tau$, $\overline{\cal M}_{TT}^a$ and $\overline{\cal M}_{LT}$ $(\times10^6)$ on $\nu$ for two different values of $Q_2^2$, the virtuality $Q_1^2=0.352 \GeV^2$ being fixed. The results correspond to the lattice ensemble G8. Note that at fixed photon virtualities, the form factors are completely determined. The black line corresponds to the total contribution and each colored line represents a single-meson contribution.}

	\label{fig:nu_G8_part1}
	
\end{figure}

\clearpage}

\afterpage{

\begin{figure}[p]
	
	\begin{minipage}[c]{0.49\linewidth}
	\centering 
	\includegraphics*[width=0.89\linewidth]{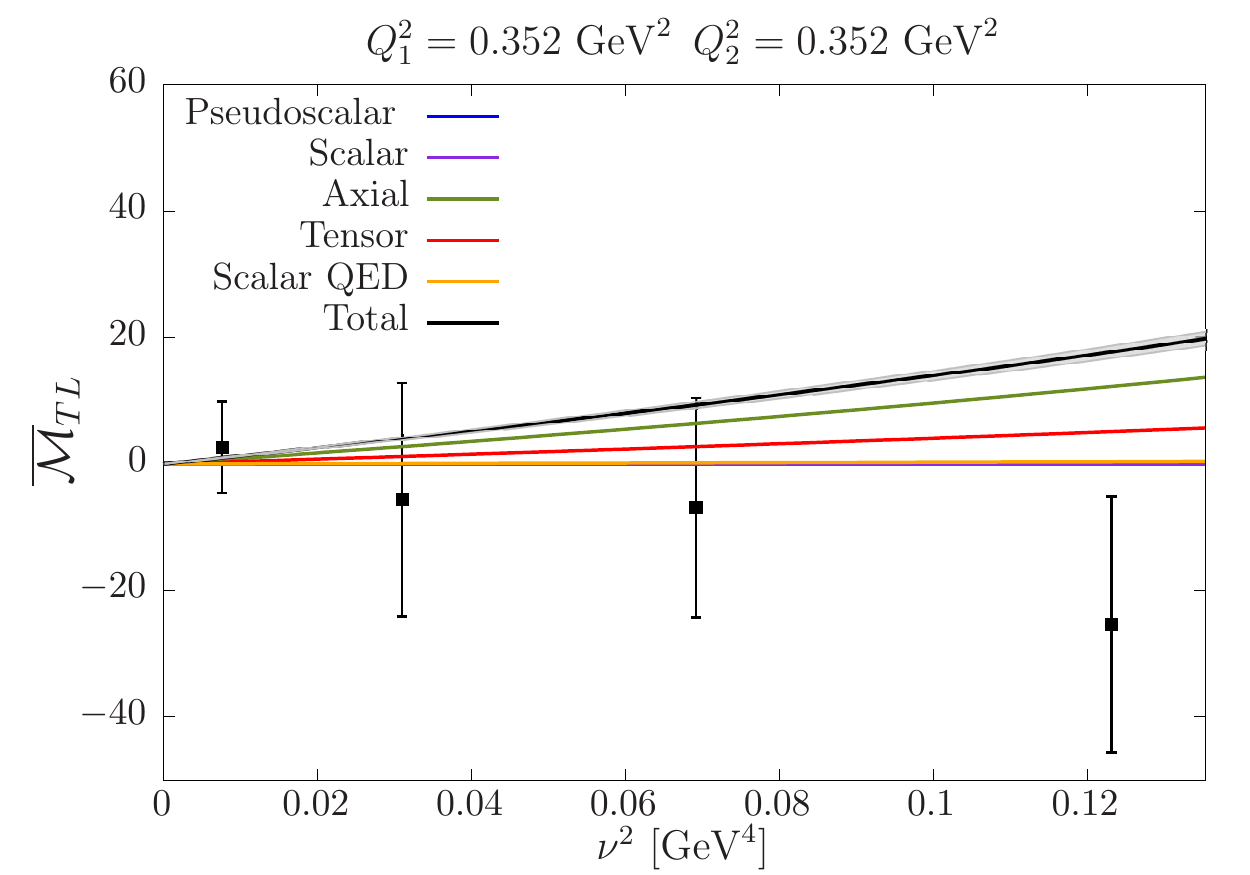}
	\end{minipage}
	\begin{minipage}[c]{0.49\linewidth}
	\centering 
	\includegraphics*[width=0.89\linewidth]{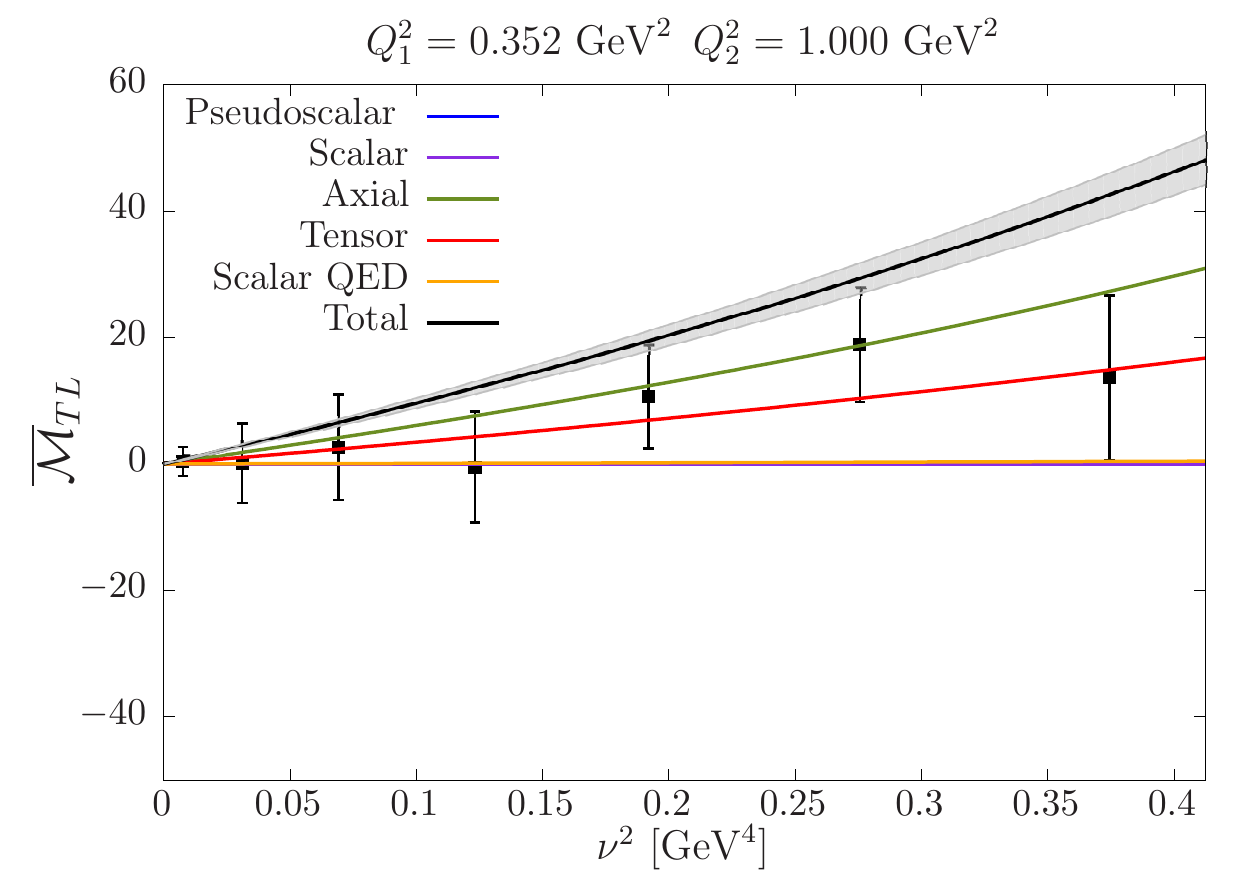}
	\end{minipage}	
	
	\vskip 0.15in
	\begin{minipage}[c]{0.49\linewidth}
	\centering 
	\includegraphics*[width=0.89\linewidth]{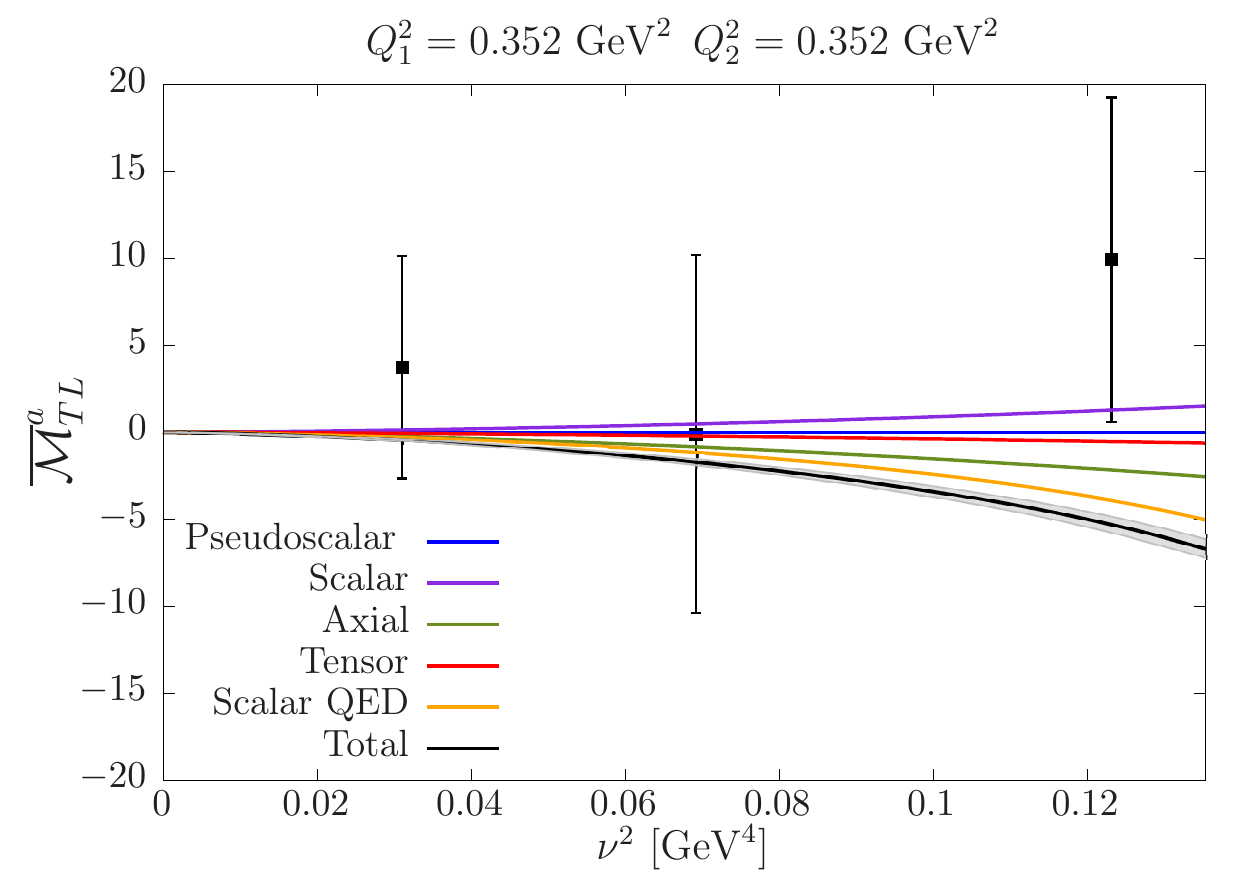}
	\end{minipage}
	\begin{minipage}[c]{0.49\linewidth}
	\centering 
	\includegraphics*[width=0.89\linewidth]{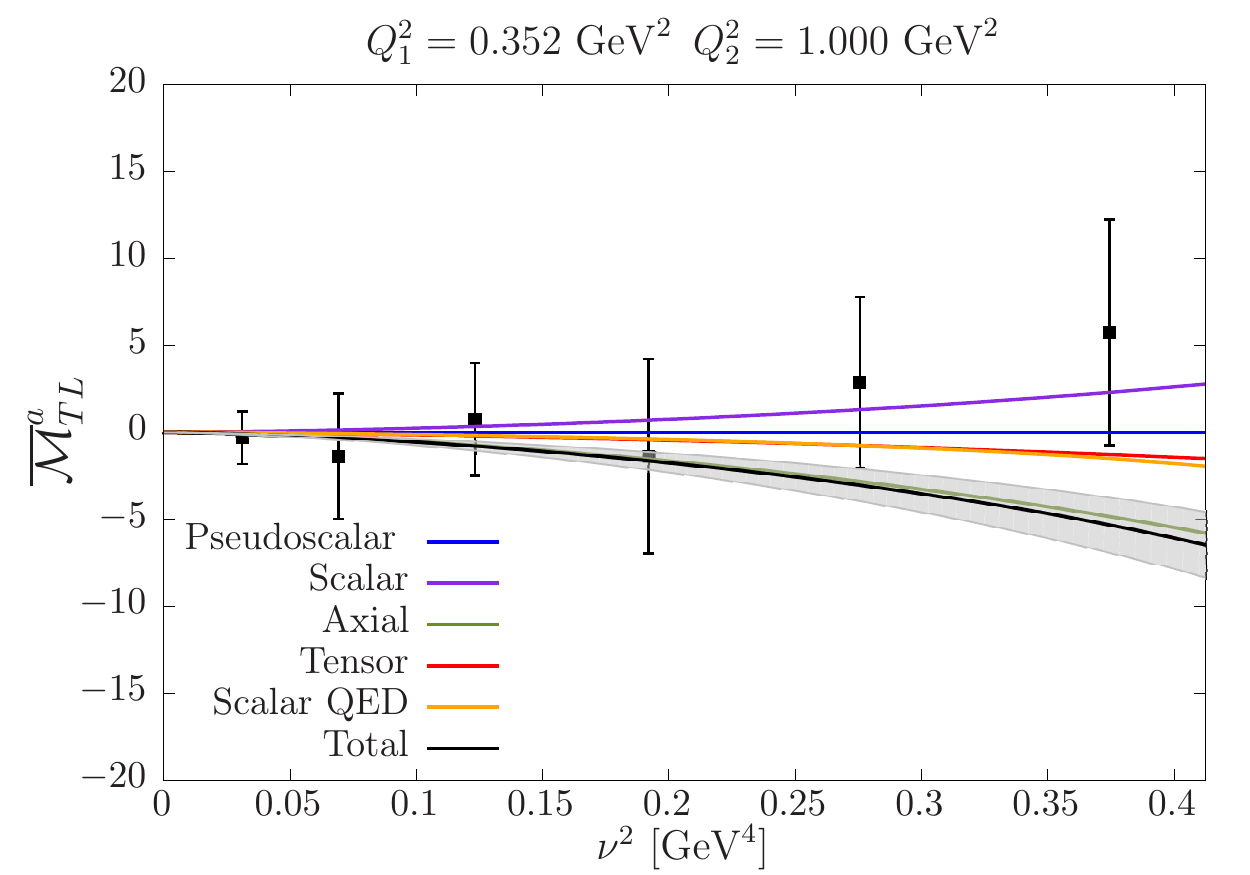}
	\end{minipage}
	
	\vskip 0.15in
	\begin{minipage}[c]{0.49\linewidth}
	\centering 
	\includegraphics*[width=0.89\linewidth]{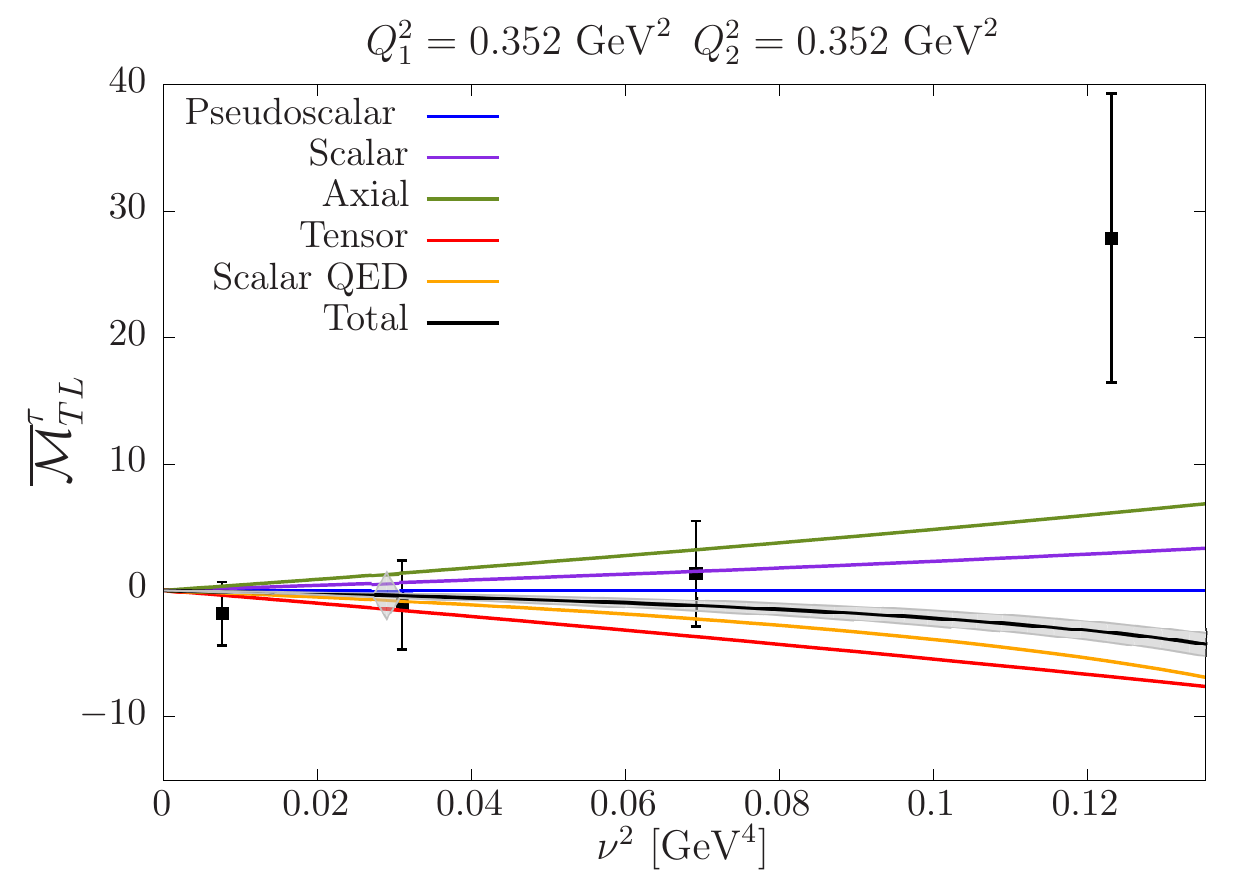}
	\end{minipage}
	\begin{minipage}[c]{0.49\linewidth}
	\centering 
	\includegraphics*[width=0.89\linewidth]{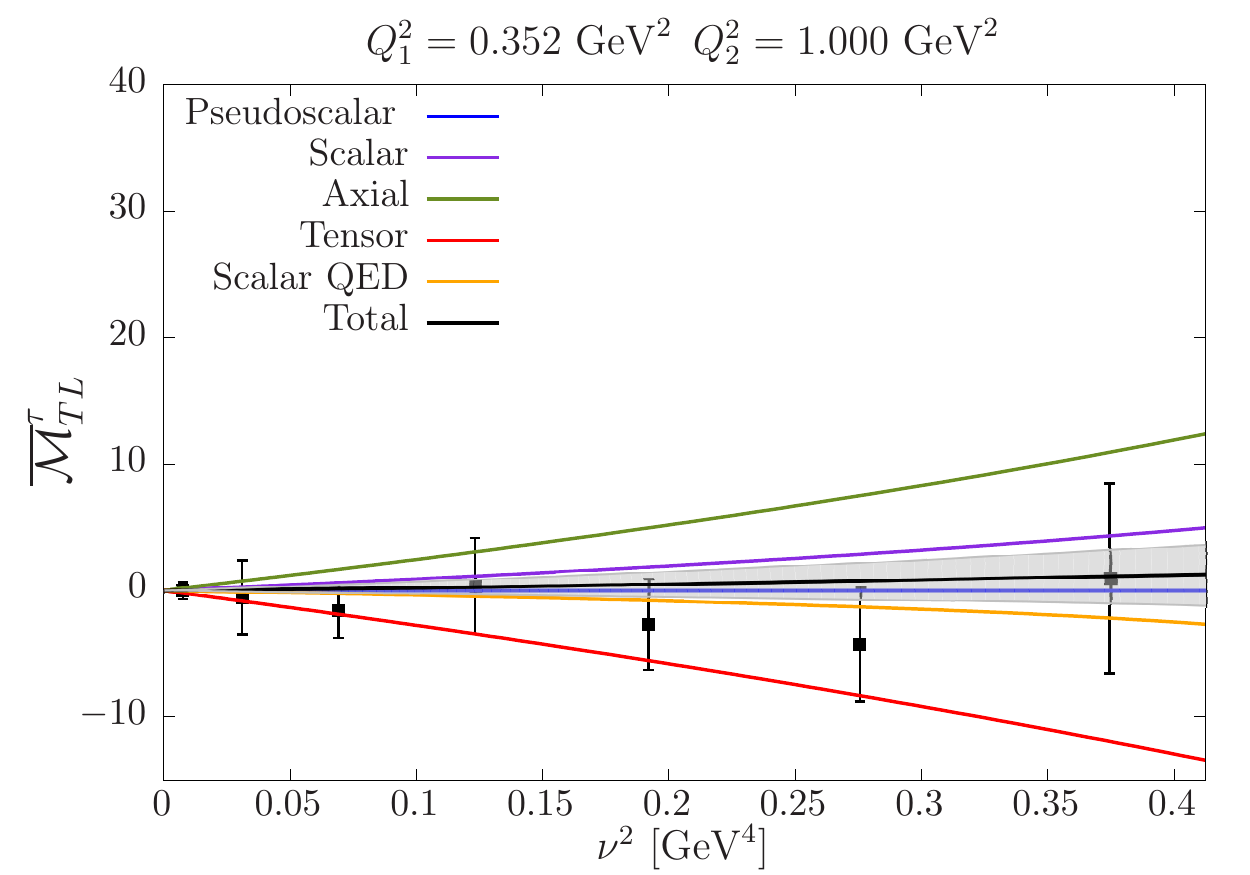}
	\end{minipage}
	
	\vskip 0.15in	
	\begin{minipage}[c]{0.49\linewidth}
	\centering 
	\includegraphics*[width=0.89\linewidth]{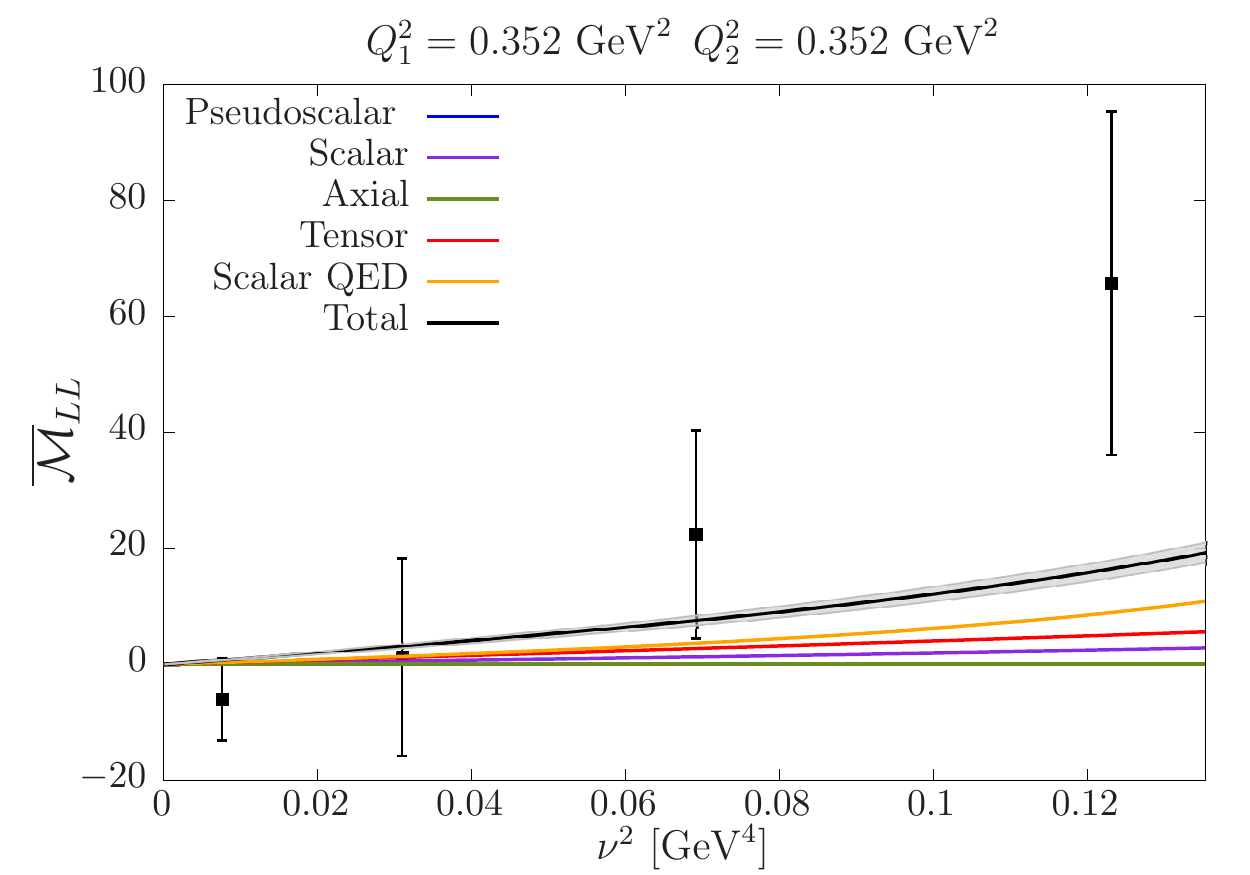}
	\end{minipage}
	\begin{minipage}[c]{0.49\linewidth}
	\centering 
	\includegraphics*[width=0.89\linewidth]{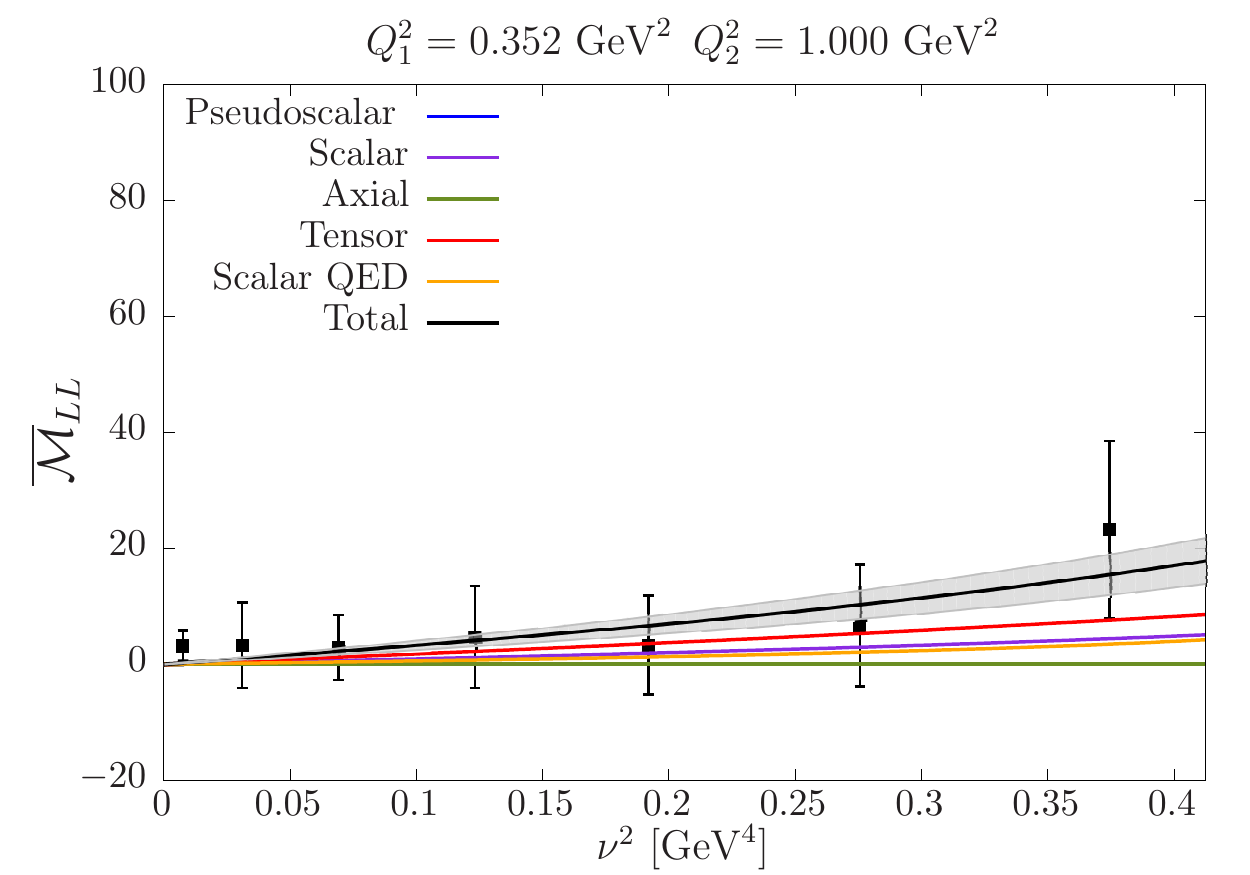}
	\end{minipage}
	
	\caption{The dependence of the amplitudes $\overline{\cal M}_{TL}$, $\overline{\cal M}^a_{TL}$, $\overline{\cal M}^\tau_{TL}$, $\overline{\cal M}_{LL}$ $(\times10^6)$ on $\nu$ for two different values of $Q_2^2$, the virtuality $Q_1^2=0.352 \GeV^2$ being fixed. The results correspond to the lattice ensemble G8. Note that at fixed photon virtualities, the form factors are completely determined. The black line corresponds to the total contribution and each colored line represents a single-meson contribution.}

	\label{fig:nu_G8_part2}
\end{figure}

\clearpage}

%---------------------------------------------------------------
\subsection{Influence of the non-fitted model parameters}
%---------------------------------------------------------------

In the previous fit, only the monopole and dipole masses entering the
form factors were considered as fit parameters. The other parameters
($p=\Gamma, \GammaGG, \delta m, \dots$) were fixed using phenomenology
as described in Sec.~\ref{sec:model}. However, these parameters are
sometimes associated with relatively large experimental errors
($\delta p$) or modelled (like the global mass shift in the spectrum
where we assume $m_X = m_X^{\exp} + \delta m$ with $\delta m =
m_{\rho}^{\rm lat} - m_{\rho}^{\exp}$).  Therefore, we perform exactly
the same fit as in the previous section but using $p \pm \delta p$
instead of $p$ (and varying only one parameter at a time). In this
way, we can see the influence of these parameters on the monopole and
dipole masses obtained in the previous section. The results are
summarized in Table~\ref{tab:fit_syst} for the ensemble F6. In this
table, $\delta m$ corresponds to the global mass shift applied to the
spectrum (see Eq.\ (\ref{eq:deltam})), and is multiplied or divided by
a factor of two.

We observe that the experimental error on the total decay widths of
the particles have a negligible effect. Increasing the two-photon
width (or equivalently, the normalization of the form factor) tends to
reduce the associated monopole or dipole mass.  Finally, increasing
the global mass shift by a factor two leads to a noticeable change in
the monopole and dipole masses with little change in the $\chi^2$.

Varying the normalization of the form factor $F_{{\cal T}\gamma^*\gamma^*}^{(0,L)}$ leads to
negligible changes in all parameters but $M_T^{(0,L)}$; this
particular correlation is studied in more detail in the next
subsection.

\begin{table}[t!]
\caption{Fit variations for F6. The first row corresponds to the results obtained in the previous section. Then, each row corresponds to a new fit using $p \pm \delta p$ and varying only one parameter at a time: the quoted number is the shift observed for the monopole/dipole mass, in units of GeV. A cross indicates that the parameter remains unchanged to all indicated digits of the  central values in the first row. For instance, using $\GammaGG(a_0) + \delta \GammaGG(a_0)$ instead of $\GammaGG(a_0)$, the scalar monopole mass is shifted by $-0.09~\GeV$, the other monopole/dipole masses being unaffected. In the last row, the mass shift $\delta m$ applied to the spectrum (see Eq.\ (\ref{eq:deltam})) is varied by a factor of two.}
\vskip 0.1in
\begin{tabular}{l@{\hskip 01em}c@{\hskip 01em}c@{\hskip 01em}c@{\hskip 01em}c@{\hskip 01em}c@{\hskip 01em}c@{\hskip 01em}c}
	\hline
	&	$M_{S}$	&	$M_{A}$	&	$M^{(2)}_{T}$	&	$M^{(0,T)}_{T}$	&	$M^{(1)}_{T}$	& $M^{(0,L)}_{T}$ & $\chi^2/\dof$ \\
	\hline
	Principal  	&	$1.12(14)$	&	1.44(5)	&	1.66(9)	&	2.17(5)		&	1.85(14)	&	0.91(7)	& 1.15	\\
	\hline
	\multirow{2}{*}{$\Gamma(a_0)$}	&	$\times$	&	$\times$	&	$\times$	&	$\times$	&	$\times$	&	$\times$	&	1.15	\\
								&	$\times$	&	$\times$	&	$\times$	&	$\times$	&	$\times$	&	$\times$	&	1.15	\\
	\hline
	\multirow{2}{*}{$\GammaGG(a_0)$}	&	$-0.09$	&	$\times$	&	$\times$	&	$\times$	&	$\times$	&	$\times$	&	1.14	\\
								&	$+0.12$	&	$\times$	&	$\times$	&	$\times$	&	$\times$	&	$\times$	&	1.15	\\
	\hline
	\multirow{2}{*}{$\Gamma(a_1)$}	&	$-0.01$	&	$+0.03$	&	$\times$	&	$+0.01$	&	$\times$	&	$\times$ &	1.14\\
								&	$-0.01$	&	$-0.02$	&	$+0.01$	&	$\times$	&	$+0.01$	&	$\times$ &	1.15	\\
	\hline
	\multirow{2}{*}{$\tilde{\Gamma}_{\gamma\gamma}(a_1)$}	&	$\times$	&	$-0.10$	&	$+0.02$	&	$-0.01$	&	$+0.02$	&	$\times$ &	1.17	\\
								&	$+0.03$	&	$+0.19$	&	$-0.01$	&	$+0.02$	&	$-0.01$	&	$\times$ &	1.12	\\
	\hline
	\multirow{2}{*}{$\Gamma(a_2)$}	&	$\times$	&	$\times$	&	$\times$	&	$\times$	&	$\times$	&	$\times$	&	1.15	\\
								&	$\times$	&	$\times$	&	$\times$	&	$\times$	&	$\times$	&	$\times$	&	1.15	\\
	\hline
	\multirow{2}{*}{$F_{{\cal T}\gamma^*\gamma^*}^{(2)}$}		&	$-0.01$	&	$\times$	&	$-0.06$	&	$+0.01$	&	$\times$	&	$\times$ &	1.15	\\
								&	$\times$	&	$\times$	&	$+0.08$	&	$\times$	&	$+0.01$	&	$\times$ &	1.14	\\
	\hline
	\multirow{2}{*}{$F_{{\cal T}\gamma^*\gamma^*}^{(0,T)}$}		&	$-0.08$	&	$\times$	&	$-0.01$	&	$-0.09$	&	$-0.01$	&	$\times$&	1.13	\\
								&	$+0.08$	&	$\times$	&	$+0.02$	&	$+0.11$	&	$+0.02$	&	$\times$&	1.17	\\
	\hline
	\multirow{2}{*}{$F_{{\cal T}\gamma^*\gamma^*}^{(1)}$}		&	$\times$	&	$\times$	&	$\times$	&	$\times$	&	$-0.14$	&	$\times$ &	1.14	\\
								&	$-0.01$	&	$\times$	&	$\times$	&	$\times$	&	$+0.21$	&	$\times$ &	1.15	\\
	\hline
	\multirow{2}{*}{$\delta m^{\times 2}_{\times 0.5}$}		&	$+0.20$	&	$+0.13$	&	$+0.11$	&	$+0.13$	&	$+0.14$	&	$+0.10$ &	1.17	\\
								&	$-0.10$	&	$-0.06$	&	$-0.05$	&	$-0.08$	&	$-0.06$	& $-0.05$ 	&	1.15\\
	\hline
 \end{tabular} 
\label{tab:fit_syst}
\end{table}

%-----------------------------------------------------------------------------------
\subsection{Bounds for the tensor form factor $F_{{\cal T}\gamma^*\gamma^*}^{(0,L)}$}
%-----------------------------------------------------------------------------------

\begin{figure}[t]

	\begin{minipage}[c]{0.49\linewidth}
	\centering 
	\includegraphics*[width=\linewidth]{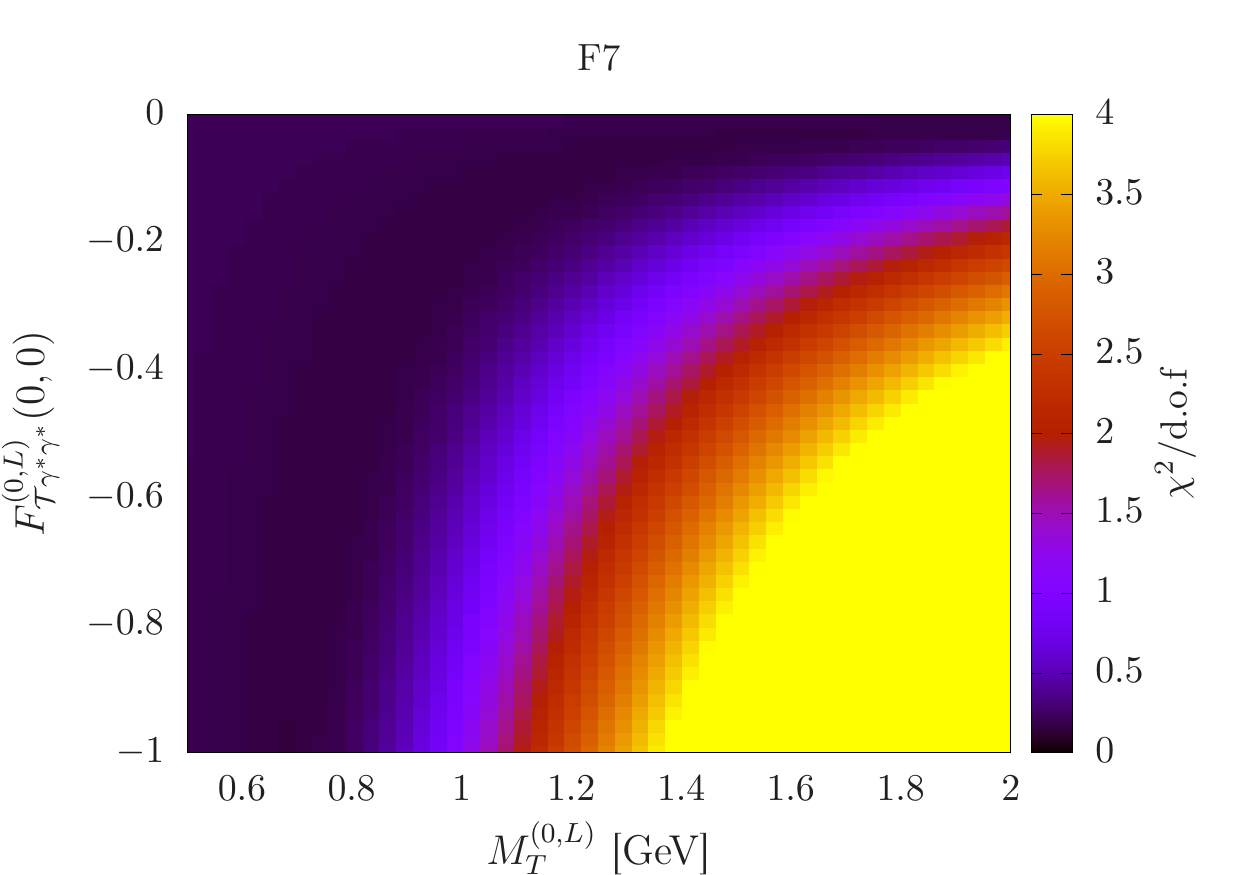}
	\end{minipage}
	\begin{minipage}[c]{0.49\linewidth}
	\centering 
	\includegraphics*[width=\linewidth]{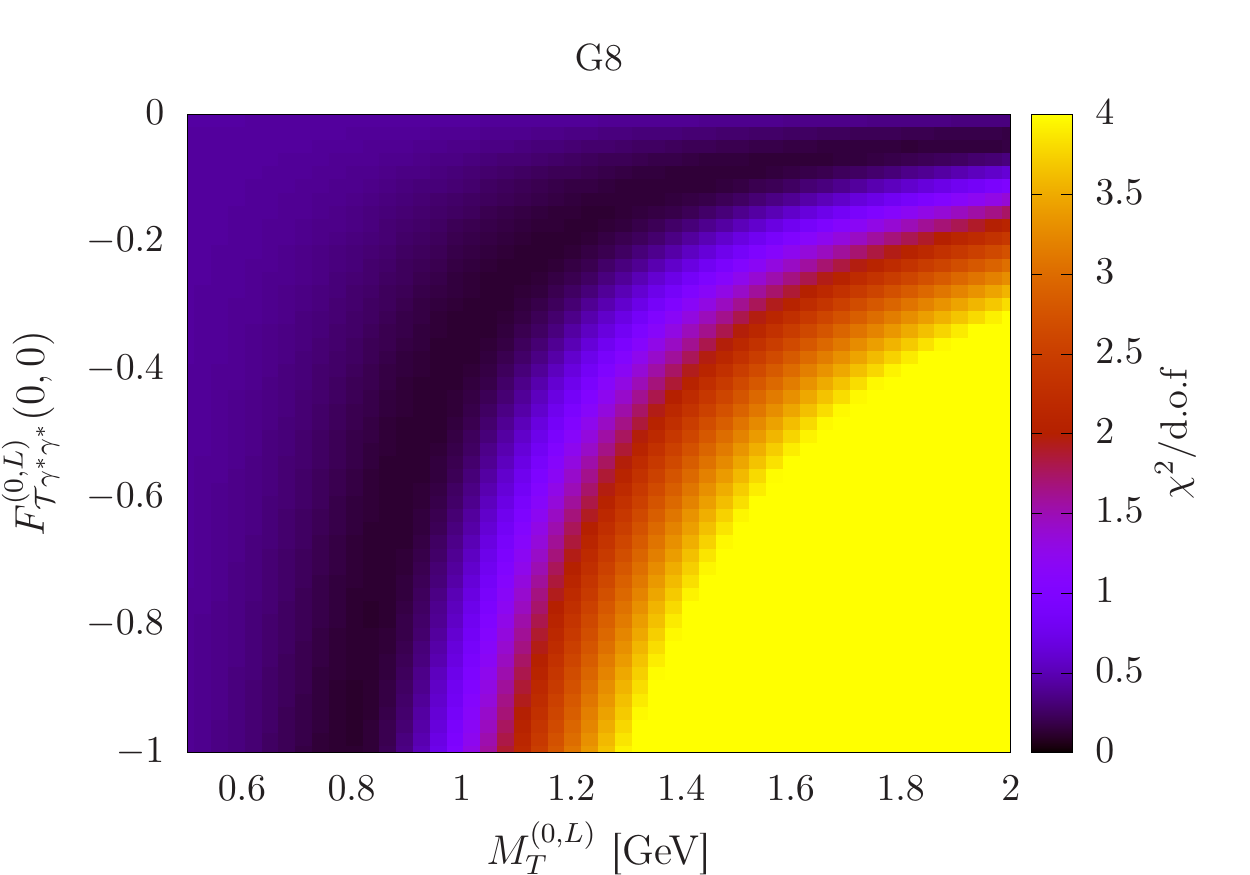}
	\end{minipage}
	
\caption{  Value of $\chi^2/\dof$ for different dipole masses and form factor normalizations 
 (tensor form factor, helicity $\Lambda = (0,L)$). Left: ensemble F7, right: ensemble G8. }	
	\label{fig:tensor_scan}
\end{figure}

The transition form factor $F_{{\cal T}\gamma^*\gamma^*}^{(0,L)}$ of the tensor meson enters only the amplitudes $\MTLt$,
$\MTLa$ and $\MLL$, which are less precisely determined on the
lattice. In particular the fit is not able to determine both the
dipole mass and the normalization independently, and they are highly
correlated. To illustrate this point, we use the previously obtained best fit parameters and
compute the $\chi^2/\dof$ along a scan in the plane ($M_T^{(0,L)}$,
$F_{{\cal T}\gamma^*\gamma^*}^{(0,L)}(0,0)$). The results are shown in
Fig.~\ref{fig:tensor_scan}: for a dipole mass of $1~\GeV$, a
normalization $F_{{\cal T}\gamma^*\gamma^*}^{(0,L)}(0,0) \approx -0.4$ is favored but the
results show a strong dependence on $M_T^{(0,L)}$.

%-------------------------------------------------------------------
\subsection{Chiral extrapolations}
%-------------------------------------------------------------------

\begin{figure}[h!]

	\begin{minipage}[c]{0.49\linewidth}
	\centering 
	\includegraphics*[width=\linewidth]{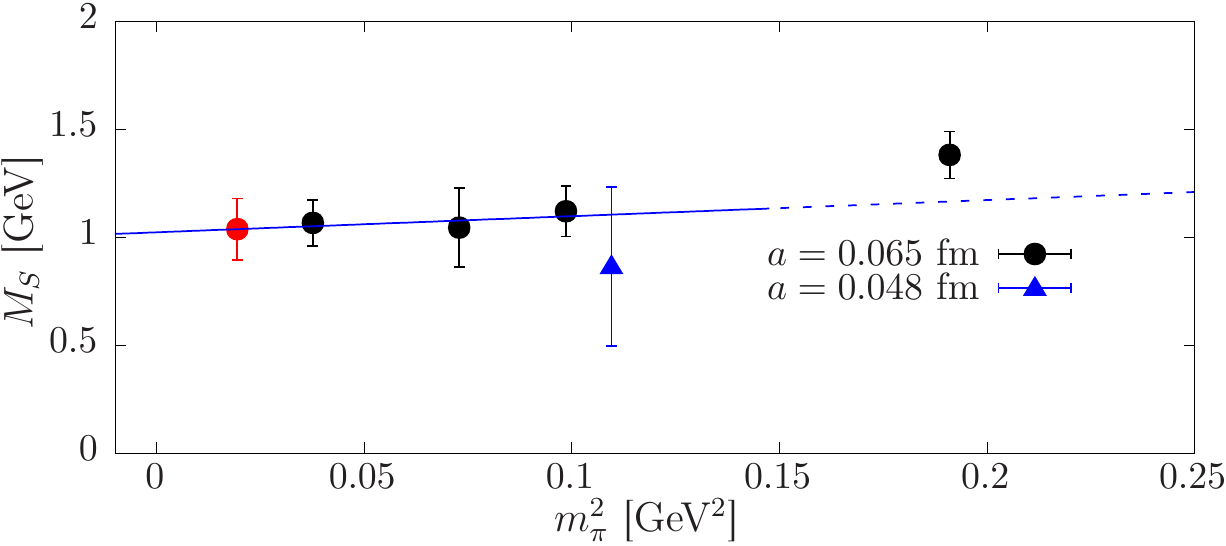}
	\end{minipage}
	\begin{minipage}[c]{0.49\linewidth}
	\centering 
	\includegraphics*[width=\linewidth]{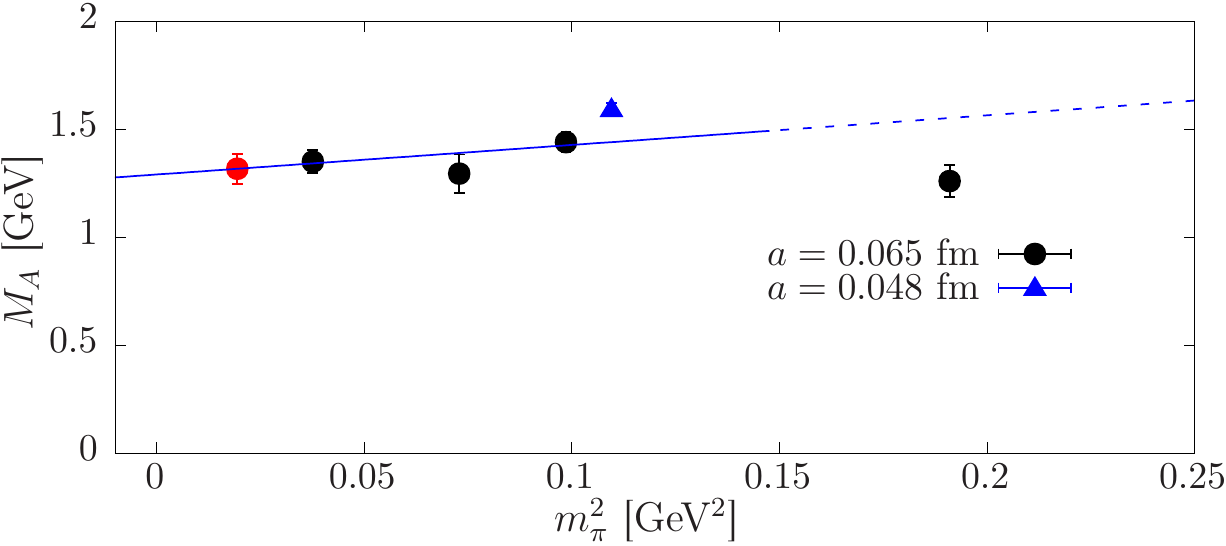}
	\end{minipage}
	
	\begin{minipage}[c]{0.49\linewidth}
	\centering 
	\includegraphics*[width=\linewidth]{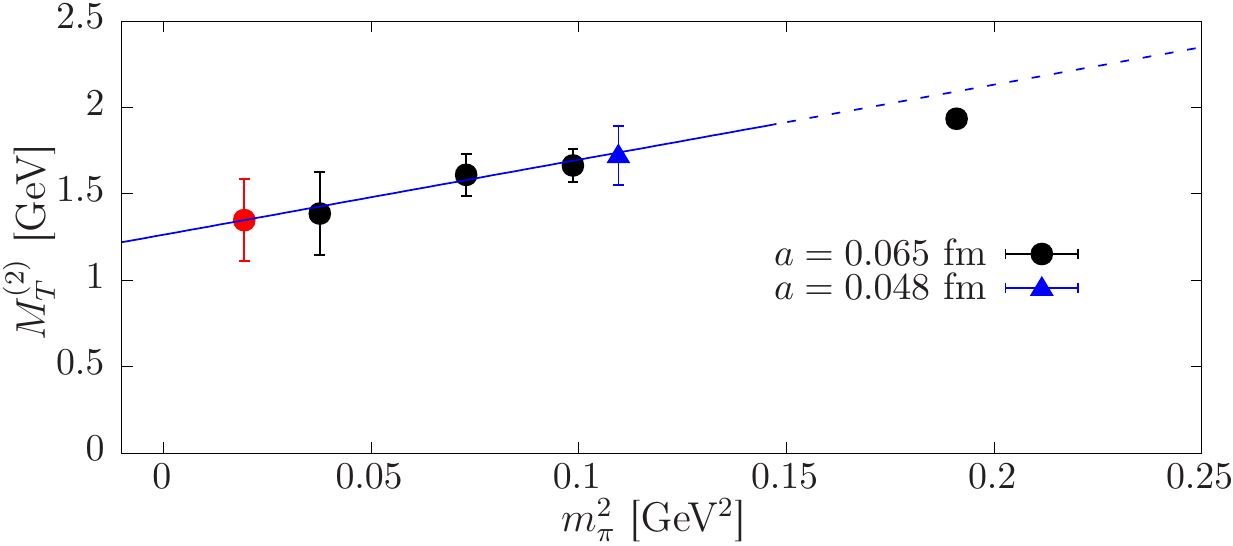}
	\end{minipage}
	\begin{minipage}[c]{0.49\linewidth}
	\centering 
	\includegraphics*[width=\linewidth]{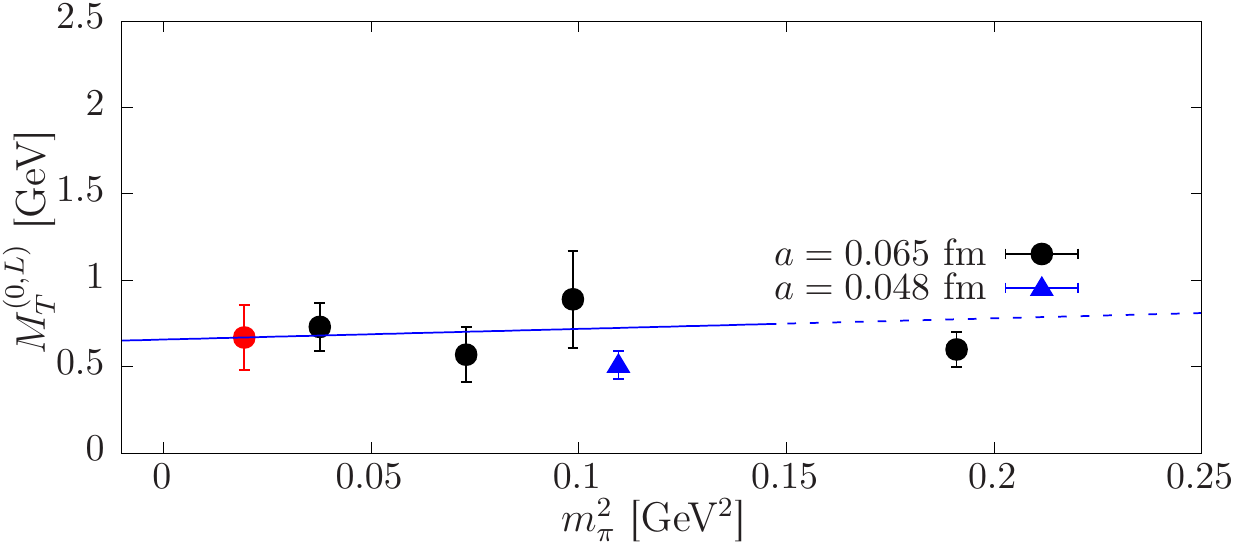}
	\end{minipage}
	
	\begin{minipage}[c]{0.49\linewidth}
	\centering 
	\includegraphics*[width=\linewidth]{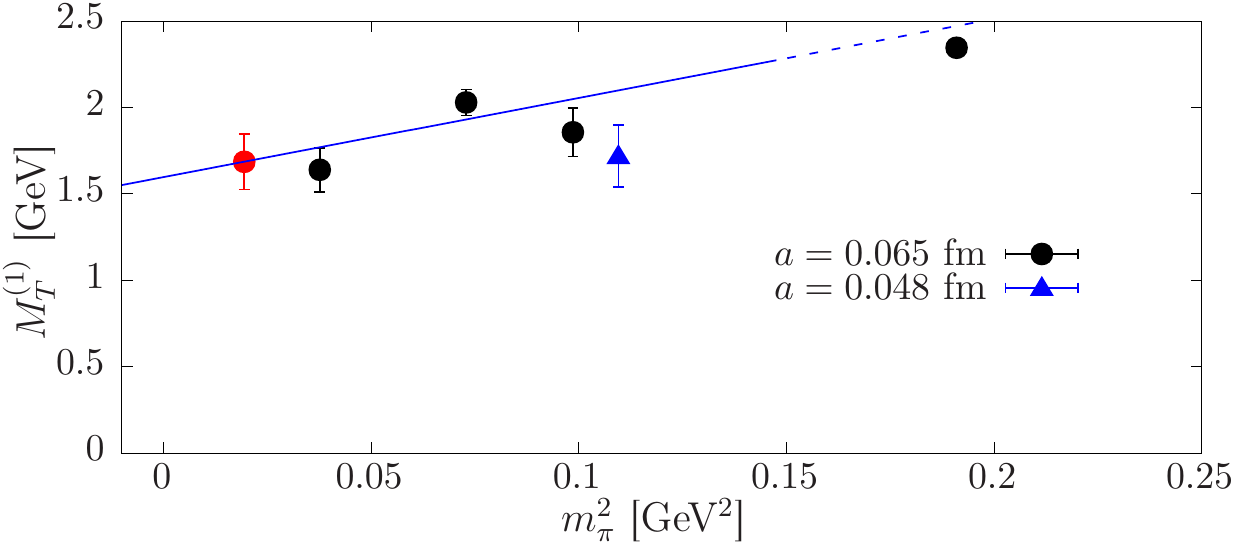}
	\end{minipage}
	\begin{minipage}[c]{0.49\linewidth}
	\centering 
	\includegraphics*[width=\linewidth]{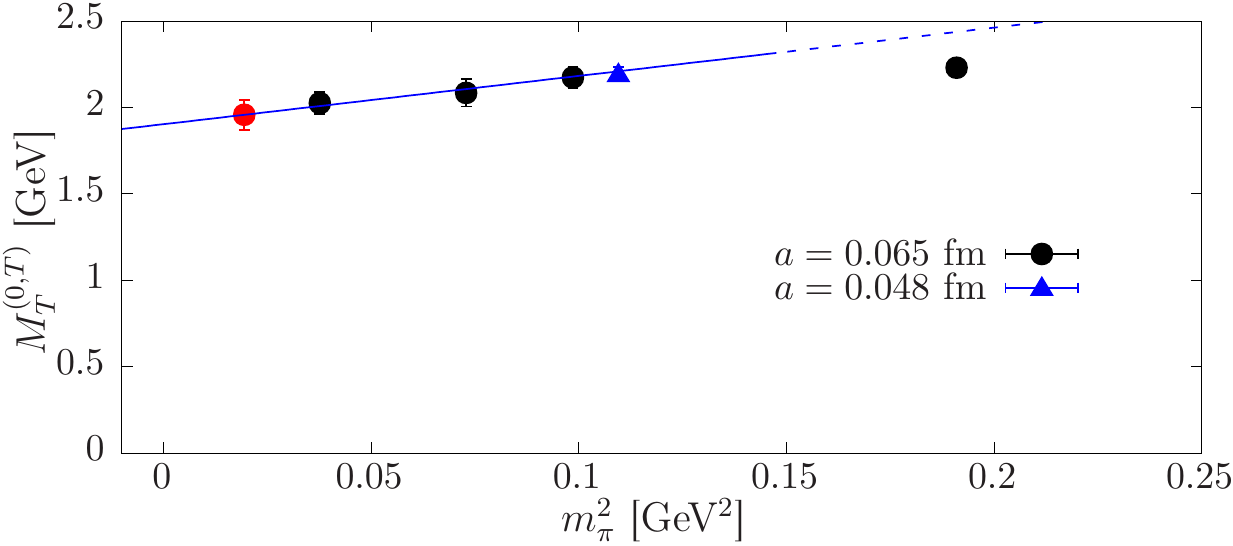}
	\end{minipage}
	
	\caption{Chiral extrapolations for each monopole and dipole mass, excluding the ensemble E5 with the largest
pion mass. The blue point corresponds to the lattice ensemble N6 and gives an indication about discretisation effects. 
Only the statistical error is displayed.}	
	\label{fig:extrap}
	
\end{figure}

Finally, we perform a chiral extrapolation for the six monopole and dipole masses that we have fitted to the lattice four-point function.
For each of these parameters, we assume a linear dependence on $m_{\pi}^2$. We perform two sets of fits, either including 
or excluding the ensemble E5, which has the largest pion mass.
The lattice results are given in Table~\ref{tab:ch_extrap} 
and depicted in Fig.~\ref{fig:extrap} together with the fits excluding E5. The displayed errors are purely statistical. 
The blue points in Fig.~\ref{fig:extrap} represent the ensemble N6 and therefore correspond to a finer lattice spacing
than the other data points. We remind the reader that we have included only isovector mesons in the description of the fully 
connected amplitude, therefore all our fitted form factor parameters correspond to isovector mesons.
While the results are quite stable under including or excluding ensemble E5, we consider the latter to be our final results,
mainly because the  $\chi^2$ per degree of freedom of the global fit was unacceptably large for E5.
We make the following observations:
\bi
\item The monopole mass of the scalar meson transition form factor  does not depend strongly on the pion mass. 
After a mild extrapolation, we obtain $ M_S = 1.04(14)~\GeV$ at the physical pion mass.
The result lies  above the experimental result $M_S = 0.796(54)~\GeV$
from the Belle Collaboration for the \emph{isoscalar} scalar meson~\cite{Masuda:2015yoh}.
\item
The axial dipole mass is also very weakly dependent on the pion mass. 
We obtain $M_A = 1.32(7)~\GeV$ at the physical pion mass. 
\begin{table}
\caption{Results of the chiral extrapolation for the scalar monopole 
mass $M_S$, the axial dipole mass $M_A$ and the four tensor dipole masses
corresponding to different helicities.
All results are given in units of GeV and correspond to isovector mesons
at the physical value of the pion mass. Results including or excluding 
the ensemble (E5) with the largest pion mass are given. We consider 
the latter to be our final results (last column of the table).}
\vskip 0.2in

\begin{tabular}{l@{~~~~~}c@{~~~~~}c}
\hline
    &   Including E5  &  Excluding E5 \\
\hline
$M_S$  &   0.94(12)     &   1.04(14) \\
$M_A$  &    1.40(07)     &   1.32(07)  \\
$M_T^{(2)}$ & 1.39(12)  &   1.35(24)  \\
$M_T^{(1)}$ & 1.67(10)   &  1.69(16)   \\
$M_T^{(0,T)}$ & 2.01(07)  &  1.96(09)  \\
$M_T^{(0,L)}$ & 0.74(14) &  0.67(19) \\
\hline
\end{tabular}
\label{tab:ch_extrap}
\end{table}
The finer ensemble N6 suggests a value $10\%$ larger. More ensembles would be needed to confirm whether $M_A$
is afflicted by large discretisation effects.
For comparison, the L3 Collaboration obtained a dipole mass $M_A = 1.040(80)~\GeV$ for the \emph{isoscalar} partner $f_1(1285)$;
the measurement relied on single-virtual measurements only~\cite{Achard:2001uu,Achard:2007hm}. 
The difference in the kinematics at which the form factor was probed could be part of the reason we found a larger dipole mass,
in addition to a potential genuine difference between the isospin partners. We also recall that the 
transition form factors have been parametrized in a fairly simplistic way (see Eq.\ (\ref{eq:axialFF}) and above).
\item
Finally, for the tensor meson $a_2$, linear extrapolations in $m_\pi^2$ yield the  results given in Table~\ref{tab:ch_extrap}.
Fits to experimental data on the  single-virtual form factor \cite{Masuda:2015yoh,Danilkin:2016hnh}
yielded smaller values for the $f_2(1270)$ meson. For instance, our result for the helicity-2 transition form factor, $M_T^{(2)}=1.35(24)~\GeV$,
is only slightly larger than the value $1.222(66)~\GeV$ obtained phenomenologically.
On the other hand, our values of $M_T^{(1)}=1.69(16)~\GeV$ and $M_T^{(0,T)}=1.96(9)~\GeV$ are 
almost a factor two larger than the corresponding phenomenological $f_2$ results, $M_T^{(1)}=0.916(20)~\GeV$ 
and $M_T^{(0,T)}=1.051(36)~\GeV$. Especially $M_T^{(0,T)}$ is statistically well constrained by the lattice data
and only weakly dependent on the lattice spacing and the pion mass. Finally, our value for $M_T^{(0,L)}=0.67(19)~\GeV$ is in agreement 
with the estimate $0.877(66)~\GeV$ obtained in~\cite{Danilkin:2016hnh} within the large uncertainties.
\ei
To summarize, in all cases except $M_T^{(0,L)}$, we obtain larger
monopole and dipole masses for the isovector mesons than in
phenomenology for the isoscalar mesons. The strongest difference is in $M_T^{(1)}$ and $M_T^{(0,T)}$, where 
we find that the form factors fall off far more slowly; this discrepancy could be due to the use of the 
factorization assumption for the dependence on the photon virtualities.
On the other hand, we find agreement within the uncertainties for the scalar monopole mass and the helicity-two 
form factor of the tensor meson.

\section{Study of disconnected diagrams\label{sec:disc}}

In this section, we test whether the hadronic model of section \ref{sec:model} together with 
the arguments summarized in section \ref{sec:34ov9} is consistent with the (2+2) disconnected diagrams,
which we have computed on two lattice ensembles.
The arguments, 
based on the large-$N$ motivated idea that an isolated vector current insertion in a fermion loop gives a suppressed contribution,  
lead to the conclusion that 
the (2+2) disconnected class of diagrams contains all of the contributions from flavor-singlet meson poles,
while the mesons in the adjoint representation of the flavor symmetry group contribute with a negative
weight factor; the latter is $(-25/9)$ in the SU(2)$_{\rm flavor}$  case and $(-2)$ in the SU(3)$_{\rm flavor}$ case.
The generic large-$N$ expectations would further lead to the stronger conclusion that, in each $J^{PC}$ sector, the non-singlet resonances 
cancel the contribution of the flavor-singlet resonances. One channel, however,
where the degeneracy is badly broken is the pseudoscalar sector, since the 
pion is much lighter than the $\eta'$ meson. Therefore, in the two-flavor theory we expect the (2+2) disconnected class of diagrams
to be given to a good approximation by 
\be\la{eq:pi0etap}
{\cal M}^{(2+2)}_{..} \approx -\frac{25}{9} {\cal M}^{(\pi^0)}_{..} + {\cal M}^{(\eta')}_{..}\;.
\ee
We have calculated the $\pi^0\to\gamma^*\gamma^*$  transition form factor on the same lattice ensembles
as used here in a previous publication~\cite{Gerardin:2016cqj}.
For the $\eta'$, the two-photon decay width is fairly well known experimentally; thus,
assuming a vector-meson-dominance model for the virtuality dependence of the $\eta'$ transition form factor
and using the known $\rho$ mass on each lattice ensemble,
Eq.\ (\ref{eq:pi0etap}) provides a prediction for the ${\cal M}^{(2+2)}_{..}$ amplitudes.
We use the vector mass given in Table \ref{tabsim}.
In Fig.\ \ref{fig:22_F6} we display the prediction for the three largest subtracted amplitudes $\overline{\cal M}^{(2+2)}_{TT}$,
$\overline{\cal M}^{\tau,(2+2)}_{TT}$ and $\overline{\cal M}^{a,(2+2)}_{TT}$ together with the direct lattice calculation.
We find that Eq.\ (\ref{eq:pi0etap}) predicts the overall size of the amplitudes well, within the fairly large uncertainties.
The agreement is most compelling in the $\overline{\cal M}^{\tau,(2+2)}_{TT}$ amplitude; this is also one of the amplitudes
where the pseudoscalar poles make a large contribution. In fact, in this channel, $\overline{\cal M}^{\tau,(2+2)}_{TT}$
amounts to about $-90\%$ of the fully connected contribution $\overline{\cal M}^{\tau,(4)}_{TT}$.

\subsection{Estimate of the contribution of the (2+2) disconnected class of diagrams to $a_\mu^{\rm HLbL}$}

We have obtained some evidence from our $N_f=2$ lattice data that the (2+2) class of diagrams is dominated by the pseudoscalar exchanges
with the weight factors derived in section \ref{sec:34ov9}.
Using the results from \cite{Nyffeler:2016gnb}, we can now estimate the importance of the $(2+2)$ disconnected class of diagrams
in $a_\mu^{\rm HLbL}$ in two limits:
\bi
\item $m_s=\infty$, which corresponds to the two-flavor theory;
\item  $m_s=m_{ud}$, which corresponds to the SU(3)-flavor symmetric theory.
\ei
We expect the real world to lie between these two predictions.
In these two limits, we obtain
\be\la{eq:amu2p2}
 a_\mu^{\rm HLbL,(2+2)} \approx \left\{  \begin{array}{l@{~~~~}l}
-\frac{25}{9} a_\mu^{{\rm HLbL},\pi^0} + a_\mu^{{\rm HLbL},\eta'} = -(162\pm27)\cdot 10^{-11} & m_s=\infty, \phantom{\Big|}
\\
- 2 ( a_\mu^{{\rm HLbL},\pi^0} + a_\mu^{{\rm HLbL},\eta} ) + a_\mu^{{\rm HLbL},\eta'} = -(142\pm19)\cdot 10^{-11} & m_s=m_{ud}. \phantom{\Big|}
\end{array}
\right.
\ee
We have used the LMD+V result for the pion ($62.9\cdot 10^{-11}$) and the VMD results for the $\eta$ and $\eta'$ 
(respectively $14.5\cdot 10^{-11}$ and $12.5\cdot 10^{-11}$) quoted in~\cite{Nyffeler:2016gnb} (see also References therein)
and assigned to each contribution an uncertainty of $15\%$. For comparison, the $N_f=2$ lattice calculation~\cite{Gerardin:2016cqj} 
of the pion transition form factor and its parametrization by the LMD+V model led to the value $a_\mu^{{\rm HLbL},\pi^0}=(65.0\pm8.3)\cdot 10^{-11}$.

Taking in addition the result $a_\mu^{\rm HLbL}\approx (102\pm39)\cdot 10^{-11}$ 
from a model calculation~\cite{Jegerlehner:2015stw}, the generic large-$N$ based expectations
imply the following estimate for the fully connected class of diagrams,
\be\la{eq:amu4}
a_{\mu,{\rm model}}^{\rm HLbL,(4)} \approx \left\{  \begin{array}{l@{~~~~}l}
(264\pm51) \cdot 10^{-11} & m_s=\infty, \phantom{\Big|}
\\
(244\pm46) \cdot 10^{-11} & m_s=m_{ud}. \phantom{\Big|}
\end{array}
\right.
\ee
These estimates give an idea of what to expect in forthcoming lattice calculations.
We remark, as also pointed out in~\cite{Nyffeler:2016gnb}, that 
the VMD model for the $\eta$ and $\eta'$ transition form factors is not tested in the doubly virtual case;
and that the VMD form factor falls off as $(Q^2)^{-2}$ in the limit of two large spacelike virtualities $Q_1^2=Q_2^2=Q^2$,
whereas the operator-product expansion predicts a $1/Q^2$ fall-off. Thus the $\eta$ and $\eta'$ contributions above 
could be somewhat underestimated due to the use of the VMD model. In the case of the pion, the `bias' from using the VMD
is $-10\%$, relative to using the more sophisticated LMD+V model.

The only lattice calculation~\cite{Blum:2016lnc} to have presented results for $a_{\mu}^{\rm HLbL,(4)}$
and $a_\mu^{\rm HLbL,(2+2)}$ found respectively $116.0(9.6)$ and $-62.5(8.0)$ in units of $10^{-11}$.
We conclude that either 
these lattice results are severely underestimated, which could be due to discretization and finite-volume effects;
or the hadronic model based on resonance exchanges is not viable;
or the large-$N$ inspired approximations made to estimate (\ref{eq:amu2p2}) and (\ref{eq:amu4}) are inadequate;
or a combination of the above.
A new high-statistics lattice calculation of $a_\mu^{\rm HLbL,(2+2)}$ in a large volume would be particularly illuminating 
to resolve the issue, since the prediction (\ref{eq:amu2p2}) is relatively clear-cut.

\section{Conclusion \la{sec:concl}}

With the hadronic light-by-light contribution to 
the muon anomalous magnetic moment $a_\mu^{\rm HLbL}$ in mind,
we have studied the eight forward light-by-light amplitudes for spacelike photons
in $N_f=2$ lattice QCD.
Via dispersive sum rules, we have tested whether the type of hadronic
models used to estimate $a_\mu^{\rm HLbL}$ provides a good description
of lattice results. All in all, we found that by fitting
the virtuality dependence of six meson transition form factors, we were able
to describe the lattice data within statistical uncertainties. The
monopole and dipole masses parametrizing the transition form factors 
compare reasonably well in magnitude with
phenomenological determinations for the $I=0$ isospin partner, 
with the notable exception of the dipole masses of the tensor meson
for helicities $\Lambda=1$ and $\Lambda=(0,T)$, where we find that the form factors
fall off far more slowly. The simultaneous fit
to all eight amplitudes allowed us to test the individual relevance of
the various resonance contributions, given that they appear with
different weights and signs in different amplitudes. Thus our study
provides evidence, by a completely independent method, that the
resonance-exchange model widely used in calculating $a_\mu^{\rm HLbL}$
is not missing a large contribution.

The (2+2) disconnected class of diagrams was computed on two lattice
ensembles.  We found that a parameter-free prediction based on a
specific large-$N$ argument presented in detail in section
\ref{sec:34ov9} (see also the earlier \cite{Bijnens:2016hgx}), which
expresses this set of diagrams in terms of the pseudoscalar mesons
alone, was compatible with the lattice data, albeit within large
relative errors.  Motivated by this observation, we estimated what
values a lattice calculation would have to obtain for the fully
connected and (2+2) set of disconnected diagrams if it is to reproduce
the current model estimates of $a_\mu^{\rm HLbL}$.

While we laid out many technical details of the method, we regard the
present calculation as exploratory, and leave a more quantitative
comparison of monopole and dipole masses, including an estimate of
systematic errors, for the future. Indeed we were only able to perform
stable fits by making model assumptions, for instance about the masses
of the lightest resonances in the scalar, axial-vector and tensor
sectors in $N_f=2$ QCD at non-physical quark masses. In addition to
neglecting the three classes of diagrams containing at least one
isolated vector current insertion in a quark loop, we had to assume
various relations between the two-photon decay widths of
isospin-partner resonances that are justified only for a large number
of colors $N$.  Also, the employed parametrization of the
axial-vector resonance form factors is a further vulnerable
assumption.

In the future, it would be useful to repeat the calculation of the
forward light-by-light amplitudes with higher statistics, on ensembles
including also the dynamical strange quark effects, and with a lighter
pion mass.  Especially at virtualities $\lesssim0.1\,{\rm GeV}^2$,
which can contribute significantly to $a_\mu^{\rm
  HLbL}$~\cite{Nyffeler:2016gnb}, smaller statistical errors would be
beneficial to test the hadronic model more stringently. Finite-volume
effects could not be addressed in any detail here, and a dedicated
study would be important to carry out, given the long-range nature of
the neutral pion contribution~\cite{Asmussen:2016lse,Asmussen:2017bup}.

\begin{acknowledgments}
\noindent We are thankful to I.\ Danilkin, A.\ Nyffeler and M.\ Vanderhaeghen  
for helpful discussions. We acknowledge the use
of CLS lattice ensembles and of QDP++ software~\cite{Edwards:2004sx} with
the deflated SAP+GCR solver from openQCD~\cite{CLScode21}. 
The correlation functions were computed at the
`{Clover}' cluster at the Helmholtz-Institut Mainz and the `Mogon' cluster of 
the University of Mainz.
This work was partially supported by the Deutsche Forschungsgemeinschaft (DFG) 
through the Collaborative Research Center ``The Low-Energy Frontier of the Standard Model'' (SFB 1044).
\end{acknowledgments}

\appendix

\small

%%%%%%%%%%%%%%%%%%%%%%%%%%%%%%%%%%%%%%%
\section{Cross sections $\gamma^* \gamma^* \to X$}
\label{app:CS}
%%%%%%%%%%%%%%%%%%%%%%%%%%%%%%%%%%%%%%%

This appendix is based on the Appendix of Ref.~\cite{Pascalutsa:2012pr}. We collect the relevant formulae needed to evaluate the sum rules in the general case with two virtual photons. 

%--------------------------------------------------------------------------------------------------------
\subsection{Notations}
%--------------------------------------------------------------------------------------------------------

The metric tensor of the subspace orthogonal to $q_1$ and $q_2$ is given by
\begin{equation}
R^{\mu \nu} (q_1, q_2) = - g^{\mu \nu} + \frac{1}{X} \, \bigl \{ (q_1 \cdot q_2) \left( q_1^\mu \, q_2^\nu + q_2^\mu \, q_1^\nu \right) - q_1^2 \, q_2^\mu \, q_2^\nu  - q_2^2 \, q_1^\mu \, q_1^\nu \bigr \} \,,
\end{equation}
such that $R_{\mu\nu} q_i^{\nu} = 0$ for $i=1,2$. It satisfies $R^{\mu\nu} = R^{\nu\mu}$, $R^{\mu}_{\ \mu}=2$ and $R^{\mu}_{\ \alpha} R^{\alpha \nu} = - R^{\mu\nu}$. We use the `mostly minus' metric convention. The virtual photon flux factor is defined through $X = (q_1 \cdot q_2)^2 - q_1^2 q_2^2 = \nu^2 - Q_1^2 Q_2^2$ with the crossing-symmetric variable $\nu$ given by $\nu=q_1 \cdot q_2$ 

The vectors $k_i$ are defined by
\begin{equation}
k_1 = \sqrt{ \frac{-q_1^2}{X}} \left( q_2 - \frac{q_1 \cdot q_2}{q_1^2} q_1 \right) \,, \quad k_2 = \sqrt{ \frac{-q_2^2}{X} } \left( q_1 - \frac{q_1 \cdot q_2}{q_1^2} q_2 \right) \,,
\end{equation}
and satisfy $k_i^2=1$, $k_i \cdot q_i = 0$. 

Finally, the helicity amplitudes for the $\gamma^\ast(\lambda_1,q_1) \gamma^\ast(\lambda_2,q_2) \to X(p_X)$ fusion process are related to the Feynman amplitudes by
\begin{equation}
{\cal M}(\lambda_1,\lambda_2) = {\cal M}_{\mu\nu} \ \epsilon_1^{\mu}(\lambda_1) \ \epsilon_2^{\nu}(\lambda_2) \,.
\end{equation}

%--------------------------------------------------------------------------------------------------------
\subsection{Pseudoscalar mesons}
%--------------------------------------------------------------------------------------------------------

The transition $\gamma^\ast (q_1, \lambda_1) + \gamma^\ast(q_2, \lambda_2) \to {\cal P}$, where ${\cal P}$ is a pseudoscalar state, is described by the following amplitude:
\begin{equation} 
{\cal M}(\lambda_1, \lambda_2) = - i \, e^2 \, \varepsilon_{\mu \nu \alpha \beta} \, \varepsilon^\mu(q_1, \lambda_1) \, \varepsilon^\nu(q_2, \lambda_2) \, q_1^\alpha \, q_2^{\beta} \,  F_{{\cal P} \gamma^\ast \gamma^\ast} (Q_1^2, Q_2^2) \,, 
\end{equation}
where $ \varepsilon^\mu(q_1, \lambda_1)$ and $\varepsilon^\nu(q_2, \lambda_2)$ are the polarization vectors of the virtual photons with helicities $\lambda_1, \lambda_2 = 0, \pm 1$. The only non-zero helicity amplitudes, which we define in the rest frame of the produced meson, are given by~:
\begin{eqnarray}
{\cal M}(+1, +1) = - {\cal M}(-1, -1)  = e^2 \, \sqrt{X} \, F_{{\cal P} \gamma^\ast \gamma^\ast}(Q_1^2, Q_2^2) \, . 
\end{eqnarray}
The two-photon decay width is given by 
\begin{equation}
\GammaGG   = \frac{\pi \alpha^2}{4} m_P^3 \left[ F_{{\cal P} \gamma^\ast \gamma^\ast}(0, 0) \right]^2 \,, 
\end{equation}
and from Eqs.~(\ref{eq:abspart}) and (\ref{eq:vcross}) 
\begin{gather}
\sigma_0 = \sigma_{\perp} = 2 \sigma_{TT} = 2 \tau_{TT}^a = -\tau_{TT} = 16 \pi^2 \delta(s-m_P^2)  \frac{ \GammaGG }{ m_P } \frac{ 2\sqrt{X} }{ m_P^2 }  \left[ \frac{  F_{{\cal P} \gamma^\ast \gamma^\ast}(Q_1^2, Q_2^2) }{ F_{{\cal P} \gamma^\ast \gamma^\ast}(0, 0)  } \right]^2 \,, \nonumber \\
\sigma_{LL} = \sigma_{TL} = \sigma_{LT} = \tau_{TL} = \tau_{TL}^a = 0 \,.
\end{gather}

%--------------------------------------------------------------------------------------------------------
\subsection{Scalar mesons}
%--------------------------------------------------------------------------------------------------------

The transition $\gamma^\ast (q_1, \lambda_1) + \gamma^\ast(q_2, \lambda_2) \to {\cal S}$ where ${\cal S}$ is a scalar state can be parameterized by one transverse ($F^{T}_{{\cal A} \gamma^\ast \gamma^\ast}$) and one longitudinal ($F^{L}_{{\cal A} \gamma^\ast \gamma^\ast}$) form factor and is described by the following matrix element
\begin{eqnarray*} 
{\cal M}(\lambda_1, \lambda_2) &=&  e^2 \, \varepsilon_\mu(q_1, \lambda_1) \, \varepsilon_\nu(q_2, \lambda_2) \, 
\, \nonumber \\
&\times& \left( \frac{\nu}{m_S} \right) \left\{ - R^{\mu \nu} (q_1, q_2) F^T_{{\cal S} \gamma^\ast \gamma^\ast} (Q_1^2, Q_2^2)  \,+\,
\frac{\nu}{X} \left( q_1^\mu + \frac{Q_1^2}{\nu} q_2^{\mu} \right) \left( q_2^\nu + \frac{Q_2^2}{\nu} q_1^{\nu} \right)  
F^L_{{\cal S} \gamma^\ast \gamma^\ast} (Q_1^2, Q_2^2) 
\right\}.
\label{eq:sff}
\end{eqnarray*}
The only non-zero helicity amplitudes are given by
\begin{eqnarray}
{\cal M}(+1, +1) &=& {\cal M}(-1, -1) =  e^2 \, \frac{\nu}{m_S} \, F^T_{{\cal S} \gamma^\ast \gamma^\ast}(Q_1^2, Q_2^2) \, , \nonumber \\
 {\cal M}(0, 0)  &=& {- e^2 \, \frac{Q_1 Q_2}{m_S} \,  F^L_{{\cal S} \gamma^\ast \gamma^\ast}(Q_1^2, Q_2^2) \, . }
\end{eqnarray}
The two-photon decay width is given by 
\begin{equation}
\GammaGG = \frac{\pi \alpha^2}{4} m_S \left[ F^T_{{\cal S} \gamma^\ast \gamma^\ast}(0, 0) \right]^2 \,, 
\end{equation}
and from Eqs.~(\ref{eq:abspart}) and (\ref{eq:vcross}) 
\begin{gather}
\sigma_0 = \sigma_{\parallel} = 2 \sigma_{TT} = 2 \tau_{TT}^a = \tau_{TT} = 16 \pi^2 \delta(s-m_S^2) \frac{ \GammaGG }{ m_S }  \frac{ 2\nu^2 }{ m_S^2 \sqrt{X} }  \left[ \frac{  F^T_{{\cal S} \gamma^\ast \gamma^\ast}(Q_1^2, Q_2^2) }{ F^T_{{\cal S} \gamma^\ast \gamma^\ast}(0, 0)  } \right]^2 \,, \nonumber \\
\sigma_{LL} = 16 \pi^2 \delta(s-m_S^2)  \frac{ \GammaGG }{ m_S }  \frac{ 2 Q_1^2 Q_2^2 }{ m_S^2 \sqrt{X} }  \left[ \frac{  F^L_{{\cal S} \gamma^\ast \gamma^\ast}(Q_1^2, Q_2^2) }{ F^T_{{\cal S} \gamma^\ast \gamma^\ast}(0, 0)  } \right]^2  \,, \nonumber \\
\tau_{TL} = \tau_{TL}^a = -16 \pi^2 \delta(s-m_S^2)  \frac{ \GammaGG }{ m_S }  \frac{ Q_1  Q_2  }{ m_S } \frac{\nu}{m_S \sqrt{X}}   \frac{  F^T_{{\cal S} \gamma^\ast \gamma^\ast}(Q_1^2, Q_2^2)  F^L_{{\cal S} \gamma^\ast \gamma^\ast}(Q_1^2, Q_2^2) }{ \left[F^T_{{\cal S} \gamma^\ast \gamma^\ast}(0, 0) \right]^2  }  \,. 
\end{gather}

%--------------------------------------------------------------------------------------------------------
\subsection{Axial mesons}
%--------------------------------------------------------------------------------------------------------

The transition $\gamma^\ast (q_1, \lambda_1) + \gamma^\ast(q_2, \lambda_2) \to {\cal A}(p_A, \Lambda)$, where ${\cal A}$ is an axial-vector state, can be parameterized by two form factors $F^{(0)}_{{\cal A} \gamma^\ast \gamma^\ast}$ and $F^{(1)}_{{\cal A} \gamma^\ast \gamma^\ast}$, where the superscript indicates the helicity state ($\Lambda$) of the axial-vector meson  
\begin{eqnarray} 
{\cal M}(\lambda_1, \lambda_2; \Lambda) &=& e^2 \, 
\varepsilon_\mu(q_1, \lambda_1) \, \varepsilon_\nu(q_2, \lambda_2) \, 
\varepsilon^{\alpha \ast}(p_f, \Lambda) \, \nonumber \\
&\times& i \, \varepsilon_{\rho \sigma \tau \alpha} \,  \left\{ 
R^{\mu \rho} (q_1, q_2) R^{\nu \sigma} (q_1, q_2) \, 
(q_1 - q_2)^\tau \, \frac{\nu}{m_A^2} \, F^{(0)}_{{\cal A} \gamma^\ast \gamma^\ast}(Q_1^2, Q_2^2)
\right. \nonumber \\
&&\hspace{1cm} + \, R^{\nu \rho}(q_1, q_2) \left( q_1^\mu + \frac{Q_1^2}{\nu} q_2^{\mu} \right) 
 q_1^\sigma \, q_2^\tau \,  \frac{1}{m_A^2} \, F_{{\cal A} \gamma^\ast \gamma^\ast}^{(1)}(Q_1^2, Q_2^2) \nonumber \\
&&\left. \hspace{1cm} + \, R^{\mu \rho}(q_1, q_2) \left( q_2^\nu + \frac{Q_2^2}{\nu} q_1^{\nu} \right) 
 q_2^\sigma \, q_1^\tau \, \frac{1}{m_A^2} \, F^{(1)}_{{\cal A} \gamma^\ast \gamma^\ast}(Q_2^2, Q_1^2) 
\right\}.
\label{eq:aff}
\end{eqnarray}
The only non-zero helicity amplitudes are given by
\begin{gather}
{\cal M}(+1, +1; \Lambda = 0) = - {\cal M}(-1, -1; \Lambda = 0) =  e^2 \, (Q_1^2 - Q_2^2) \, \frac{\nu}{m_A^3} \, F^{(0)}_{{\cal A} \gamma^\ast \gamma^\ast}(Q_1^2, Q_2^2) \, , \nonumber \\
{\cal M}(0, +1; \Lambda = -1) = - \, e^2 \, Q_1 \, \left( \frac{X}{\nu m_A^2} \right) \, F^{(1)}_{{\cal A} \gamma^\ast \gamma^\ast}(Q_1^2, Q_2^2) \, , \nonumber \\
{\cal M}(-1, 0; \Lambda = -1) = - \, e^2 \, Q_2 \, \left( \frac{X}{\nu m_A^2} \right) \, F^{(1)}_{{\cal A} \gamma^\ast \gamma^\ast}(Q_2^2, Q_1^2) \, . 
\end{gather}
In this case, the equivalent two-photon width is defined by
\begin{equation}
\GammaGGt \equiv \lim_{Q_1^2 \to 0} \frac{m_A^2}{Q_1^2} \frac{1}{2} \Gamma(\mathcal{A} \to \gamma_L^* \gamma_T) = \frac{\pi \alpha^2}{4} \frac{m_A}{3} \left[ F^{(1)}_{{\cal A} \gamma^\ast \gamma^\ast}(0, 0) \right]^2 \,, 
\end{equation}
and from Eqs.~(\ref{eq:abspart}) and (\ref{eq:vcross}) 
\begin{gather}
\sigma_0 = \sigma_{\perp} = 2 \sigma_{TT} = 2 \tau_{TT}^a = - \tau_{TT} = 16 \pi^2 \delta(s-m_A^2)  \frac{ 3 \GammaGGt }{ m_A }  \frac{ (Q_1^2-Q_2^2)^2 }{ m_A^4 } \frac{2\nu^2}{m_A^2 \sqrt{X}}  \left[ \frac{  F^{(0)}_{{\cal A} \gamma^\ast \gamma^\ast}(Q_1^2, Q_2^2) }{ F^{(1)}_{{\cal A} \gamma^\ast \gamma^\ast}(0, 0)  } \right]^2 \,, \nonumber \\
\sigma_{LT} = 16 \pi^2 \delta(s-m_A^2)   \frac{ 3 \GammaGGt }{ m_A } \frac{ 2X \sqrt{X} }{ \nu^2 m_A^2 }  \frac{Q_1^2}{m_A^2}  \left[ \frac{  F^{(1)}_{{\cal A} \gamma^\ast \gamma^\ast}(Q_1^2, Q_2^2) }{ F^{(1)}_{{\cal A} \gamma^\ast \gamma^\ast}(0, 0)  } \right]^2 \,, \nonumber \\
\sigma_{TL} = 16 \pi^2 \delta(s-m_A^2)   \frac{ 3 \GammaGGt }{ m_A } \frac{ 2X \sqrt{X} }{ \nu^2 m_A^2 }   \frac{Q_2^2}{m_A^2}  \left[ \frac{  F^{(1)}_{{\cal A} \gamma^\ast \gamma^\ast}(Q_2^2, Q_1^2) }{ F^{(1)}_{{\cal A} \gamma^\ast \gamma^\ast}(0, 0)  } \right]^2 \,, \nonumber \\
\tau_{TL} = -\tau^a_{TL} =  16 \pi^2 \delta(s-m_A^2)   \frac{ 3 \GammaGGt }{ m_A } \frac{Q_1 Q_2}{m_A^2}  \frac{ X \sqrt{X} }{ \nu^2 m_A^2 }  \left[ \frac{  F^{(1)}_{{\cal A} \gamma^\ast \gamma^\ast}(Q_1^2, Q_2^2) }{ F^{(1)}_{{\cal A} \gamma^\ast \gamma^\ast}(0, 0)  }     \frac{  F^{(1)}_{{\cal A} \gamma^\ast \gamma^\ast}(Q_2^2, Q_1^2) }{ F^{(1)}_{{\cal A} \gamma^\ast \gamma^\ast}(0, 0)  } \right] \,, \nonumber \\
\sigma_{LL} = 0 \,.
\label{eq:axialff}
\end{gather}

%--------------------------------------------------------------------------------------------------------
\subsection{Tensor mesons}
%--------------------------------------------------------------------------------------------------------

The transition $\gamma^\ast (q_1, \lambda_1) + \gamma^\ast(q_2, \lambda_2) \to {\cal T}(\Lambda)$ where ${\cal T}$ is a tensor state 
with helicity $\Lambda = \pm2,\pm1,0$ can be parameterized by four form factors $T^{(\Lambda)}$,
\begin{eqnarray}
{\cal M}(\lambda_1, \lambda_2; \Lambda) &=& e^2 \, 
\varepsilon_\mu(q_1, \lambda_1) \, \varepsilon_\nu(q_2, \lambda_2) \, 
\varepsilon^\ast_{\alpha \beta}(p_f, \Lambda) \, \nonumber \\
&\times& \left\{ 
 \left[ R^{\mu \alpha} (q_1, q_2) R^{\nu \beta} (q_1, q_2) 
+ \frac{s}{8 X} \, R^{\mu \nu}(q_1, q_2) 
(q_1 - q_2)^\alpha \, (q_1 - q_2)^\beta \right] \, \frac{\nu}{m_T} \,F^{(2)}_{{\cal T} \gamma^\ast \gamma^\ast}(Q_1^2, Q_2^2)
\right. \nonumber \\
&&+ \, R^{\nu \alpha}(q_1, q_2) (q_1 - q_2)^\beta  \left( q_1^\mu + \frac{Q_1^2}{\nu} q_2^{\mu} \right) 
\, \frac{1}{m_T} \, F^{(1)}_{{\cal T} \gamma^\ast \gamma^\ast}(Q_1^2, Q_2^2) \nonumber \\
&&+ R^{\mu \alpha}(q_1, q_2) (q_2 - q_1)^\beta   \left( q_2^\nu + \frac{Q_2^2}{\nu} q_1^{\nu} \right) 
\,  \frac{1}{m_T} \, F^{(1)}_{{\cal T} \gamma^\ast \gamma^\ast}(Q_2^2, Q_1^2)
\nonumber \\ 
&&+ \, R^{\mu \nu}(q_1, q_2) (q_1 - q_2)^\alpha \, (q_1 - q_2)^\beta \, \frac{1}{m_T} \, 
 F^{(0, T)}_{{\cal T} \gamma^\ast \gamma^\ast}(Q_1^2, Q_2^2) \, \nonumber \\
&&\left. + \,  \left( q_1^\mu + \frac{Q_1^2}{\nu} q_2^{\mu} \right) \left( q_2^\nu + \frac{Q_2^2}{\nu} q_1^{\nu} \right)  
(q_1 - q_2)^\alpha  (q_1 - q_2)^\beta 
\, \frac{1}{m_T^3} \, F^{(0, L)}_{{\cal T} \gamma^\ast \gamma^\ast}(Q_1^2, Q_2^2)
\right\},
\label{eq:tensorff}
\end{eqnarray}
where $\varepsilon_{\alpha \beta}(p_f, \Lambda)$ is the polarization tensor for the tensor meson with 
four-momentum $p_f$ and helicity $\Lambda$.
The different non-vanishing helicity amplitudes are
\begin{gather}
{\cal M}(+1, -1; \Lambda = +2) =  {\cal M}(-1, +1; \Lambda = -2) = e^2 \, \frac{\nu}{m_T} \, F^{(2)}_{{\cal T} \gamma^\ast \gamma^\ast}(Q_1^2, Q_2^2) \, , \nonumber \\
{\cal M}(0, +1; \Lambda = -1) = - e^2 \, Q_1 \,  \frac{1}{\sqrt{2}} \, \left( \frac{2 X}{\nu m_T^2} \right) \, F^{(1)}_{{\cal T} \gamma^\ast \gamma^\ast}(Q_1^2, Q_2^2) \, , \nonumber \\
{\cal M}(-1, 0; \Lambda = -1) = - e^2 \, Q_2 \, \frac{1}{\sqrt{2}} \, \left( \frac{2 X}{\nu m_T^2} \right) \, F^{(1)}_{{\cal T} \gamma^\ast \gamma^\ast}(Q_2^2, Q_1^2) \, , \nonumber \\
{\cal M}(+1, +1; \Lambda = 0) =  {\cal M}(-1, -1; \Lambda = 0) = - e^2 \, \sqrt{\frac{2}{3}} \, \left( \frac{4 X}{m_T^3} \right) \, F^{(0, T)}_{{\cal T} \gamma^\ast \gamma^\ast}(Q_1^2, Q_2^2) \, , \nonumber \\
{\cal M}(0, 0; \Lambda = 0) =  - e^2 \, Q_1 Q_2 \, \sqrt{\frac{2}{3}} \,  \left( \frac{4 X^2}{\nu^2 m_T^5} \right) \, F^{(0, L)}_{{\cal T} \gamma^\ast \gamma^\ast}(Q_1^2, Q_2^2)\, . 
\end{gather}
The two-photon decay widths for helicities $\Lambda=0,2$ are respectively given by 
\begin{align}
\GammaGG^{(0)} &= \pi \alpha^2 \, m_T   \, \frac{2}{15} \,  \left[ F^{(0, T)}_{{\cal T} \gamma^\ast \gamma^\ast} (0,0) \right]^2 \,, \nonumber \\
\GammaGG^{(2)}  &= \frac{\pi \alpha^2}{4}  \, m_T \, \frac{1}{5} \, \left[ F^{(2)}_{{\cal T} \gamma^\ast \gamma^\ast} (0,0) \right]^2 \,. 
\end{align}
and from Eqs.~(\ref{eq:abspart}) and (\ref{eq:vcross}) 
\begin{gather*}
\sigma_0 =  16 \pi^2 \delta(s-m_T^2)  \frac{ 5 \GammaGGt^{(0)} }{ m_T }  \frac{ 8 X \sqrt{X} }{m_T^6}  \left[ \frac{  F^{(0,T)}_{{\cal T} \gamma^\ast \gamma^\ast}(Q_1^2, Q_2^2) }{ F^{(0,T)}_{{\cal T} \gamma^\ast \gamma^\ast}(0, 0)  } \right]^2 \,, \\
\sigma_2 =  16 \pi^2 \delta(s-m_T^2)  \frac{ 5 \GammaGGt^{(2)} }{ m_T }  \frac{2 \nu^2 }{m_T^2 \sqrt{X} }  \left[ \frac{  F^{(2)}_{{\cal T} \gamma^\ast \gamma^\ast}(Q_1^2, Q_2^2) }{ F^{(2)}_{{\cal T} \gamma^\ast \gamma^\ast}(0, 0)  } \right]^2 \,, \\
\sigma_{\parallel} = \sigma_0 + \frac{\sigma_2}{2} \,,\\
\sigma_{\perp} = \frac{\sigma_2}{2} \,,\\
\sigma_{LT} =  16 \pi^3 \delta(s-m_T^2) \alpha^2   \frac{ Q_1^2 }{ m_T^2 }  \frac{ X \sqrt{X} }{ \nu^2 m_T^2 }  \left[  F^{(1)}_{{\cal T} \gamma^\ast \gamma^\ast}(Q_1^2, Q_2^2)   \right]^2 \,, \\
\sigma_{TL} =  16 \pi^3 \delta(s-m_T^2) \alpha^2   \frac{ Q_2^2 }{ m_T^2 }  \frac{ X \sqrt{X} }{ \nu^2 m_T^2 }  \left[  F^{(1)}_{{\cal T} \gamma^\ast \gamma^\ast}(Q_2^2, Q_1^2)   \right]^2 \,, \\
\tau_{TL} =  16 \pi^3 \delta(s-m_T^2) \alpha^2   \frac{ X \sqrt{X} }{ \nu^2 m_T^2 }  \frac{ Q_1 Q_2 }{ m_T^2 }  \left[ \frac{2}{3} \frac{4X}{m_T^4}  F^{(0,T)}_{{\cal T} \gamma^\ast \gamma^\ast}(Q_1^2, Q_2^2) F^{(0,L)}_{{\cal T} \gamma^\ast \gamma^\ast}(Q_1^2, Q_2^2) - \frac{1}{2}  F^{(1)}_{{\cal T} \gamma^\ast \gamma^\ast}(Q_1^2, Q_2^2)  F^{(1)}_{{\cal T} \gamma^\ast \gamma^\ast}(Q_2^2, Q_1^2)  \right] \,, \\
\tau_{TL}^a = 16 \pi^3 \delta(s-m_T^2) \alpha^2   \frac{ X \sqrt{X} }{ \nu^2 m_T^2 }  \frac{ Q_1 Q_2 }{ m_T^2 }  \left[ \frac{2}{3} \frac{4X}{m_T^4}  F^{(0,T)}_{{\cal T} \gamma^\ast \gamma^\ast}(Q_1^2, Q_2^2) F^{(0,L)}_{{\cal T} \gamma^\ast \gamma^\ast}(Q_1^2, Q_2^2) + \frac{1}{2}  F^{(1)}_{{\cal T} \gamma^\ast \gamma^\ast}(Q_1^2, Q_2^2)  F^{(1)}_{{\cal T} \gamma^\ast \gamma^\ast}(Q_2^2, Q_1^2)  \right] \,, \\
\sigma_{LL} = 16 \pi^3 \delta(s-m_T^2) \alpha^2   \frac{ Q_1^2 Q_2^2 }{ m_T^4 } \frac{16}{3}  \frac{ X^3 \sqrt{X} }{ \nu^4 m_T^6 }  \left[  F^{(0,L)}_{{\cal T} \gamma^\ast \gamma^\ast}(Q_2^2, Q_1^2)   \right]^2 \,.
\end{gather*}

\newpage

\section{Additional material: tables and figures \label{sec:addmat}}

We present tables of results for the forward HLbL scattering amplitudes 
on three lattice ensembles at a few values of the kinematic variables.
Figure \ref{fig:amps_G8_part1} displays the results on ensemble G8 as a 
function of $Q_2^2$.

\begin{table}[h!]
\caption{Forward HLbL scattering amplitudes $(\times10^6$) on the ensembles G8, F7, F6. The variable $\nu$ is given in $\GeV^2$ units.
Results are given for two sets of  virtualities, $A$ corresponding to $(Q_1^2=0.352\,\GeV^2$, $Q_2^2=0.352\,\GeV^2)$
and $B$ to $(Q_1^2=0.352\,\GeV^2$, $Q_2^2=1.000\,\GeV^2)$.}
\vskip 0.3in
\centerline{
\begin{tabular}{l@{\quad}c@{\quad}S@{\quad}S@{\quad}S@{\quad}S@{\quad}S@{\quad}S@{\quad}S@{\quad}S}
	\hline
G8		& $\nu$	& $\MsTT$	& $\MsTTt$	& $\MsTTa$		&	$\MsTL$	& $\MsLT$	& $\MsTLa$	& $\MsTLt$	& $\MsLL$	 	 \\
	\hline
\multirow{4}{*}{$A$} 
	& 0.087 & \num{8.5 \pm 6.0} 	& \num{1.0 \pm 5.1} & $\times$ & \num{2.6 \pm 7.2} & \num{-7.9 \pm 6.0} & $\times$ & \num{-1.8 \pm 2.5} & \num{-6.1 \pm 7.1} 	\\ 
	& 0.176 & \num{43.0 \pm 11.2} 	& \num{-20.3 \pm 15.9} & \num{-0.9 \pm 8.1} & \num{-5.7 \pm 18.5} & \num{-26.8 \pm 15.4} & \num{3.7 \pm 6.4} & \num{-1.1 \pm 3.5} & \num{1.2 \pm 17.0} 	\\ 
	& 0.263 & \num{35.0 \pm 10.5} 	& \num{-12.6 \pm 10.6} & \num{5.6 \pm 13.1} & \num{-6.9 \pm 17.4} & \num{-0.2 \pm 14.8} & \num{-0.1 \pm 10.3} & \num{1.3 \pm 4.2} & \num{22.4 \pm 18.0} 	\\ 
	& 0.351 & \num{141.5 \pm 22.0} & \num{-69.2 \pm 22.7} & \num{77.0 \pm 20.3} & \num{-25.4 \pm 20.2} & \num{-17.9 \pm 21.7} & \num{9.9 \pm 9.3} & \num{27.9 \pm 11.4} & \num{65.8 \pm 29.6} 	\\ 
	\hline
\multirow{7}{*}{$B$}
	& 0.087 & \num{-1.7 \pm 1.8} & \num{0.0 \pm 2.1} & $\times$ & \num{0.4 \pm 2.3} & \num{2.2 \pm 2.8} & $\times$ & \num{-0.0 \pm 0.7} & \num{3.1 \pm 2.6} 	\\ 
	& 0.176 & \num{1.7 \pm 5.7} & \num{0.6 \pm 4.4} & \num{1.5 \pm 2.7} & \num{0.1 \pm 6.3} & \num{3.9 \pm 5.5} & \num{-0.3 \pm 1.5} & \num{-0.6 \pm 2.9} & \num{3.3 \pm 7.4} 	\\ 
	& 0.263 & \num{10.6 \pm 3.9} & \num{1.0 \pm 3.6} & \num{5.2 \pm 5.0} & \num{2.7 \pm 8.4} & \num{5.6 \pm 4.9} & \num{-1.4 \pm 3.6} & \num{-1.6 \pm 2.2} & \num{2.9 \pm 5.6} 	\\ 
	& 0.351 & \num{15.5 \pm 5.4} & \num{6.1 \pm 6.2} & \num{8.5 \pm 6.9} & \num{-0.5 \pm 8.8} & \num{6.6 \pm 7.9} & \num{0.8 \pm 3.2} & \num{0.4 \pm 3.8} & \num{4.7 \pm 8.8} 	\\ 
	& 0.438 & \num{28.6 \pm 5.7} & \num{-0.1 \pm 3.5} & \num{11.6 \pm 7.6} & \num{10.6 \pm 8.1} & \num{16.7 \pm 5.2} & \num{-1.4 \pm 5.6} & \num{-2.7 \pm 3.6} & \num{3.3 \pm 8.6} 	\\ 
	& 0.525 & \num{46.4 \pm 7.5} & \num{-5.6 \pm 4.9} & \num{22.2 \pm 8.1} & \num{18.8 \pm 9.0} & \num{22.8 \pm 7.2} & \num{2.9 \pm 4.9} & \num{-4.3 \pm 4.5} & \num{6.7 \pm 10.5} 	\\ 
	& 0.612 & \num{76.6 \pm 9.7} & \num{-27.6 \pm 10.3} & \num{40.2 \pm 16.2} & \num{13.6 \pm 13.1} & \num{17.4 \pm 9.5} & \num{5.7 \pm 6.5} & \num{0.9 \pm 7.5} & \num{23.2 \pm 15.2}	\\ 
	\hline
 \end{tabular}}

\vskip 0.6in
\centerline{\begin{tabular}{l@{\quad}c@{\quad}S@{\quad}S@{\quad}S@{\quad}S@{\quad}S@{\quad}S@{\quad}S@{\quad}S}
	\hline
F7		& $\nu$	& $\MsTT$	& $\MsTTt$	& $\MsTTa$		&	$\MsTL$	& $\MsLT$	& $\MsTLa$	& $\MsTLt$	& $\MsLL$	 	 \\
	\hline
\multirow{3}{*}{$A$}	
	& 0.117 & \num{3.1 \pm 2.8} & \num{5.0 \pm 2.0} & $\times$ & \num{0.3 \pm 6.3} & \num{0.7 \pm 4.2} & $\times$ & \num{-2.9 \pm 1.5} & \num{4.3 \pm 6.4} 	\\ 
	& 0.234 & \num{23.4 \pm 10.0} & \num{-0.6 \pm 7.9} & \num{-2.2 \pm 3.7} & \num{8.1 \pm 14.0} & \num{-0.5 \pm 9.8} & \num{1.0 \pm 3.0} & \num{-2.7 \pm 3.3} & \num{-9.4 \pm 16.5} 	\\ 
	& 0.351 & \num{72.1 \pm 20.5} & \num{-79.4 \pm 17.8} & \num{46.9 \pm 7.5} & \num{15.3 \pm 22.5} & \num{16.2 \pm 16.0} & \num{-6.3 \pm 10.4} & \num{6.6 \pm 9.5} & \num{-21.8 \pm 39.7} 	\\  
	\hline
\multirow{5}{*}{$B$}
	& 0.117 & \num{-0.3 \pm 1.7} & \num{-0.1 \pm 1.3} & $\times$ & \num{3.4 \pm 2.7} & \num{1.8 \pm 1.8} & $\times$ & \num{-0.8 \pm 0.7} & \num{-2.1 \pm 2.7} 	\\ 
	& 0.234 & \num{0.9 \pm 3.4} & \num{0.3 \pm 3.1} & \num{-0.5 \pm 1.6} & \num{10.4 \pm 5.1} & \num{6.9 \pm 3.8} & \num{-0.4 \pm 1.0} & \num{-2.2 \pm 1.2} & \num{-9.1 \pm 5.3} 	\\ 
	& 0.351 & \num{13.6 \pm 3.8} & \num{0.3 \pm 4.1} & \num{1.3 \pm 3.9} & \num{24.4 \pm 7.4} & \num{2.8 \pm 4.8} & \num{-1.7 \pm 2.3} & \num{-3.0 \pm 1.9} & \num{-10.4 \pm 7.1} 	\\ 
	& 0.467 & \num{31.0 \pm 4.7} & \num{-4.8 \pm 4.6} & \num{8.0 \pm 3.6} & \num{28.5 \pm 9.7} & \num{13.2 \pm 5.1} & \num{1.8 \pm 3.3} & \num{-6.3 \pm 2.0} & \num{-17.0 \pm 11.2} \\ 
	& 0.583 & \num{74.1 \pm 9.0} & \num{-25.7 \pm 9.0} & \num{28.9 \pm 5.0} & \num{35.4 \pm 11.2} & \num{15.1 \pm 7.7} & \num{6.2 \pm 5.4} & \num{-8.3 \pm 4.4} & \num{-4.2 \pm 16.6} 	\\ 
	\hline
 \end{tabular} }

\vskip 0.6in
\centerline{
\begin{tabular}{l@{\quad}c@{\quad}S@{\quad}S@{\quad}S@{\quad}S@{\quad}S@{\quad}S@{\quad}S@{\quad}S}
	\hline
F6		& $\nu$	& $\MsTT$	& $\MsTTt$	& $\MsTTa$		&	$\MsTL$	& $\MsLT$	& $\MsTLa$	& $\MsTLt$	& $\MsLL$	 	 \\
	\hline
\multirow{3}{*}{$A$}	
	& 0.117 & \num{7.0 \pm 1.8} & \num{6.2 \pm 1.5} & $\times$ & \num{-0.9 \pm 3.4} & \num{-0.5 \pm 2.1} & $\times$ & \num{1.1 \pm 1.0} & \num{3.4 \pm 3.7} 	\\ 
	& 0.234 & \num{28.8 \pm 5.9} & \num{3.1 \pm 5.3} & \num{-0.2 \pm 2.1} & \num{15.0 \pm 9.6} & \num{5.1 \pm 6.3} & \num{-2.2 \pm 1.7} & \num{4.9 \pm 2.4} & \num{-8.2 \pm 9.5} 	\\ 
	& 0.351 & \num{77.2 \pm 8.9} & \num{-63.9 \pm 7.7} & \num{40.0 \pm 6.4} & \num{32.3 \pm 11.0} & \num{20.8 \pm 8.4} & \num{-1.0 \pm 4.3} & \num{-3.4 \pm 7.2} & \num{-19.2 \pm 17.3} \\
	\hline
\multirow{5}{*}{$B$}
	& 0.117 & \num{2.1 \pm 1.1} & \num{0.6 \pm 0.7} & $\times$ & \num{0.0 \pm 1.5} & \num{1.4 \pm 1.0} & $\times$ & \num{-0.1 \pm 0.2} & \num{0.9 \pm 1.7} 	\\ 
	& 0.234 & \num{8.8 \pm 2.3} & \num{0.3 \pm 1.7} & \num{-0.7 \pm 0.9} & \num{4.7 \pm 3.7} & \num{3.3 \pm 2.5} & \num{-0.5 \pm 0.7} & \num{0.1 \pm 0.8} & \num{3.1 \pm 4.4} 	\\ 
	& 0.351 & \num{18.2 \pm 3.5} & \num{-3.5 \pm 2.3} & \num{1.0 \pm 2.8} & \num{13.0 \pm 5.8} & \num{5.8 \pm 4.3} & \num{-1.0 \pm 1.8} & \num{-0.1 \pm 1.3} & \num{9.4 \pm 6.5} 	\\ 
	& 0.467 & \num{36.8 \pm 4.9} & \num{ 8.4 \pm 2.3} & \num{7.6 \pm 2.4} & \num{20.7 \pm 6.4} & \num{8.2 \pm 4.3} & \num{1.4 \pm 1.9} & \num{-1.3 \pm 1.5} & \num{0.3 \pm 7.3} 	\\ 
	& 0.583 & \num{79.3 \pm 6.1} & \num{-33.7 \pm 5.7} & \num{25.3 \pm 3.8} & \num{30.5 \pm 9.1} & \num{12.0 \pm 5.7} & \num{3.0 \pm 3.7} & \num{-5.4 \pm 3.8} & \num{2.3 \pm 11.4} 	\\
	\hline
 \end{tabular} }
\vskip 1.2in
\end{table}

\afterpage{

\begin{figure}[p]

	\begin{minipage}[c]{0.49\linewidth}
	\centering 
	\includegraphics*[width=0.99\linewidth]{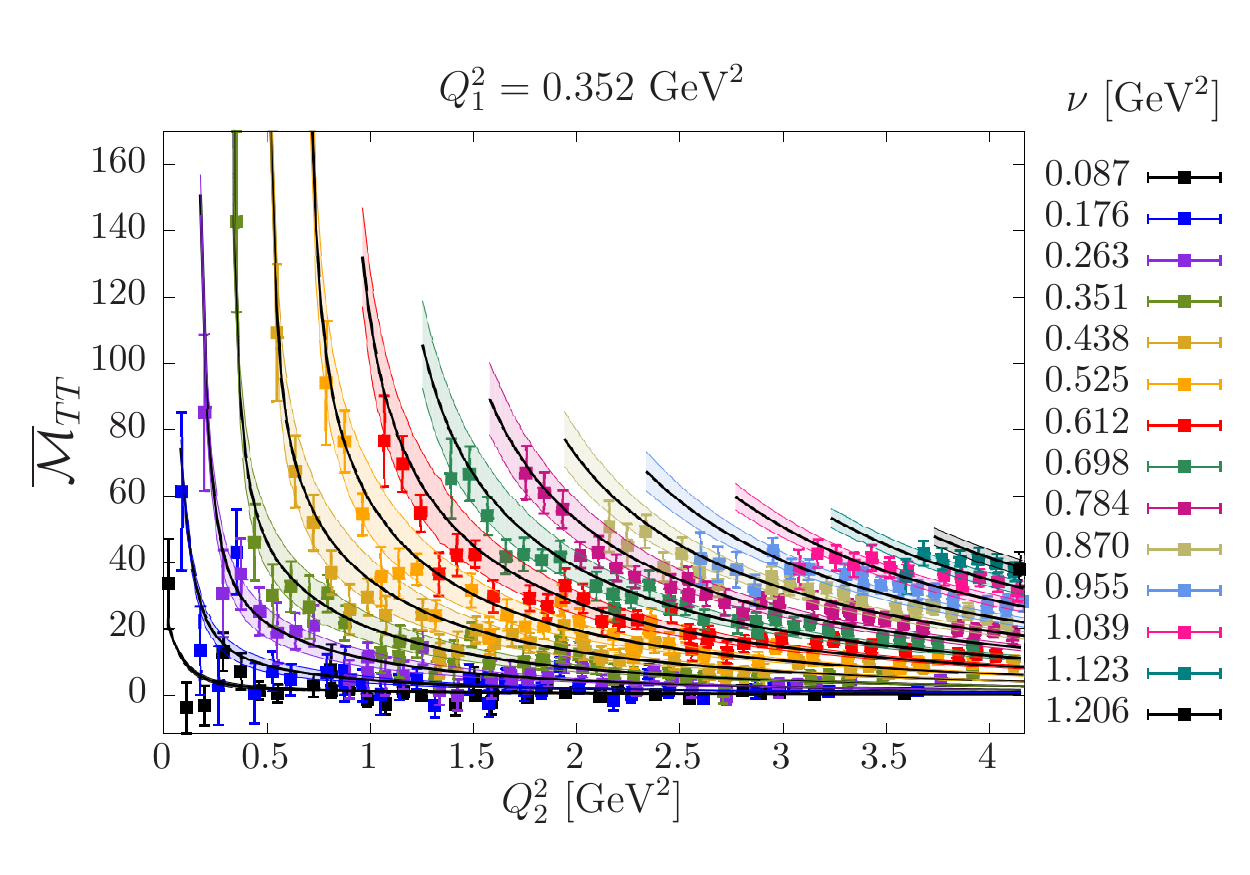}
	\end{minipage}
	\begin{minipage}[c]{0.49\linewidth}
	\centering 
	\includegraphics*[width=0.99\linewidth]{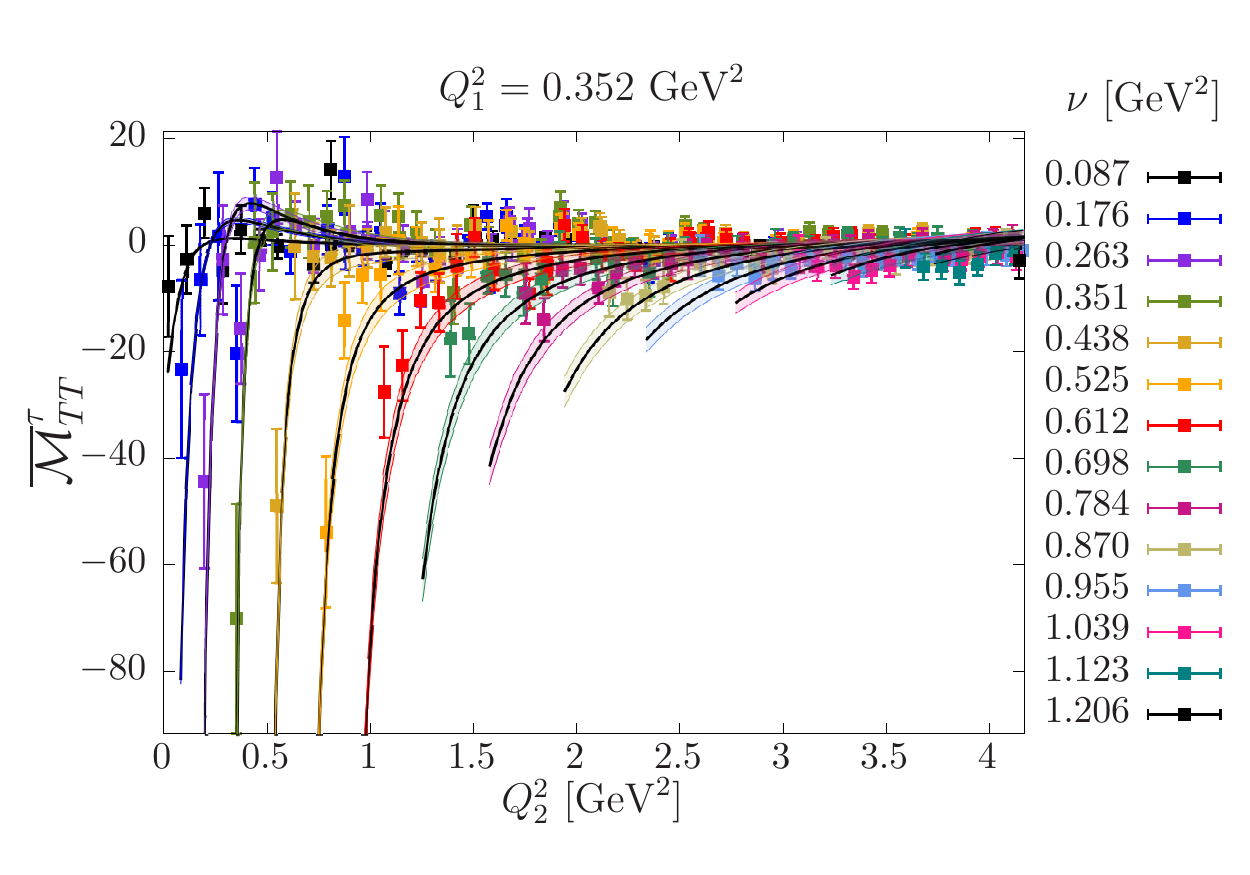}
	\end{minipage}
	
	\begin{minipage}[c]{0.49\linewidth}
	\centering 
	\includegraphics*[width=0.99\linewidth]{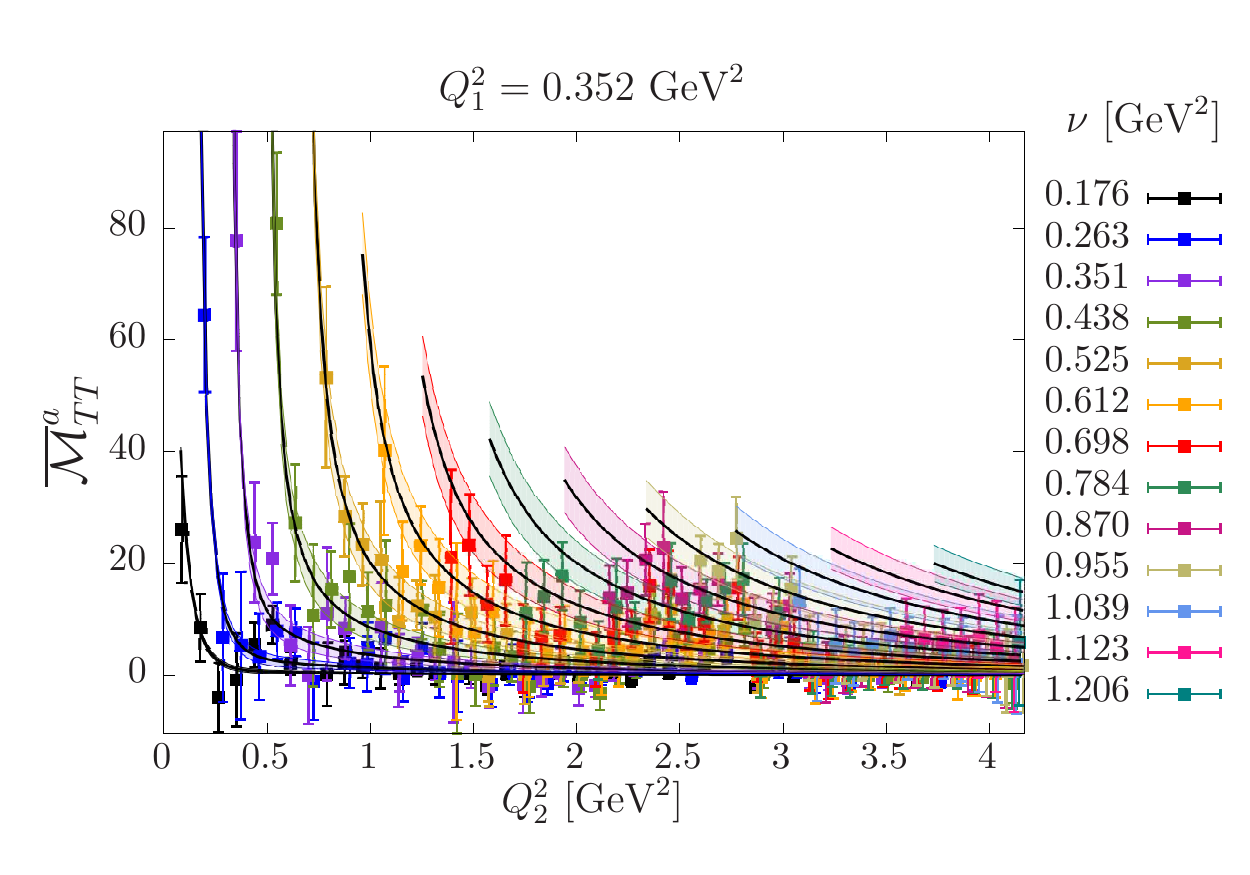}
	\end{minipage}
	\begin{minipage}[c]{0.49\linewidth}
	\centering 
	\includegraphics*[width=0.99\linewidth]{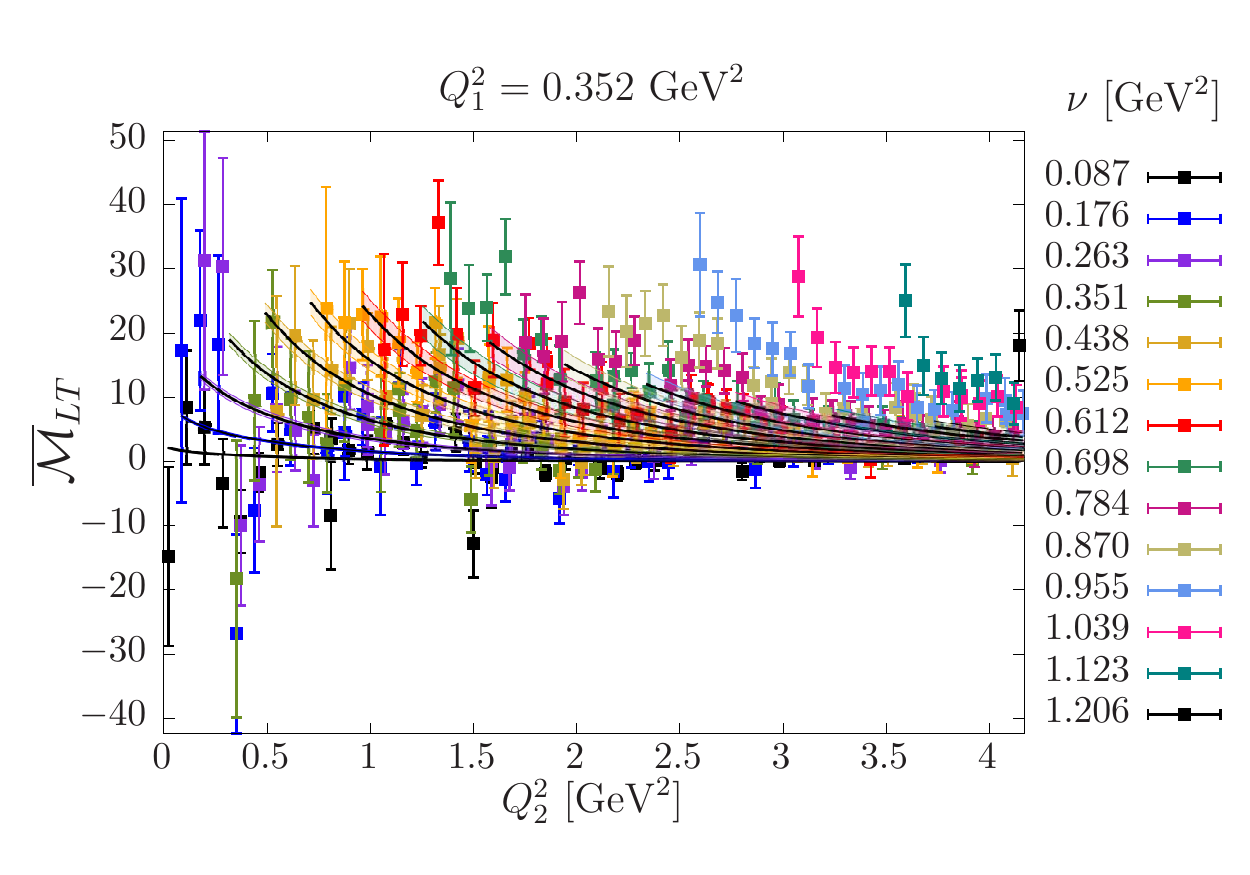}
	\end{minipage}
	
	\begin{minipage}[c]{0.49\linewidth}
	\centering 
	\includegraphics*[width=0.99\linewidth]{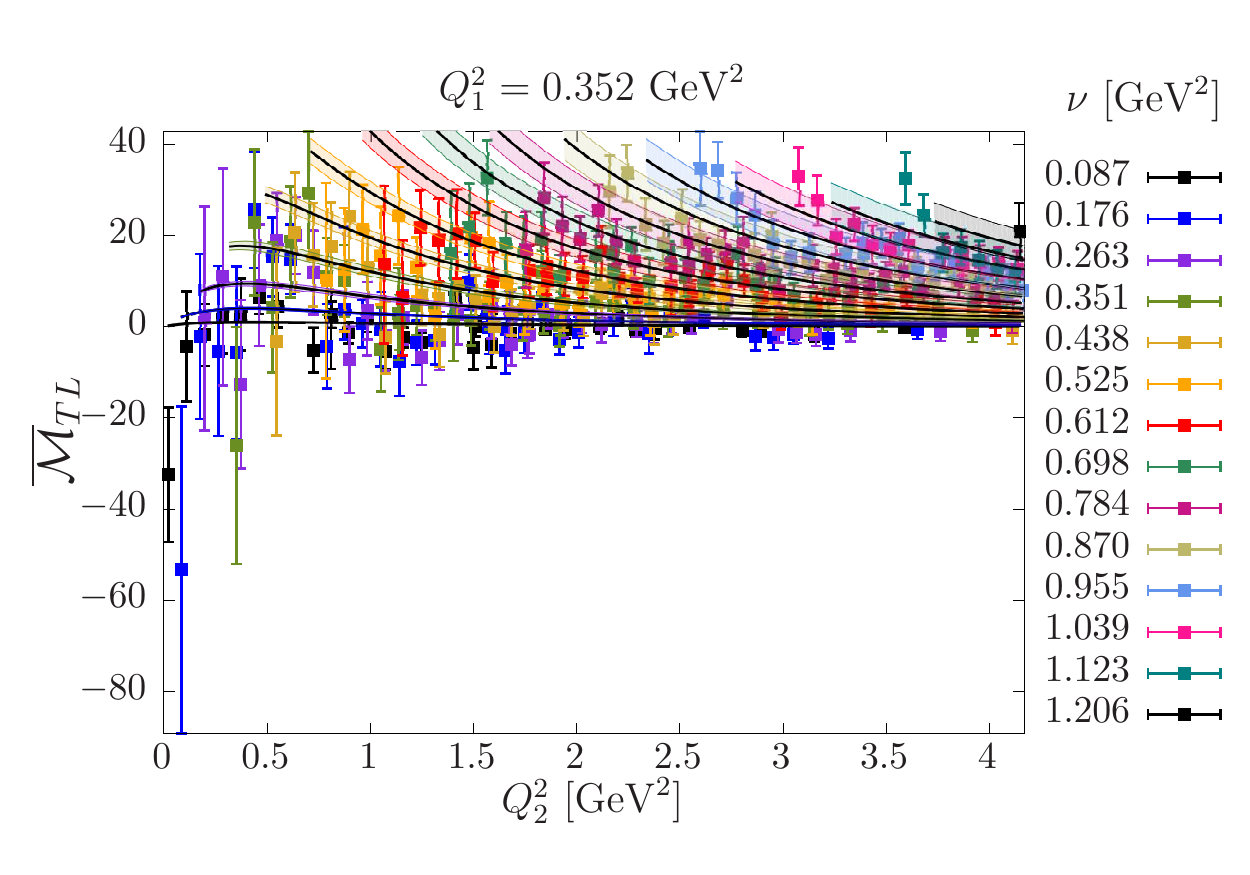}
	\end{minipage}
	\begin{minipage}[c]{0.49\linewidth}
	\centering 
	\includegraphics*[width=0.99\linewidth]{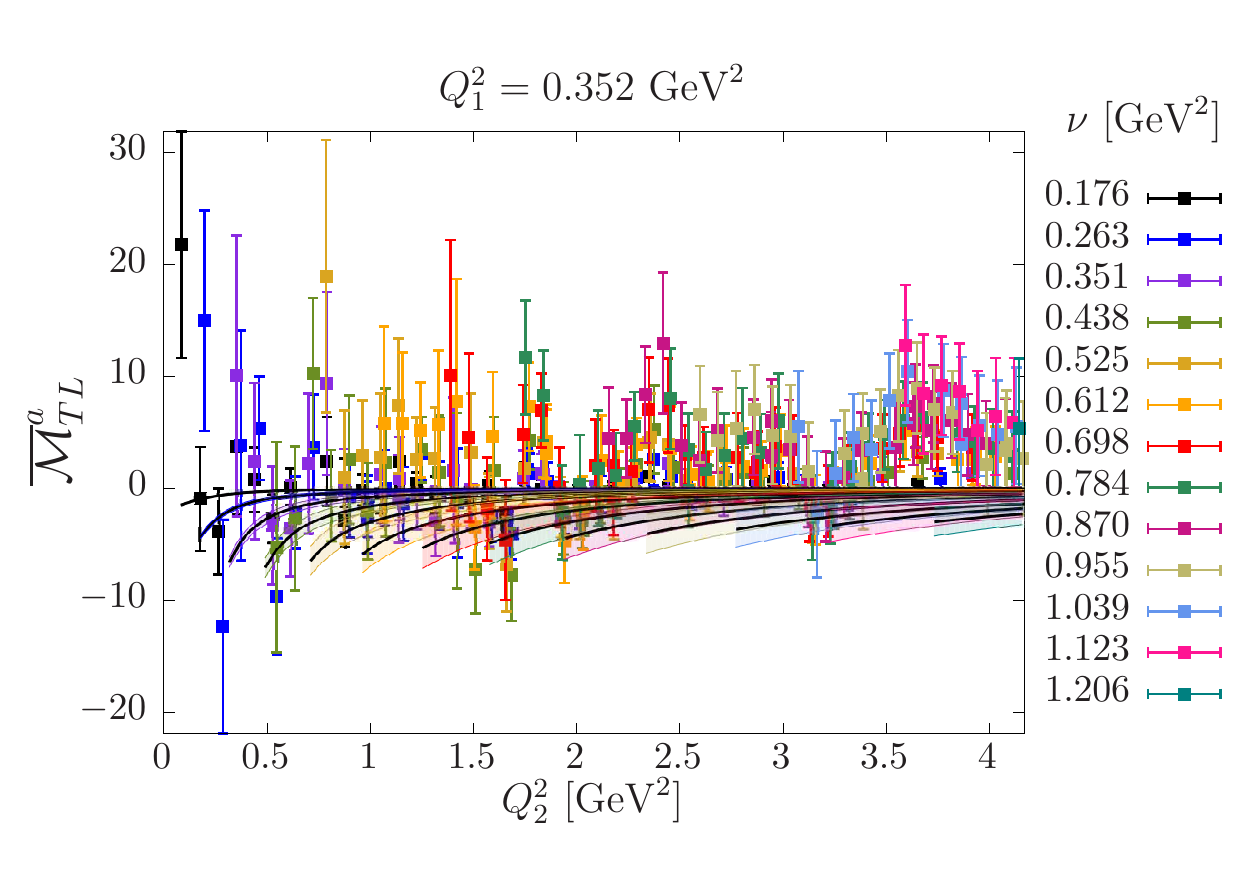}
	\end{minipage}
	
	\begin{minipage}[c]{0.49\linewidth}
	\centering 
	\includegraphics*[width=0.99\linewidth]{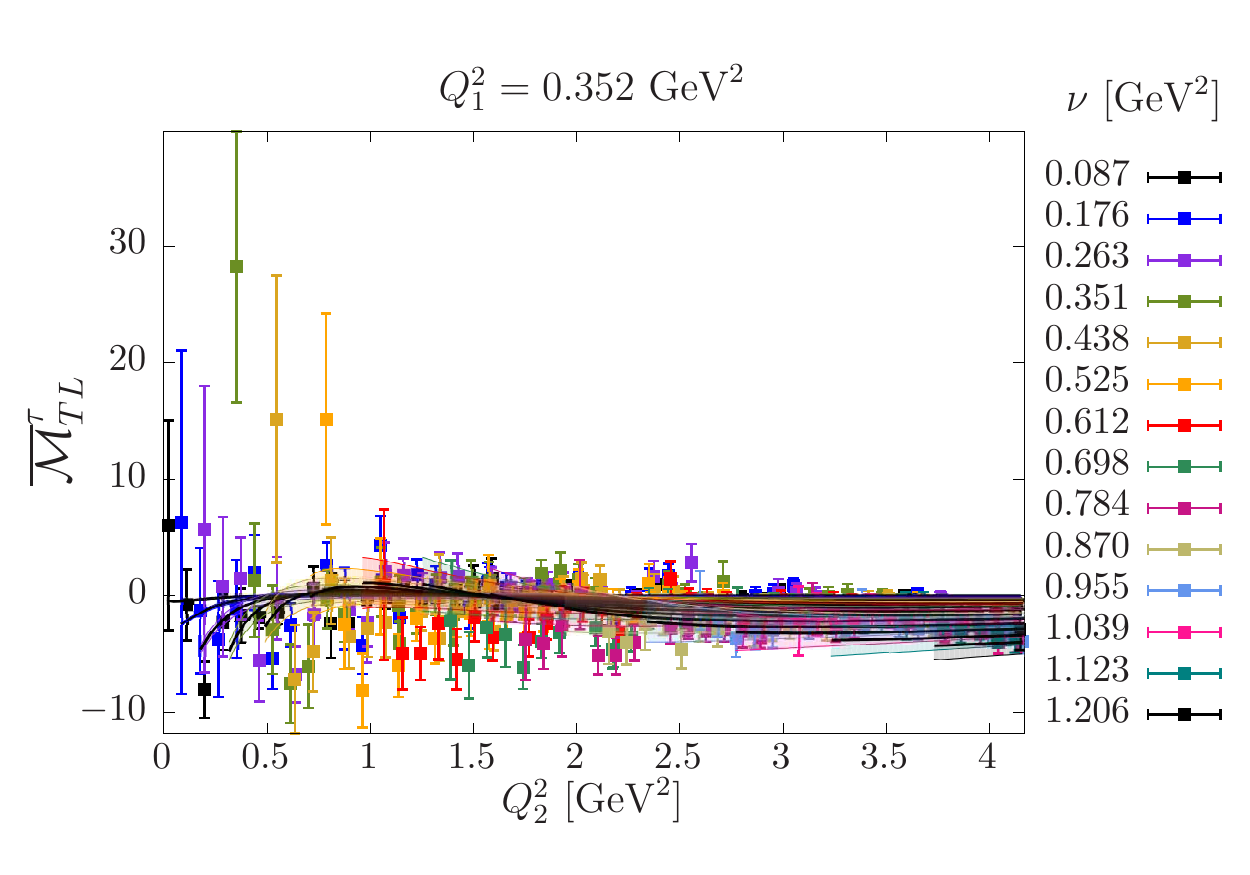}
	\end{minipage}
	\begin{minipage}[c]{0.49\linewidth}
	\centering 
	\includegraphics*[width=0.99\linewidth]{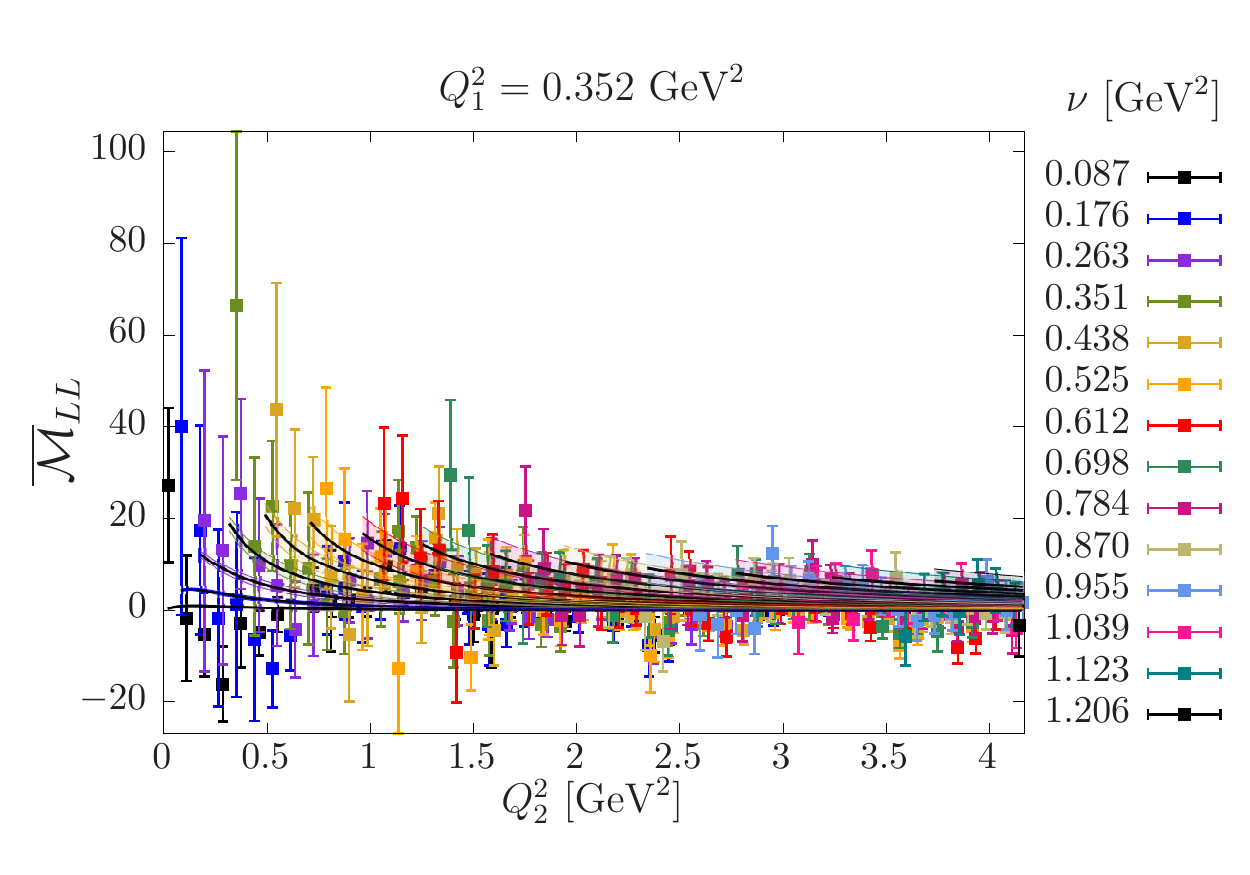}
	\end{minipage}

	\caption{The eight amplitudes $(\times10^6)$ for the ensemble G8 and for  $Q_1^2=0.352~\GeV^2$. 
   The curves with error-bands represent the fit results discussed in Sec.~\ref{sec:fit}.}	
	\label{fig:amps_G8_part1}
	
\end{figure}

\clearpage}

\bibliographystyle{utphys-noitalics} 
\bibliography{viscobib}

\providecommand{\href}[2]{#2}\begingroup\raggedright\begin{thebibliography}{10}

\bibitem{Heisenberg1936}
W.~Heisenberg and H.~Euler, ``Folgerungen aus der Diracschen Theorie des
  Positrons,'' \href{http://dx.doi.org/10.1007/BF01343663}{Zeitschrift f{\"u}r
  Physik {\bfseries 98} no.~11, (Nov, 1936) 714--732}.
  \url{https://doi.org/10.1007/BF01343663}.

\bibitem{Karplus:1951uo}
R.~Karplus and M.~Neuman, ``The scattering of light by light,''
  \href{http://dx.doi.org/10.1103/PhysRev.83.776}{Phys. Rev. {\bfseries 83}
  no.~4, (1951) 776--784}.

\bibitem{Aaboud:2017bwk}
{\bfseries ATLAS} Collaboration, M.~Aaboud {\em et~al.}, ``{Evidence for
  light-by-light scattering in heavy-ion collisions with the ATLAS detector at
  the LHC},'' \href{http://dx.doi.org/10.1038/nphys4208}{Nature Phys. no.~13,
  (2017) 852},
\href{http://arxiv.org/abs/1702.01625}{{\ttfamily arXiv:1702.01625 [hep-ex]}}.
%%CITATION = ARXIV:1702.01625;%%.

\bibitem{Pascalutsa:2010sj}
V.~Pascalutsa and M.~Vanderhaeghen, ``{Sum rules for light-by-light
  scattering},''
  \href{http://dx.doi.org/10.1103/PhysRevLett.105.201603}{Phys.Rev.Lett.
  {\bfseries 105} (2010) 201603},
\href{http://arxiv.org/abs/1008.1088}{{\ttfamily arXiv:1008.1088 [hep-ph]}}.
%%CITATION = ARXIV:1008.1088;%%.

\bibitem{Pascalutsa:2012pr}
V.~Pascalutsa, V.~Pauk, and M.~Vanderhaeghen, ``{Light-by-light scattering sum
  rules constraining meson transition form factors},''
  \href{http://dx.doi.org/10.1103/PhysRevD.85.116001}{Phys.Rev. {\bfseries D85}
  (2012) 116001},
\href{http://arxiv.org/abs/1204.0740}{{\ttfamily arXiv:1204.0740 [hep-ph]}}.
%%CITATION = ARXIV:1204.0740;%%.

\bibitem{Danilkin:2016hnh}
I.~Danilkin and M.~Vanderhaeghen, ``{Light-by-light scattering sum rules in
  light of new data},''
  \href{http://dx.doi.org/10.1103/PhysRevD.95.014019}{Phys. Rev. {\bfseries
  D95} no.~1, (2017) 014019},
\href{http://arxiv.org/abs/1611.04646}{{\ttfamily arXiv:1611.04646 [hep-ph]}}.
%%CITATION = ARXIV:1611.04646;%%.

\bibitem{Dai:2017cvz}
L.-Y. Dai and M.~Pennington, ``{Pascalutsa-Vanderhaeghen light-by-light sum
  rule from photon-photon collisions},''
  \href{http://dx.doi.org/10.1103/PhysRevD.95.056007}{Phys. Rev. {\bfseries
  D95} no.~5, (2017) 056007},
\href{http://arxiv.org/abs/1701.04460}{{\ttfamily arXiv:1701.04460 [hep-ph]}}.
%%CITATION = ARXIV:1701.04460;%%.

\bibitem{Green:2015sra}
J.~Green, O.~Gryniuk, G.~von Hippel, H.~B. Meyer, and V.~Pascalutsa, ``{Lattice
  QCD calculation of hadronic light-by-light scattering},''
  \href{http://dx.doi.org/10.1103/PhysRevLett.115.222003}{Phys. Rev. Lett.
  {\bfseries 115} no.~22, (2015) 222003},
\href{http://arxiv.org/abs/1507.01577}{{\ttfamily arXiv:1507.01577 [hep-lat]}}.
%%CITATION = ARXIV:1507.01577;%%.

\bibitem{Blum:2013xva}
T.~Blum, A.~Denig, I.~Logashenko, E.~de~Rafael, B.~Lee~Roberts, {\em et~al.},
  ``{The Muon (g-2) Theory Value: Present and Future},''
\href{http://arxiv.org/abs/1311.2198}{{\ttfamily arXiv:1311.2198 [hep-ph]}}.
%%CITATION = ARXIV:1311.2198;%%.

\bibitem{Venanzoni:2014ixa}
{\bfseries Fermilab E989} Collaboration, G.~Venanzoni, ``{The New Muon (g-2)
  experiment at Fermilab},''
  \href{http://dx.doi.org/10.1016/j.nuclphysbps.2015.09.087}{Nucl. Part. Phys.
  Proc. {\bfseries 273-275} (2016) 584--588},
\href{http://arxiv.org/abs/1411.2555}{{\ttfamily arXiv:1411.2555
  [physics.ins-det]}}.
%%CITATION = ARXIV:1411.2555;%%.

\bibitem{Otani:2015lra}
{\bfseries E34} Collaboration, M.~Otani, ``{Design of the J-PARC MUSE H-line
  for the Muon g-2/EDM Experiment at J-PARC (E34)},''
\href{http://dx.doi.org/10.7566/JPSCP.8.025010}{JPS Conf. Proc. {\bfseries 8}
  (2015) 025010}.
%%CITATION = INSPIRE-1400449;%%.

\bibitem{Colangelo:2014dfa}
G.~Colangelo, M.~Hoferichter, M.~Procura, and P.~Stoffer, ``{Dispersive
  approach to hadronic light-by-light scattering},''
  \href{http://dx.doi.org/10.1007/JHEP09(2014)091}{JHEP {\bfseries 09} (2014)
  091},
\href{http://arxiv.org/abs/1402.7081}{{\ttfamily arXiv:1402.7081 [hep-ph]}}.
%%CITATION = ARXIV:1402.7081;%%.

\bibitem{Colangelo:2014pva}
G.~Colangelo, M.~Hoferichter, B.~Kubis, M.~Procura, and P.~Stoffer, ``{Towards
  a data-driven analysis of hadronic light-by-light scattering},''
  \href{http://dx.doi.org/10.1016/j.physletb.2014.09.021}{Phys.Lett. {\bfseries
  B738} (2014) 6--12},
\href{http://arxiv.org/abs/1408.2517}{{\ttfamily arXiv:1408.2517 [hep-ph]}}.
%%CITATION = ARXIV:1408.2517;%%.

\bibitem{Colangelo:2015ama}
G.~Colangelo, M.~Hoferichter, M.~Procura, and P.~Stoffer, ``{Dispersion
  relation for hadronic light-by-light scattering: theoretical foundations},''
  \href{http://dx.doi.org/10.1007/JHEP09(2015)074}{JHEP {\bfseries 09} (2015)
  074},
\href{http://arxiv.org/abs/1506.01386}{{\ttfamily arXiv:1506.01386 [hep-ph]}}.
%%CITATION = ARXIV:1506.01386;%%.

\bibitem{Pauk:2014rfa}
V.~Pauk and M.~Vanderhaeghen, ``{Anomalous magnetic moment of the muon in a
  dispersive approach},''
  \href{http://dx.doi.org/10.1103/PhysRevD.90.113012}{Phys.Rev. {\bfseries D90}
  no.~11, (2014) 113012},
\href{http://arxiv.org/abs/1409.0819}{{\ttfamily arXiv:1409.0819 [hep-ph]}}.
%%CITATION = ARXIV:1409.0819;%%.

\bibitem{Hagelstein:2017obr}
F.~Hagelstein and V.~Pascalutsa, ``{Dissecting the hadronic contributions to
  $(g-2)_\mu $ by Schwinger's sum rule},''
\href{http://arxiv.org/abs/1710.04571}{{\ttfamily arXiv:1710.04571 [hep-ph]}}.
%%CITATION = ARXIV:1710.04571;%%.

\bibitem{Blum:2014oka}
T.~Blum, S.~Chowdhury, M.~Hayakawa, and T.~Izubuchi, ``{Hadronic light-by-light
  scattering contribution to the muon anomalous magnetic moment from lattice
  QCD},''
  \href{http://dx.doi.org/10.1103/PhysRevLett.114.012001}{Phys.Rev.Lett.
  {\bfseries 114} no.~1, (2015) 012001},
\href{http://arxiv.org/abs/1407.2923}{{\ttfamily arXiv:1407.2923 [hep-lat]}}.
%%CITATION = ARXIV:1407.2923;%%.

\bibitem{Blum:2015gfa}
T.~Blum, N.~Christ, M.~Hayakawa, T.~Izubuchi, L.~Jin, and C.~Lehner, ``{Lattice
  Calculation of Hadronic Light-by-Light Contribution to the Muon Anomalous
  Magnetic Moment},'' \href{http://dx.doi.org/10.1103/PhysRevD.93.014503}{Phys.
  Rev. {\bfseries D93} no.~1, (2016) 014503},
\href{http://arxiv.org/abs/1510.07100}{{\ttfamily arXiv:1510.07100 [hep-lat]}}.
%%CITATION = ARXIV:1510.07100;%%.

\bibitem{Blum:2016lnc}
T.~Blum, N.~Christ, M.~Hayakawa, T.~Izubuchi, L.~Jin, C.~Jung, and C.~Lehner,
  ``{Connected and Leading Disconnected Hadronic Light-by-Light Contribution to
  the Muon Anomalous Magnetic Moment with a Physical Pion Mass},''
  \href{http://dx.doi.org/10.1103/PhysRevLett.118.022005}{Phys. Rev. Lett.
  {\bfseries 118} no.~2, (2017) 022005},
\href{http://arxiv.org/abs/1610.04603}{{\ttfamily arXiv:1610.04603 [hep-lat]}}.
%%CITATION = ARXIV:1610.04603;%%.

\bibitem{Green:2015mva}
J.~Green, N.~Asmussen, O.~Gryniuk, G.~von Hippel, H.~B. Meyer, A.~Nyffeler, and
  V.~Pascalutsa, ``{Direct calculation of hadronic light-by-light
  scattering},''
\newblock PoS LATTICE {\bf 2015} (2016) 109,
% in {\em {Proceedings, 33rd International Symposium on Lattice   Field Theory (Lattice 2015)}}.
% \newblock 2015.
\newblock \href{http://arxiv.org/abs/1510.08384}{{\ttfamily arXiv:1510.08384
  [hep-lat]}}.
% \newblock
% \url{http://inspirehep.net/record/1401230/files/arXiv:1510.08384.pdf}.
% \newblock
%%CITATION = ARXIV:1510.08384;%%.

\bibitem{Asmussen:2016lse}
N.~Asmussen, J.~Green, H.~B. Meyer, and A.~Nyffeler, ``{Position-space approach
  to hadronic light-by-light scattering in the muon $g-2$ on the lattice},''
\newblock PoS LATTICE {\bf 2016} (2016) 164,
\href{http://arxiv.org/abs/1609.08454}{{\ttfamily arXiv:1609.08454 [hep-lat]}}.
%%CITATION = ARXIV:1609.08454;%%.

\bibitem{Asmussen:2017bup}
N.~Asmussen, A.~G\'erardin, H.~B. Meyer, and A.~Nyffeler, ``{Exploratory
  studies for the position-space approach to hadronic light-by-light scattering
  in the muon $g-2$},''
\href{http://arxiv.org/abs/1711.02466}{{\ttfamily arXiv:1711.02466 [hep-lat]}}.
%%CITATION = ARXIV:1711.02466;%%.

\bibitem{Nyffeler:2017ohp}
A.~Nyffeler, ``{Hadronic light-by-light scattering in the muon g-2},'' in {\em
  {International Workshop on e+e- Collisions from Phi to Psi (PHIPSI17) Mainz,
  Germany, June 26-29, 2017}}.
\newblock 2017.
\newblock \href{http://arxiv.org/abs/1710.09742}{{\ttfamily arXiv:1710.09742
  [hep-ph]}}.
% \newblock
% \url{http://inspirehep.net/record/1632784/files/arXiv:1710.09742.pdf}.
% \newblock
%%CITATION = ARXIV:1710.09742;%%.

\bibitem{Gerardin:2016cqj}
A.~G\'erardin, H.~B. Meyer, and A.~Nyffeler, ``{Lattice calculation of the pion
  transition form factor $\pi^0 \to \gamma^* \gamma^*$},''
  \href{http://dx.doi.org/10.1103/PhysRevD.94.074507}{Phys. Rev. {\bfseries
  D94} no.~7, (2016) 074507},
\href{http://arxiv.org/abs/1607.08174}{{\ttfamily arXiv:1607.08174 [hep-lat]}}.
%%CITATION = ARXIV:1607.08174;%%.

\bibitem{Bijnens:2015jqa}
J.~Bijnens, ``{Hadronic light-by-light contribution to $a_\mu$: extended
  Nambu-Jona-Lasinio, chiral quark models and chiral Lagrangians},''
  \href{http://dx.doi.org/10.1051/epjconf/201611801002}{EPJ Web Conf.
  {\bfseries 118} (2016) 01002},
\href{http://arxiv.org/abs/1510.05796}{{\ttfamily arXiv:1510.05796 [hep-ph]}}.
%%CITATION = ARXIV:1510.05796;%%.

\bibitem{Bijnens:2016hgx}
J.~Bijnens and J.~Relefors, ``{Pion light-by-light contributions to the muon
  $g-2$},'' \href{http://dx.doi.org/10.1007/JHEP09(2016)113}{JHEP {\bfseries
  09} (2016) 113},
\href{http://arxiv.org/abs/1608.01454}{{\ttfamily arXiv:1608.01454 [hep-ph]}}.
%%CITATION = ARXIV:1608.01454;%%.

\bibitem{Jin:2016rmu}
L.~Jin, T.~Blum, N.~Christ, M.~Hayakawa, T.~Izubuchi, C.~Jung, and C.~Lehner,
  ``{The connected and leading disconnected diagrams of the hadronic
  light-by-light contribution to muon $g - 2$},'' PoS {\bfseries LATTICE2016}
  (2016) 181,
\href{http://arxiv.org/abs/1611.08685}{{\ttfamily arXiv:1611.08685 [hep-lat]}}.
%%CITATION = ARXIV:1611.08685;%%.

\bibitem{Gerasimov:1973ja}
S.~B. Gerasimov and J.~Moulin, ``{Check of Sum Rules for Photon Interaction
  Cross-Sections in Quantum Electrodynamics and Mesodynamics},''
  \href{http://dx.doi.org/10.1016/0550-3213(75)90438-1}{Yad. Fiz. {\bfseries
  23} (1976) 142--153}.
[Nucl. Phys. {\bfseries B98}, 349 (1975)].
%%CITATION = YAFIA,23,142;%%.

\bibitem{Budnev:1971sz}
V.~Budnev, V.~Chernyak, and I.~Ginzburg, ``{Kinematics of $\gamma \gamma$
  scattering},''
\href{http://dx.doi.org/10.1016/0550-3213(71)90340-3}{Nucl.Phys. {\bfseries
  B34} (1971) 470--476}.
%%CITATION = NUPHA,B34,470;%%.

\bibitem{Budnev:1974de}
V.~Budnev, I.~Ginzburg, G.~Meledin, and V.~Serbo, ``{The Two photon particle
  production mechanism. Physical problems. Applications. Equivalent photon
  approximation},''
\href{http://dx.doi.org/10.1016/0370-1573(75)90009-5}{Phys.Rept. {\bfseries 15}
  (1975) 181--281}.
%%CITATION = PRPLC,15,181;%%.

\bibitem{Ji:2001wha}
X.~Ji and C.~Jung, ``{Studying hadronic structure of the photon in lattice
  QCD},'' \href{http://dx.doi.org/10.1103/PhysRevLett.86.208}{Phys.Rev.Lett.
  {\bfseries 86} (2001) 208},
\href{http://arxiv.org/abs/hep-lat/0101014}{{\ttfamily arXiv:hep-lat/0101014
  [hep-lat]}}.
%%CITATION = HEP-LAT/0101014;%%.

\bibitem{Gulpers:2013uca}
V.~G\"ulpers, G.~von Hippel, and H.~Wittig, ``{Scalar pion form factor in
  two-flavor lattice QCD},''
  \href{http://dx.doi.org/10.1103/PhysRevD.89.094503}{Phys. Rev. {\bfseries
  D89} no.~9, (2014) 094503},
\href{http://arxiv.org/abs/1309.2104}{{\ttfamily arXiv:1309.2104 [hep-lat]}}.
%%CITATION = ARXIV:1309.2104;%%.

\bibitem{Green:2015wqa}
J.~Green, S.~Meinel, M.~Engelhardt, S.~Krieg, J.~Laeuchli, J.~Negele,
  K.~Orginos, A.~Pochinsky, and S.~Syritsyn, ``{High-precision calculation of
  the strange nucleon electromagnetic form factors},''
  \href{http://dx.doi.org/10.1103/PhysRevD.92.031501}{Phys. Rev. {\bfseries
  D92} no.~3, (2015) 031501},
\href{http://arxiv.org/abs/1505.01803}{{\ttfamily arXiv:1505.01803 [hep-lat]}}.
%%CITATION = ARXIV:1505.01803;%%.

\bibitem{Blum:2015you}
T.~Blum, P.~A. Boyle, T.~Izubuchi, L.~Jin, A.~Juettner, C.~Lehner, K.~Maltman,
  M.~Marinkovic, A.~Portelli, and M.~Spraggs, ``{Calculation of the hadronic
  vacuum polarization disconnected contribution to the muon anomalous magnetic
  moment},'' \href{http://dx.doi.org/10.1103/PhysRevLett.116.232002}{Phys. Rev.
  Lett. {\bfseries 116} no.~23, (2016) 232002},
\href{http://arxiv.org/abs/1512.09054}{{\ttfamily arXiv:1512.09054 [hep-lat]}}.
%%CITATION = ARXIV:1512.09054;%%.

\bibitem{DellaMorte:2017dyu}
M.~Della~Morte, A.~Francis, V.~G\"ulpers, G.~Herdo\'iza, G.~von Hippel, H.~Horch,
  B.~J\"ager, H.~B. Meyer, A.~Nyffeler, and H.~Wittig, ``{The hadronic vacuum
  polarization contribution to the muon $g-2$ from lattice QCD},''
  \href{http://dx.doi.org/10.1007/JHEP10(2017)020}{JHEP {\bfseries 10} (2017)
  020},
\href{http://arxiv.org/abs/1705.01775}{{\ttfamily arXiv:1705.01775 [hep-lat]}}.
%%CITATION = ARXIV:1705.01775;%%.

\bibitem{Chetyrkin:1994js}
K.~G. Chetyrkin, J.~H. Kuhn, and A.~Kwiatkowski, ``{QCD corrections to the
  $e^{+} e^{-}$ cross-section and the $Z$ boson decay rate},''
Phys.\ Rept.\  {\bf 277} (1996) 189,
  \href{http://arxiv.org/abs/hep-ph/9503396}{{\ttfamily arXiv:hep-ph/9503396
  [hep-ph]}}.
% [Phys. Rept.277,189(1996)].
%%CITATION = HEP-PH/9503396;%%.

\bibitem{Olive:2016xmw}
{\bfseries Particle Data Group} Collaboration, C.~Patrignani {\em et~al.},
  ``{Review of Particle Physics},''
\href{http://dx.doi.org/10.1088/1674-1137/40/10/100001}{Chin. Phys. {\bfseries
  C40} no.~10, (2016) 100001}.
%%CITATION = CHPHD,C40,100001;%%.

\bibitem{Larin:2010kq}
{\bfseries PrimEx} Collaboration, I.~Larin {\em et~al.}, ``{A New Measurement
  of the $\pi^0$ Radiative Decay Width},''
  \href{http://dx.doi.org/10.1103/PhysRevLett.106.162303}{Phys. Rev. Lett.
  {\bfseries 106} (2011) 162303},
\href{http://arxiv.org/abs/1009.1681}{{\ttfamily arXiv:1009.1681 [nucl-ex]}}.
%%CITATION = ARXIV:1009.1681;%%.

\bibitem{Boucaud:2008xu}
{\bfseries ETM} Collaboration, P.~Boucaud {\em et~al.}, ``{Dynamical Twisted
  Mass Fermions with Light Quarks: Simulation and Analysis Details},''
  \href{http://dx.doi.org/10.1016/j.cpc.2008.06.013}{Comput. Phys. Commun.
  {\bfseries 179} (2008) 695--715},
\href{http://arxiv.org/abs/0803.0224}{{\ttfamily arXiv:0803.0224 [hep-lat]}}.
%%CITATION = ARXIV:0803.0224;%%.

\bibitem{Foster:1998vw}
{\bfseries UKQCD} Collaboration, M.~Foster and C.~Michael, ``{Quark mass
  dependence of hadron masses from lattice QCD},''
  \href{http://dx.doi.org/10.1103/PhysRevD.59.074503}{Phys. Rev. D {\bfseries
  59} (1999) 074503},
\href{http://arxiv.org/abs/hep-lat/9810021}{{\ttfamily arXiv:hep-lat/9810021
  [hep-lat]}}.
%%CITATION = HEP-LAT/9810021;%%.

\bibitem{Wilcox:1999ab}
W.~Wilcox, \href{http://dx.doi.org/10.1007/978-3-642-58333-9_10}{``{Noise
  methods for flavor singlet quantities},''} in {\em Numerical Challenges in
  Lattice Quantum Chromodynamics}, A.~Frommer, T.~Lippert, B.~Medeke, and
  K.~Schilling, eds., vol.~15 of {\em Lecture Notes in Computational Science
  and Engineering}, pp.~127--141.
\newblock Springer Berlin Heidelberg, 2000.
\newblock
\href{http://arxiv.org/abs/hep-lat/9911013}{{\ttfamily arXiv:hep-lat/9911013}}.
\newblock
%%CITATION = HEP-LAT/9911013;%%.

\bibitem{Foley:2005ac}
J.~Foley, K.~Jimmy~Juge, A.~O'Cais, M.~Peardon, S.~M. Ryan, {\em et~al.},
  ``{Practical all-to-all propagators for lattice QCD},''
  \href{http://dx.doi.org/10.1016/j.cpc.2005.06.008}{Comput.Phys.Commun.
  {\bfseries 172} (2005) 145--162},
  \href{http://arxiv.org/abs/hep-lat/0505023}{{\ttfamily arXiv:hep-lat/0505023
  [hep-lat]}}.

\bibitem{Stathopoulos:2013aci}
A.~Stathopoulos, J.~Laeuchli, and K.~Orginos, ``{Hierarchical probing for
  estimating the trace of the matrix inverse on toroidal lattices},''
  \href{http://dx.doi.org/10.1137/120881452}{SIAM J. Sci. Comput. {\bfseries
  35(5)} (2013) S299--S322},
\href{http://arxiv.org/abs/1302.4018}{{\ttfamily arXiv:1302.4018 [hep-lat]}}.
%%CITATION = ARXIV:1302.4018;%%.

\bibitem{Fritzsch:2012wq}
P.~Fritzsch, F.~Knechtli, B.~Leder, M.~Marinkovic, S.~Schaefer, {\em et~al.},
  ``{The strange quark mass and Lambda parameter of two flavor QCD},''
  \href{http://dx.doi.org/10.1016/j.nuclphysb.2012.07.026}{Nucl.Phys.
  {\bfseries B865} (2012) 397--429},
\href{http://arxiv.org/abs/1205.5380}{{\ttfamily arXiv:1205.5380 [hep-lat]}}.
%%CITATION = ARXIV:1205.5380;%%.

\bibitem{Wilson:1974sk}
K.~G. Wilson, ``{Confinement of quarks},''
\href{http://dx.doi.org/10.1103/PhysRevD.10.2445}{Phys. Rev. {\bfseries D10}
  (1974) 2445--2459}.
%%CITATION = PHRVA,D10,2445;%%.

\bibitem{Sheikholeslami:1985ij}
B.~Sheikholeslami and R.~Wohlert, ``{Improved Continuum Limit Lattice Action
  for QCD with Wilson Fermions},''
\href{http://dx.doi.org/10.1016/0550-3213(85)90002-1}{Nucl. Phys. {\bfseries
  B259} (1985) 572}.
%%CITATION = NUPHA,B259,572;%%.

\bibitem{Jansen:1998mx}
{\bfseries ALPHA} Collaboration, K.~Jansen and R.~Sommer, ``{O($a$)
  improvement of lattice QCD with two flavors of Wilson quarks},''
  \href{http://dx.doi.org/10.1016/S0550-3213(98)00396-4,
  10.1016/S0550-3213(02)00624-7}{Nucl. Phys. {\bfseries B530} (1998) 185--203},
  \href{http://arxiv.org/abs/hep-lat/9803017}{{\ttfamily arXiv:hep-lat/9803017
  [hep-lat]}}.
[Erratum: Nucl. Phys.B643,517(2002)].
%%CITATION = HEP-LAT/9803017;%%.

\bibitem{Blum:2012uh}
T.~Blum, T.~Izubuchi, and E.~Shintani, ``{New class of variance-reduction
  techniques using lattice symmetries},''
  \href{http://dx.doi.org/10.1103/PhysRevD.88.094503}{Phys. Rev. {\bfseries
  D88} no.~9, (2013) 094503},
\href{http://arxiv.org/abs/1208.4349}{{\ttfamily arXiv:1208.4349 [hep-lat]}}.
%%CITATION = ARXIV:1208.4349;%%.

\bibitem{Dai:2014zta}
  L.~Y.~Dai and M.~R.~Pennington,
  ``Comprehensive amplitude analysis of $\gamma\gamma \rightarrow \pi^+\pi^-, \pi^0\pi^0$ and $\overline{K} K$ below 1.5 GeV,''
  \href{http://dx.doi.org/10.1103/PhysRevD.90.036004}{Phys.\ Rev.\ {\bf D90}  no.~3, (2014) 036004},
  % doi:10.1103/PhysRevD.90.036004
  \href{http://arxiv.org/abs/1404.7524}{{\ttfamily arXiv:1404.7524 [hep-ph]}}.
  %%CITATION = doi:10.1103/PhysRevD.90.036004;%%

% \bibitem{Danilkin:2016hnh}
%   I.~Danilkin and M.~Vanderhaeghen,
%   ``Light-by-light scattering sum rules in light of new data,''
%   \href{http://dx.doi.org/10.1103/PhysRevD.95.014019}{Phys.\ Rev.\ {\bf D95}  no.1, (2017) 014019},
%  doi:10.1103/PhysRevD.95.014019
%   \href{http://arxiv.org/abs/1611.04646}{{\ttfamily arXiv:1611.04646 [hep-ph]}}.
  %%CITATION = doi:10.1103/PhysRevD.95.014019;%%

\bibitem{Behrend:1990sr}
{\bfseries CELLO} Collaboration, H.~J. Behrend {\em et~al.}, ``{A Measurement
  of the $\pi^0$, $\eta$ and $\eta'$ electromagnetic form-factors},''
\href{http://dx.doi.org/10.1007/BF01549692}{Z. Phys. {\bfseries C49} (1991)
  401--410}.
%%CITATION = ZEPYA,C49,401;%%.

\bibitem{Gronberg:1997fj}
{\bfseries CLEO} Collaboration, J.~Gronberg {\em et~al.}, ``{Measurements of
  the meson-photon transition form-factors of light pseudoscalar mesons at
  large momentum transfer},''
  \href{http://dx.doi.org/10.1103/PhysRevD.57.33}{Phys. Rev. {\bfseries D57}
  (1998) 33--54},
\href{http://arxiv.org/abs/hep-ex/9707031}{{\ttfamily arXiv:hep-ex/9707031
  [hep-ex]}}.
%%CITATION = HEP-EX/9707031;%%.

\bibitem{Aubert:2009mc}
{\bfseries BaBar} Collaboration, B.~Aubert {\em et~al.}, ``{Measurement of the
  $\gamma \gamma^* \to \pi^0$ transition form factor},''
  \href{http://dx.doi.org/10.1103/PhysRevD.80.052002}{Phys. Rev. {\bfseries
  D80} (2009) 052002},
\href{http://arxiv.org/abs/0905.4778}{{\ttfamily arXiv:0905.4778 [hep-ex]}}.
%%CITATION = ARXIV:0905.4778;%%.

\bibitem{Uehara:2012ag}
{\bfseries Belle} Collaboration, S.~Uehara {\em et~al.}, ``{Measurement of
  $\gamma \gamma^* \to \pi^0$ transition form factor at Belle},''
  \href{http://dx.doi.org/10.1103/PhysRevD.86.092007}{Phys. Rev. {\bfseries
  D86} (2012) 092007},
\href{http://arxiv.org/abs/1205.3249}{{\ttfamily arXiv:1205.3249 [hep-ex]}}.
%%CITATION = ARXIV:1205.3249;%%.

\bibitem{Lepage:1979zb}
G.~P. Lepage and S.~J. Brodsky, ``{Exclusive Processes in Quantum
  Chromodynamics: Evolution Equations for Hadronic Wave Functions and the
  Form-Factors of Mesons},''
\href{http://dx.doi.org/10.1016/0370-2693(79)90554-9}{Phys. Lett. {\bfseries
  87B} (1979) 359--365}.
%%CITATION = PHLTA,87B,359;%%.

\bibitem{Lepage:1980fj}
G.~P. Lepage and S.~J. Brodsky, ``{Exclusive Processes in Perturbative Quantum
  Chromodynamics},''
\href{http://dx.doi.org/10.1103/PhysRevD.22.2157}{Phys. Rev. {\bfseries D22}
  (1980) 2157}.
%%CITATION = PHRVA,D22,2157;%%.

\bibitem{Brodsky:1981rp}
S.~J. Brodsky and G.~P. Lepage, ``{Large Angle Two Photon Exclusive Channels in
  Quantum Chromodynamics},''
\href{http://dx.doi.org/10.1103/PhysRevD.24.1808}{Phys. Rev. {\bfseries D24}
  (1981) 1808}.
%%CITATION = PHRVA,D24,1808;%%.

\bibitem{Nesterenko:1982dn}
V.~A. Nesterenko and A.~V. Radyushkin, ``{Comparison of the {QCD} Sum Rule
  Approach and Perturbative {QCD} Analysis for $\gamma^* \gamma^* \to \pi^0$
  Process},'' Sov. J. Nucl. Phys. {\bfseries 38} (1983) 284.
[Yad. Fiz.38,476(1983)].
%%CITATION = SJNCA,38,284;%%.

\bibitem{Novikov:1983jt}
V.~A. Novikov, M.~A. Shifman, A.~I. Vainshtein, M.~B. Voloshin, and V.~I.
  Zakharov, ``{Use and Misuse of QCD Sum Rules, Factorization and Related
  Topics},''
\href{http://dx.doi.org/10.1016/0550-3213(84)90006-3}{Nucl. Phys. {\bfseries
  B237} (1984) 525--552}.
%%CITATION = NUPHA,B237,525;%%.

\bibitem{Masuda:2015yoh}
{\bfseries Belle} Collaboration, M.~Masuda {\em et~al.}, ``{Study of $\pi^0$
  pair production in single-tag two-photon collisions},''
  \href{http://dx.doi.org/10.1103/PhysRevD.93.032003}{Phys. Rev. {\bfseries
  D93} no.~3, (2016) 032003},
\href{http://arxiv.org/abs/1508.06757}{{\ttfamily arXiv:1508.06757 [hep-ex]}}.
%%CITATION = ARXIV:1508.06757;%%.

\bibitem{Achard:2001uu}
{\bfseries L3} Collaboration, P.~Achard {\em et~al.}, ``{$f_1(1285)$ formation
  in two photon collisions at LEP},''
  \href{http://dx.doi.org/10.1016/S0370-2693(01)01477-0}{Phys. Lett. {\bfseries
  B526} (2002) 269--277},
\href{http://arxiv.org/abs/hep-ex/0110073}{{\ttfamily arXiv:hep-ex/0110073
  [hep-ex]}}.
%%CITATION = HEP-EX/0110073;%%.

\bibitem{Achard:2007hm}
{\bfseries L3} Collaboration, P.~Achard {\em et~al.}, ``{Study of resonance
  formation in the mass region 1400-MeV to 1500-MeV through the reaction $\gamma
  \gamma \to K_0(S) K^\pm \pi^\mp$},''
\href{http://dx.doi.org/10.1088/1126-6708/2007/03/018}{JHEP {\bfseries 03}
  (2007) 018}.
%%CITATION = JHEPA,0703,018;%%.

\bibitem{Dai:2014lza}
L.-Y. Dai and M.~R. Pennington, ``{Two photon couplings of the lightest
  isoscalars from BELLE data},''
  \href{http://dx.doi.org/10.1016/j.physletb.2014.07.005}{Phys. Lett.
  {\bfseries B736} (2014) 11--15},
\href{http://arxiv.org/abs/1403.7514}{{\ttfamily arXiv:1403.7514 [hep-ph]}}.
%%CITATION = ARXIV:1403.7514;%%.

\bibitem{Minuit}
 {http://lcgapp.cern.ch/project/cls/work-packages/mathlibs/minuit/index.html}
  (2006) .

\bibitem{GSL}
 {https://www.gnu.org/software/gsl/} (2017) .

\bibitem{Landau:1948kw}
L.~D. Landau, ``{On the angular momentum of a system of two photons},''
\href{http://dx.doi.org/10.1016/B978-0-08-010586-4.50070-5}{Dokl. Akad. Nauk
  Ser. Fiz. {\bfseries 60} no.~2, (1948) 207--209}.
%%CITATION = DANKA,60,207;%%.

\bibitem{Yang:1950rg}
C.-N. Yang, ``{Selection Rules for the Dematerialization of a Particle Into Two
  Photons},''
\href{http://dx.doi.org/10.1103/PhysRev.77.242}{Phys. Rev. {\bfseries 77}
  (1950) 242--245}.
%%CITATION = PHRVA,77,242;%%.

\bibitem{Nyffeler:2016gnb}
A.~Nyffeler, ``{Precision of a data-driven estimate of hadronic light-by-light
  scattering in the muon $g-2$: Pseudoscalar-pole contribution},''
  \href{http://dx.doi.org/10.1103/PhysRevD.94.053006}{Phys. Rev. {\bfseries
  D94} no.~5, (2016) 053006},
\href{http://arxiv.org/abs/1602.03398}{{\ttfamily arXiv:1602.03398 [hep-ph]}}.
%%CITATION = ARXIV:1602.03398;%%.

\bibitem{Jegerlehner:2015stw}
F.~Jegerlehner, ``{Leading-order hadronic contribution to the electron and muon
  $g-2$},'' \href{http://dx.doi.org/10.1051/epjconf/201611801016}{EPJ Web Conf.
  {\bfseries 118} (2016) 01016},
\href{http://arxiv.org/abs/1511.04473}{{\ttfamily arXiv:1511.04473 [hep-ph]}}.
%%CITATION = ARXIV:1511.04473;%%.

\bibitem{Edwards:2004sx}
{\bfseries SciDAC, LHPC, UKQCD} Collaboration, R.~G. Edwards and B.~Jo\'o, ``{The
  Chroma software system for lattice QCD},''
  \href{http://dx.doi.org/10.1016/j.nuclphysbps.2004.11.254}{Nucl. Phys. Proc.
  Suppl. {\bfseries 140} (2005) 832},
  \href{http://arxiv.org/abs/hep-lat/0409003}{{\ttfamily arXiv:hep-lat/0409003
  [hep-lat]}}.
% [,832(2004)].
%%CITATION = HEP-LAT/0409003;%%.

\bibitem{CLScode21}
 {http://luscher.web.cern.ch/luscher/openQCD} (2013) .

\end{thebibliography}\endgroup

%---------------
\end{document}